\documentclass[a4paper,11pt]{article}
\pdfoutput=1 

\usepackage[table]{xcolor}
\usepackage{lmodern}
\usepackage{jcappub} 
\usepackage[toc,page]{appendix}
\usepackage{subcaption}
\usepackage{mathtools}
\usepackage{orcidlink}
\usepackage[T1]{fontenc} 
\usepackage{bm}
\usepackage{enumitem}
\usepackage[normalem]{ulem}

\newcommand{\semibold}[1]{{\fontseries{b}\selectfont{#1}}}

\newcommand{\para}[1]{\par\vspace{2mm}\noindent\semibold{{#1.}---}\ignorespaces} 

\DeclareMathOperator{\Or}{O}

\renewcommand{\leq}{\leqslant}
\renewcommand{\geq}{\geqslant}

\definecolor{SussexCobaltBlue}{HTML}{1d4289}
\definecolor{SussexDeepAquamarine}{HTML}{007a78}
\definecolor{SussexPowderBlue}{HTML}{7da1c4}
\definecolor{SussexCornYellow}{HTML}{f2c75c}
\definecolor{SussexChinaRose}{HTML}{be84a3}
\definecolor{SussexBurntOrange}{HTML}{dc582a}
\definecolor{SussexGrape}{HTML}{59315F}
\definecolor{SussexDarkMagenta}{HTML}{AC145A}
\definecolor{SussexVividRed}{HTML}{A6192E}
\definecolor{SussexVibrantGreen}{HTML}{007A33}
\definecolor{SussexMidBlue}{HTML}{006BA6}

\hypersetup{colorlinks=true,citecolor=SussexVibrantGreen,linkcolor=SussexMidBlue,urlcolor=SussexVividRed}

\newcommand{\etal}{\emph{et al.}}

\renewcommand{\d}{\mathrm{d}}
\newcommand{\e}[1]{\mathrm{e}^{{#1}}}
\newcommand{\im}{\mathrm{i}}
\newcommand{\Mp}{M_{\mathrm{P}}}

\newcommand{\vect}[1]{\bm{\mathrm{{#1}}}}
\newcommand{\divider}{\;|\;}
\newcommand{\transpose}{\mathsf{T}}
\newcommand{\connected}{c}

\newcommand{\dimP}{\mathcal{P}}

\newcommand{\Nstar}{N_{\star}}
\newcommand{\NstarDet}{N_{\star}^{\text{det}}}

\newcommand{\GammaStar}{\Gamma_{\star}}
\newcommand{\sstar}{s_{\star}}
\newcommand{\NKramers}{N_{\text{Kramers}}}

\newcommand{\Ninit}{N_0}
\newcommand{\phiinit}{\phi_0}
\newcommand{\phistart}{\phiinit}
\newcommand{\phistop}{\phi_1}
\newcommand{\Nstart}{\Ninit}
\newcommand{\Nstop}{N_1}

\newcommand{\phiend}{\phi_{\text{end}}}

\newcommand{\phiuv}{\phi_{\text{\textsc{uv}}}}

\newcommand{\phiplus}{\phi_+}
\newcommand{\phiminus}{\phi_-}
\newcommand{\phiplusminus}{\phi_{\pm}}
\newcommand{\momplus}{\mom_+}
\newcommand{\momminus}{\mom_-}
\newcommand{\momplusminus}{\mom_{\pm}}

\newcommand{\ProbP}{\bm{\mathsf{P}}}
\newcommand{\ProbW}{\bm{\mathsf{W}}}
\newcommand{\ProbQ}{\bm{\mathsf{Q}}}
\newcommand{\ProbZ}{\bm{\mathsf{Z}}}

\newcommand{\Expect}{\bm{\mathsf{E}}}
\newcommand{\Survival}{\bm{\mathsf{S}}}
\newcommand{\Current}{\bm{\mathsf{J}}}
\newcommand{\ProbPre}{\ProbP'}
\newcommand{\CurrentRe}{\Current'}
\newcommand{\ExpectRe}{\Expect'}

\newcommand{\primedint}{\int^{\prime}}

\newcommand{\Normalization}{\mathcal{N}}
\newcommand{\NormalizationRe}{\Normalization'}

\newcommand{\LFP}{\mathcal{L}}
\newcommand{\LFPadj}{\LFP^\dag}

\newcommand{\Hamiltonian}{\mathcal{H}}
\newcommand{\HFP}{\Hamiltonian_{\text{FP}}}
\newcommand{\Xmom}{X}
\newcommand{\XEmom}{\mathsf{P}}
\newcommand{\mom}{\mathfrak{p}}

\newcommand{\SMSR}{S_{\text{MSR}}}
\newcommand{\Seff}{S_{\text{eff}}}
\newcommand{\SIR}{S_{\text{IR}}}

\newcommand{\phiSP}{\phi^{\star}}
\newcommand{\XEmomSP}{\XEmom^{\star}}

\newcommand{\FeynmanVernon}{\mathcal{F}}
\newcommand{\SpatialVolume}{\mathcal{V}}

\newcommand{\ir}{\text{IR}}
\newcommand{\vt}[1]{{\small \bm{#1}}} 

\newcommand{\cl}[1]{#1_{\text{cl}}}
\newcommand{\qu}[1]{#1_{\text{q}}}
\newcommand{\D}{D} 
\newcommand{\qa}{\varphi_1} 
\newcommand{\qb}{\varphi_2} 
\newcommand{\pa}{\XEmom_1} 
\newcommand{\pb}{\XEmom_2} 
\newcommand{\pq}{\qu{\phi}^{_\ir}}
\newcommand{\ppq}{\qu{\pi}^{_\ir}}
\newcommand{\pc}{\cl{\phi}^{_\ir}}

\newcommand{\HFPtilde}{\HFP^{\text{SP}}}
\newcommand{\phiasm}{\phi_{\infty}}
\newcommand{\Nend}{\NstarDet}
\newcommand{\epc}{\sigma} 
\newcommand{\pef}{\XEmom_\text{eff}}

\DeclareMathOperator{\sinc}{sinc}
\DeclareMathOperator{\csch}{csch}
\DeclareMathOperator{\Erfc}{erfc}

\title{Stochastic instantons and the tail of the inflationary density perturbation}

\author{Jaime Calder\'{o}n-Figueroa\,$^{\text{\orcidlink{0000-0002-9929-7603}}}$,}
\author{David Seery\,$^{\text{\orcidlink{0000-0003-3421-6080}}}$}

\affiliation{Astronomy Centre, University of Sussex,
Falmer, Brighton, BN1 9QH, UK}

\emailAdd{jrc43@sussex.ac.uk}
\emailAdd{D.Seery@sussex.ac.uk}

\abstract{
    In the ``stochastic $\delta N$ formalism'',
    the statistics
    of the inflationary
    density perturbation
    are obtained from the
    first passage distribution
    of a stochastic process.
    We develop a general framework in which to evaluate
    the rare tail of this distribution, based on
    an instanton approximation
    to a path integral representation
    for the transition probability.
    We relate our formalism to the Schwinger--Keldysh
    path integral,
    by integrating out short wavelength degrees of
    freedom to produce an influence functional.
    This provides a principled
    way to extend the calculation beyond the
    slow-roll limit,
    and to models with multiple fields.
    We argue that our framework has a number of
    advantages in comparison with existing methods.
    In particular,
    it reliably captures the tail
    behaviour
    in cases
    where existing techniques do not apply,
    including cases where the noise amplitude
    has strong time dependence.
    We demonstrate the method by computing the
    tail probability in a number of scenarios,
    including
    a beyond-slow-roll analysis of a linear potential,
    ultra-slow-roll,
    and constant-roll inflation.
    We
    find close agreement with
    results already reported in the literature.
    Finally, we discuss a scenario with exponentially decaying
    noise amplitude.
    This is a model for the stochastic evolution
    of a fixed comoving volume of spacetime
    on superhorizon scales.
    In this case we show that the tail reverts to
    a Gaussian weight.
}

\begin{document}
\maketitle
\newpage

\section{Introduction}\label{sec:introduction}
Generation of relic density perturbations from
inflation is a key component of the standard
cosmological scenario~\cite{Mukhanov:1981xt, Mukhanov:1982nu}.
Since the early days of the inflationary paradigm,
it has been
understood
that,
for fluctuations that are not too extreme,
the distribution of
these relic perturbations is close to Gaussian.
In particular,
if $\zeta_k \sim \delta \rho / \rho$
is the amplitude of the density contrast
on some lengthscale $L$, where $k \sim 2\pi / L$,
then
\begin{equation}
    \label{eq:Gaussian-PDF}
    \ProbP(\zeta_k) \sim \exp
    \bigg(
        {-\frac{1}{2}}
        \frac{\zeta_k^2}{\sigma_k^2} 
    \bigg)\, ,
\end{equation}
where $\sigma^2_k$ is the corresponding variance.
Here, $\ProbP(\zeta_k) \, \d \zeta_k$
is the probability of finding
a perturbation with amplitude in the range
$\zeta_k$ to $\zeta_k + \d \zeta_k$.

Eq.~\eqref{eq:Gaussian-PDF}
characterizes $\ProbP(\zeta_k)$,
up to small deviations from Gaussianity,
near the centre of the distribution
where $|\zeta_k|/\sigma_k$ is $\Or(1)$.
It has recently become important to understand the behaviour
of $\ProbP(\zeta_k)$
in the tails,
where $|\zeta_k| \gg \sigma_k$.
A $|\zeta_k|$ of this magnitude
represents a much more extreme density
fluctuation.
Such extreme fluctuations are very rare,
but may have
important observational consequences.
One possibility is that a population of early collapsed
objects, such as primordial black holes,
may form on the positive overdensity
tail of the distribution.
A reliable estimate of their
abundance depends sensitively on
an accurate characterization of the
weight carried by this tail.

In principle,
$\ProbP(\zeta_k)$
can be reconstructed
from knowledge of
the
connected
correlation functions,
such as
$\langle \zeta_{\vect{k}_1} \zeta_{\vect{k}_2} \rangle_{\connected}$,
$\langle \zeta_{\vect{k}_1} \zeta_{\vect{k}_2} \zeta_{\vect{k}_3} \rangle_{\connected}$,
\ldots,
and so on.
One way to do so
is to write the probability density
as a Fourier transform
of the characteristic function,%
    \footnote{A related discussion of the role
    of the characteristic function in specifying the
    tail of the PDF
    was given by Ezquiaga {\etal}~\cite{Ezquiaga:2019ftu}.}
\begin{equation}
    \label{eq:pdf-from-fourier-characteristic}
    \ProbP(\zeta_k)
    =
    \int
    \frac{\d t_k}{(2\pi)^3} \;
    \chi(t_k) \, \e{-\im t_k \zeta_k} \,,
\end{equation}
where the characteristic function
$\chi(t_k)$ is defined by
\begin{equation}
    \label{eq:characteristic-function-def}
    \chi(t_k) = \Expect(\e{\im t_k \zeta_k})
    =
    \int \, \d \zeta_k \;
    \ProbP(\zeta_k) \, \e{\im t_k \zeta_k} \,,
\end{equation}
and $\Expect$ denotes an expectation value.
When $|t_k \zeta_k| \gg 1$,
the oscillating factor in~\eqref{eq:characteristic-function-def}
rapidly damps contributions.
Hence, for a fixed $t_k$,
$\chi(t_k)$
captures information about $\ProbP(\zeta_k)$
for $|\zeta_k| \lesssim |1/t_k|$.
Information about the tail of $\ProbP(\zeta_k)$
is therefore encoded in the behaviour of
$\chi(t_k)$
near the origin~\cite{phillips1980characteristic,Strawderman:2004},
which we can regard as being specified
by the coefficients in a Taylor expansion
of $\chi(t_k)$ around $t_k = 0$.
Conversely, the tail behaviour of the characteristic
function determines the smoothness of $\ProbP(\zeta_k)$.

Eq.~\eqref{eq:characteristic-function-def}
shows that $\ln \chi(t_k)$ is
a generating function for connected correlation functions.
Therefore,
we can (at least formally)
identify the required Taylor coefficients
with these correlators,
\begin{equation}
    \ln \chi(t_k)
    =
    -
    \frac{1}{2!}
    \langle \zeta_k \zeta_k \rangle_{\connected} t_k^2
    -
    \frac{\im}{3!}
    \langle \zeta_k \zeta_k \zeta_k \rangle_{\connected} t_k^3
    +
    \cdots \,,
    \label{eq:characteristic-fn-power-series}
\end{equation}
where we have assumed $\langle \zeta_k \rangle_{\connected} = 0$.
If the series in Eq.~\eqref{eq:characteristic-fn-power-series}
were convergent, $\ln \chi(t_k)$ would be well-described near the
origin using a few low-order
connected correlation functions, which can be
computed according to the rules of perturbative quantum field
theory.
Further,
the tail could be described in terms of these
low-order correlations.
In some cases this leads to a practical
method of computation.
For example, for a normal distribution $Y \sim N(0, \sigma^2)$,
only the second-order cumulant is present,
and $\chi(t) \sim \exp(-\sigma^2 t^2/2)$.
(We use the symbol $\sim$ to mean that we are dropping an overall
normalization constant.)
Analytically continuing to the moment generating function
$M(t) = \chi(-\im t)$
allows us to apply Chernoff's bound~\cite{Chernoff1952},
\begin{equation}
    \ProbP(Y \geq a)
    \leq
    \inf_{t \geq 0} M(t) \e{-ta}
    \sim
    \inf_{t \geq 0}
    \exp
    \bigg(
        \frac{1}{2} \sigma^2 t^2
        - t a
    \bigg) \;.
\end{equation}
Optimizing over $t$ then yields a tail estimate,
\begin{equation}
    \ProbP(Y \geq a)
    \sim
    \exp
    \bigg(
        {-\frac{1}{2}} \frac{a^2}{\sigma^2}
    \bigg) \,,
\end{equation}
again up to an overall normalization that we do not retain.
Examination of the steps in this computation shows
that it is effectively the same as the saddle point
evaluation of~\eqref{eq:pdf-from-fourier-characteristic}.

Unfortunately, Dyson observed long ago that
a diagrammatic series such as~\eqref{eq:characteristic-fn-power-series}
will typically not converge~\cite{PhysRev.85.631}.
Even if it did, with some finite radius of convergence,
the saddle point associated with~\eqref{eq:pdf-from-fourier-characteristic}
(or the location of the infimum used in the Chernoff bound)
grows
as $|\zeta_k| \rightarrow \infty$.
This causes two difficulties:
First,
the location of the saddle,
and hence the detailed
tail estimate,
will depend on a precise balance among a large
number of terms in the series expansion of
$\ln \chi(t_k)$.
To calculate
this location accurately
would require knowledge of very many
correlation functions.
Second,
even if we can obtain all
correlation functions exactly,
the saddle point will eventually move outside the
radius of convergence of the series.
When this happens we must abandon the Taylor
expansion~\eqref{eq:characteristic-fn-power-series}
and replace it with something else.
A third problem (not directly related to convergence of the
series)
is that correlation functions computed in
perturbation theory know nothing about
exponentially rare fluctuations.
Therefore, there would be no guarantee that the
delicate balance between correlation functions
that fixes the location for the saddle
for $|\zeta_k| \gtrsim 1$ would
be accurate.

These considerations show that beginning with
perturbative correlation
functions, and attempting
to reconstruct the probability density from them,
is not the right strategy.
Nevertheless, there has
recently been significant progress
in evaluating these tail probabilities
using more appropriate tools.
We briefly review these efforts
in~\S\ref{sec:methodoogy-review}.
In general, they are based on four
primary computational techniques,
all apparently distinct.
\semibold{First},
if formation occurs in a single event, by
nucleation of a large density perturbation at horizon
exit,
the tail probability can be estimated by a
quantum-mechanical instanton method.
This approach was studied by Celoria \etal~\cite{Celoria:2021vjw}. 
The same methodology was later applied to a resonant non-Gaussianity model 
by Creminelli \etal~\cite{Creminelli:2024cge} (see also
Ref.~\cite{Creminelli:2025tae}).

Alternatively,
the large fluctuation may instead be generated
by superposition of many smaller events,
corresponding to the ``noise''
generated by
an entire sequence of fluctuations
exiting the horizon over an
extended interval.
In this case,
the tail probability can be estimated
using the tools
of stochastic inflation,
originally suggested by Starobinsky~\cite{Starobinsky:1986fx}.
Pattison~\etal~\cite{Pattison:2017mbe,Pattison:2019hef,Pattison:2021oen},
Ezquiaga~\etal~\cite{Ezquiaga:2019ftu},
and
Animali \& Vennin~\cite{Animali:2022otk}
developed a
\semibold{second} major
approach based on the backward Kolmogorov equation,
sometimes called the adjoint Fokker--Planck equation.
They used the spectral theory of
the Fokker--Planck differential operator
to obtain asymptotic decay rates,
which they interpreted as controlling
the tail of the probability density.
We present this approach,
in the form used by Ezquiaga {\etal} but with some minor
refinements, in~\S\ref{sec:ezquiaga-spectral-method}.
Related spectral methods have been widely applied
to similar stochastic problems in
many areas of
physics.
For early developments,
see Refs.~\cite{tomita1976eigenvalue,van1977soluble,caroli1979diffusion}.

A \semibold{third}
approach was suggested by Tomberg~\cite{Tomberg:2022mkt,Tomberg:2023kli},
working directly with a stochastic Langevin
equation rather than the backward Kolmogorov equation.
In this method,
the probability of a rare fluctuation
is estimated
from the probability of the sequence of
noise events needed to assemble it.
We describe this approach
in~\S\ref{sec:tomberg-langevin-method}.
Tomberg was able to
obtain the distribution function for $\zeta$
from this functional
by a Jacobian change-of-variables formula.
In particular,
the method of
Ref.~\cite{Tomberg:2023kli}
has a number of interesting parallels
with the approach developed in this paper,
although it is not the same.
We will comment on these similarities
and differences as we proceed.

\semibold{Fourth}
(and finally)
any perturbation in $\zeta$
originates as a disturbance in the fields
that support the inflationary phase,
which we write generically as $\phi^\alpha$.
A number of authors have noted that,
typically, the nonlinear mapping
from $\phi^\alpha$ to $\zeta$
may already induce
heavy tails, even if
the parent $\phi^\alpha$ distribution
is Gaussian~\cite{Cai:2018dkf,Atal:2019cdz,Biagetti:2021eep,
Hooshangi:2022lao,Hooshangi:2021ubn,Pi:2022ysn,
Hooshangi:2023kss,Ballesteros:2024pwn}.
Under certain circumstances,
this might constitute a valid procedure to obtain
a tail estimate for $\zeta$,
even without knowledge of the tail behaviour of the parent
distribution.
This could happen if
$\zeta$ is sufficiently large to represent a
fluctuation that collapses to (for example)
a primordial black hole, but the corresponding
$\phi^\alpha$ configuration
still lies close to the central region of its distribution.
Alternatively,
it may be possible to superpose a number of individual
noise events to produce a large, aggregate
$\phi^\alpha$
fluctuation
with a \emph{known} Gaussian distribution.%
    \footnote{Clearly, it is always possible to superpose
    fluctuations in this way. The difficulty is that usually
    the distribution of the resulting fluctuation
    is not known. Indeed, usually, it is this distribution
    (or a related one) that we are attempting to compute.}
The more complicated stochastic frameworks
described by Ezquiaga {\etal} and Tomberg would
then be unnecessary.
Jackson {\etal}~\cite{Jackson:2024aoo}
have argued
that, generically, this cannot be possible.
However, Tomberg's
method can be interpreted
as a demonstration
that
this approach does work
in certain models,
provided that
the stochastic evolution of the $\phi^\alpha$
fluctuation is linear, and the noise
is not influenced by stochastic
fluctuations of the background~\cite{Tomberg:2023kli}.
For example, as shown in Ref.~\cite{Tomberg:2023kli},
these conditions are realized in certain
`constant-roll' models.
(See~\S\ref{sec:tomberg-langevin-method} and~\S\ref{sec:CR}
below.)

In general, however,
we expect these special conditions will not apply,
and
the distribution function for
such an effective $\phi^\alpha$
fluctuation
will depend on parameters describing its assembly history.
It must then be computed using one
of the more detailed
methods,
and may develop
nontrivial tail behaviour of its own.
Nevertheless, where it can be used,
this approach is especially simple.
We will refer to it as the `Jacobian'
method, since
(apart from an estimate for the distribution for the aggregate
$\phi^\alpha$ fluctuation)
it
requires only
computation of the Jacobian needed to change the probability
density from $\phi^\alpha$ to $\zeta$.

The relation between these approaches
has remained somewhat unclear.
Some connections can be made out.
Tomberg's formalism can be regarded as a
procedure to compute the
distribution for an aggregate $\phi^\alpha$
fluctuation, followed by a Jacobian
transformation~\cite{Tomberg:2023kli}.
We will study this method in~\S\ref{sec:tomberg-langevin-method}.
However, the relationship between
the Jacobian method and that of Ezquiaga {\etal}
is opaque.
There is also no clear
relationship between the methods of
Ezquiaga {\etal} and Tomberg,
even though both are based on stochastic formulations.
Finally, all of these approaches
stand apart from the quantum-mechanical
instanton method of Celoria {\etal},
which addresses a different problem.

In this paper, we introduce a new method
to compute tail estimates
based on the use of \emph{stochastic instantons}.
Such instantons have been widely applied
in many areas of physics,
including
chemical reactions,
turbulent fluid flow,
climate modelling,
and
soft condensed matter.
For recent reviews, see Refs.~\cite{Touchette_2009,Grafke_2015,Grafke_2019}.
We use the instanton technique
to characterize rare stochastic fluctuations,
so (in the form presented here) it
represents an alternative
to the second and third 
approaches described above.
We argue that it is more flexible
than either of these existing schemes,
and can be applied to a wider range
of scenarios,
while not
introducing significant
extra technical complexity.
However, because it is an instanton
method, the relation to the Celoria {\etal}
approach is simplified.
(Indeed, although we do not do so,
one could in principle use both approaches
together.)
We also explain how it
can be obtained from
the Schwinger--Keldysh path integral of non-equilibrium
quantum field theory.
This relation makes it possible, if desired,
to systematically incoporate microscopic non-Markovian
memory effects,
and
non-local contributions
from a quantum effective action.

\subsection*{Summary and outline}
\label{sec:summary-outline}
In order to clearly
explain
the relationship between
our approaches,
in~\S\ref{sec:methodoogy-review}
we review the formalisms of Ezquiaga {\etal}~\cite{Ezquiaga:2019ftu}
and Tomberg~\cite{Tomberg:2023kli}
in a unified notation and from the point of view adopted in the remainder
of this paper.
We also use this section to fix our notation for stochastic
inflation,
and to describe
the different species of
transition
probabilities associated with stochastic processes.
This section could be omitted by readers familiar with the
approaches
of
Ezquiaga {\etal} and Tomberg.
Such readers may wish to note, however,
that the quantity calculated by Ezquiaga {\etal} is the
restricted transition probability $\ProbPre$,
which is a density with respect to the field configuration.
However, the distribution required for
a stochastic $\delta N$ computation is
the first passage distribution
$\ProbQ$, which is a density with respect to the e-folding number.
In this paper we try to distinguish clearly between $\ProbPre$
and $\ProbQ$.

In \S\ref{sec:slow-roll-instanton}
we introduce the stochastic instanton
method
and explain
how it
reproduces
the tail estimate
for a slow-roll model with linear potential,
studied by Ezquiaga {\etal}~\cite{Ezquiaga:2019ftu}.
To do this we express the transition
probability $\ProbPre$
and first passage distribution
$\ProbQ$
in terms of a constrained path integral
of Martin--Siggia--Rose (MSR) type.
For rare transitions,
we argue that the path integral can be evaluated
using a saddle point approximation.
In principle this can be done for both the
restricted and unrestricted cases, although
(for reasons to be described later)
in this
paper
we prefer to evaluate an instanton for the unrestricted
transition probability, and relate it to its
restricted counterpart using other methods.
The field space trajectory corresponding to the
rare transition
is the instanton,
and
the response field of the MSR formalism encodes the
corresponding noise realization.
Evaluation of the MSR action on this instanton
produces the tail estimate.

In \S\ref{sec:instanton-phase}, we extend our analysis
beyond the
single-field, slow-roll
regime by building an MSR path integral
in phase space,
working directly from the Schwinger--Keldysh
formulation of non-equilibrium quantum field theory.
This is achieved by integrating out short-wavelength
degrees of freedom to obtain a
Feynman--Vernon influence functional.
In its simplest form this
reproduces the dynamics of the stochastic formalism,
as already shown by a number of authors.
The Schwinger--Keldysh framework
enables us to identify the MSR response fields
with the quantum components of the fields in the
Keldysh basis.
(A full derivation of the influence functional from
the Schwinger--Keldysh path integral is presented in
Appendix~\ref{sec:Inf_func}; see also the detailed
treatments by
Moss \& Rigopoulos~\cite{Moss:2016uix},
Collins {\etal}~\cite{Collins:2017haz},
Pinol {\etal}~\cite{Pinol:2020cdp},
and Andersen {\etal}~\cite{Andersen:2021lii}.)
From the resulting MSR effective action, we obtain a system of
differential equations whose solutions determine the
instanton trajectories underlying rare fluctuations.

In \S\ref{sec:Apps}, we illustrate the application of
this formalism to a number of cases. We begin by
generalizing the result for a linear potential,
obtained in~\S\ref{sec:slow-roll-instanton},
beyond the slow-roll approximation.
We then study ultra slow-roll (USR, \S\ref{sec:USR})
and constant-roll (CR, \S\ref{sec:CR})
scenarios, and obtain their corresponding
tail estimates.
Each of these scenarios
introduces qualitatively new effects, such as the
possibility of overshooting or second crossings.
In~\S\ref{sec:CR},
our results very closely reproduce those obtained
by Tomberg (\S\ref{sec:tomberg-langevin-method}),
but with interesting differences.
Further,
we are able to partially validate our
results by relating the model to
the well-studied Ornstein--Uhlenbeck process,
for which
asymptotic estimates of the
rare first-passage distributions
are
known for certain parameter combinations.
Finally, in \S\ref{sec:decaying-noise},
we evaluate the tail of
$\ProbQ$ for a model with exponentially decaying
noise amplitude. This is a proxy for the stochastic
evolution of
a spacetime region of fixed comoving volume
after it has
passed
outside the horizon.
On superhorizon scales, this volume will sample many
copies of the horizon, and hence average over many realizations
of the noise. If all these samples are independent,
a ``central limit theorem''-like
suppression
acts to exponentially damp
the noise.
In typical stochastic calculations,
evolution during this epoch is
ignored, because it is not expected to yield large
effects.
We confirm that a heavy tail does not form outside the horizon.
This conclusion agrees with numerical simulations
reported by Figueroa~{\etal}~\cite{Figueroa:2021zah}.

We conclude in~\S\ref{sec:conclusions}.
The main text is followed by two appendices.
In Appendix~\ref{sec:Inf_func}
we describe the computation of a
Feynman--Vernon influence functional by integrating
out short wavelength degrees of freedom.
This is needed to formulate a stochastic
theory for the long-wavelength degrees if freedom
in a Schwinger--Keldysh path integral.
In Appendix~\ref{sec:FPT_LT}
we describe
the use of Laplace transforms and the renewal equation
to determine $\ProbQ$ from $\ProbP$.

\subsection*{Notation and conventions}
We work in natural units where $c = \hbar = 1$.
The reduced Planck mass is defined by $\Mp = (8\pi G)^{-1/2}$,

\section{Review of existing formalisms}
\label{sec:methodoogy-review}
To estimate probabilities in the
tail of the distribution function, we must know
something about the formation mechanism
for the rare events that populate it.
Any such events will be highly unlikely.
Therefore,
typically,
the tail probability will be dominated by
the single \emph{least unlikely} mechanism.
There are two primary mechanisms
that can assemble extreme fluctuations.
First,
one can start with a large initial condition
generated at horizon exit,
followed by a typical
superhorizon history.
Alternatively, one may start with a number of
smaller
fluctuations
(although perhaps still large enough to be moderately rare),
generated by horizon exit
of successive scales,
which are assembled in an atypical
sequence to produce
a large final result.
A similar observation
was
made
by Hooshangi {\etal}~\cite{Hooshangi:2021ubn}.

The probability to synthesize a large fluctuation at horizon exit
was discussed by Celoria {\etal}~\cite{Celoria:2021vjw}
in the context of a specific single-field model dominated by a $\dot{\zeta}^4$
interaction.
They constructed a saddle-point
approximation to the Schwinger--Keldysh path integral,
along a Euclidean contour that interpolates
between past infinity
and the real-valued horizon exit time.
In this example, they estimated
the tail probability to scale like
$\ProbP(\zeta) \sim \exp\big( -\Or(1) \times \zeta^{3/2} \big)$.

\subsection{Stochastic inflation and transition probabilities}
\label{sec:stochastic-inflation}
In this paper we focus on the second formation channel,
in which many fluctuations are co-added
over a period of time.
In
specific examples, we will see that the asymptotic tail
produced by this mechanism is heavier than that found by
Celoria {\etal}, and therefore would dominate
for sufficiently large $\zeta$.

To evaluate the
probability of synthesizing a large fluctuation in this way,
we must understand how fluctuations produced by
many different modes combine.
This is far from trivial.
First, once outside the horizon,
large-scale modes
may continue to evolve.
This evolution
must be tracked accurately,
even if the amplitude of the perturbation becomes
large.
Second,
small-scale modes
exit into a geometry perturbed by
larger-scale modes,
which can modify their properties.
Third, the accumulation of many such small-scale
modes
can back-react onto the large-scale modes,
adjusting their amplitude.
In a traditional loop expansion
of correlation functions,
this back-reaction
process is described by calculations at 1-loop
level or higher~\cite{Iacconi:2023ggt}.
These have attracted considerable interest, but
remain
highly technical, and some details continue to be disputed.

A different approach was suggested by
Starobinsky~\cite{Starobinsky:1982ee,Starobinsky:1985ibc},
which
can accurately model the first and second
effects described above.
The third effect is important, but not immediately relevant to
the computations described in this paper.

After smoothing over small-scale substructure,
a superhorizon-sized patch of the universe
will be characterized by a homogeneous value for each
relevant field.
Neglecting back-reaction,
these values
can be taken as initial conditions for the
subsequent evolution,
selecting
one of the phase space trajectories
available to the homogeneous background.
If spatial gradients are not significant,
the subsequent evolution
of the patch will follow this trajectory
just as if it were an isolated ``separate'' universe.
This is the \emph{separate universe picture}.
It was later refined by a number of authors
\cite{Lyth:1984gv,Sasaki:1995aw,Wands:2000dp,Lyth:2003im,Lyth:2004gb,Lyth:2005fi,Rigopoulos:2005xx,Seery:2012vj}.
Starobinsky
chose to continuously adjust the smoothing scale
so that it is always
larger than the current horizon by a fixed factor.
Then, the effect of an emerging perturbation
will be to displace this superhorizon-scale
patch from one background trajectory to
another~\cite{Starobinsky:1986fx,Starobinsky:1994bd}.
Temporarily assuming the background can be described by a
single-field slow-roll solution, the
evolution will be described by a Langevin
equation~\cite{Starobinsky:1985ibc},
\begin{equation}
    \label{eq:Starobinsky-Langevin}
    \d \phi = - \frac{V'}{3H^2} \, \d N + \frac{H}{2\pi} \, \d \xi \,,
\end{equation}
where $\d \xi$ is a stochastic process with zero mean
and $\Expect(\d \xi)^2 = \d N$.

We will call~\eqref{eq:Starobinsky-Langevin} the
\emph{Starobinsky equation}.
In it,
the ``drift'' term proportional to $\d N$ represents
evolution along the existing selected trajectory.
The stochastic $\d \xi$ term represents displacement
to a different trajectory due to the emerging perturbation.
To make~\eqref{eq:Starobinsky-Langevin}
well-defined
we must specify its discretization.
In this paper we work in It\^{o}
discretization,
because this enables us to make
a connection with the Schwinger--Keldysh path integral,
to be described below.
For a discussion of the different discretization
schemes in the inflationary
context, see Pinol {\etal}~\cite{Pinol:2018euk,Pinol:2020cdp} and  Tomberg~\cite{Tomberg:2024evi}.

The generalization to multiple fields was
considered in Ref.~\cite{Starobinsky:1985ibc},
and also by
Salopek \& Bond~\cite{SalopekBond1990,SalopekBond1991}.
For recent discussions,
see Refs.~\cite{Vennin:2015hra,Assadullahi:2016gkk}.
The slow-roll approximation can be dropped by working in
a phase-space framework
and writing another
Langevin equation for the momentum
field $\pi$~\cite{Grain:2017dqa,Pattison:2019hef,Pattison:2021oen,Mishra:2023lhe}.
We describe this formalism
in~\S\ref{sec:tomberg-langevin-method}
and~\S\ref{sec:instanton-phase}
below.
(See also Appendix~\ref{sec:Lang_app} for a discussion of microphysical
properties of the noise in such cases.)
However,
the details do not significantly change the main elements
of the discussion.
Eq.~\eqref{eq:Starobinsky-Langevin}
is the basis for the tail estimates made
in Refs.~\cite{Pattison:2017mbe,Ezquiaga:2019ftu,Tomberg:2023kli}.

The Starobinsky equation~\eqref{eq:Starobinsky-Langevin}
describes the evolution of an isolated spatial patch,
neglecting coupling or correlations between patches.
Suppose the patch begins with some field value $\phiinit$
at time $\Ninit$.
After averaging over the stochastic process $\xi$,
we obtain a description of the field
distribution in an ensemble of such patches.
We write this distribution
$\ProbP(\phi, N \divider \phiinit, \Ninit)$,
abbreviated to
$\ProbP(\phi, \Delta N \divider \phiinit)$
if the absolute initial and final times
of the transition are unnecessary,
where $\Delta N = N - \Ninit$.
Note that $\ProbP$ is a density with respect to the final
field configuration $\d \phi$,
but \emph{not} with respect to $N$,
which is just a parameter
describing the transition.
$\ProbP$ satisfies the forward Kolmogorov equation,
also called the Fokker--Planck equation,
associated with Eq.~\eqref{eq:Starobinsky-Langevin},
\begin{equation}
    \label{eq:kolmogorov-forwards}
    \frac{\d \ProbP}{\d N}
    =
    \frac{\partial}{\partial \phi}
    \Bigg(
        \frac{V'}{3 H^2} \ProbP
    \Bigg)
    +
    \frac{\partial^2}{\partial \phi^2}
    \Bigg(
        \frac{H^2}{8\pi^2} \ProbP
    \Bigg)
    =
    \LFP \ProbP
    .
\end{equation}
On the right-hand side we have
summarized this combination
of gradients by
introducing
a second-order
differential operator $\LFP$.

It will be important to understand how
the forward Kolmogorov equation
transports probability
in field space.
It can be recast as an
equation expressing conservation of probability,
\begin{subequations}
\begin{equation}
    \label{eq:forward-komogorov-transport}
    \frac{\d \ProbP}{\d N} + \frac{\partial \Current}{\partial \phi} = 0 \,,
\end{equation}
where $\Current$ can be regarded as a probability current,
\begin{equation}
    \label{eq:probability-current}
    \Current \equiv
    -
    \frac{V'}{3H^2} \ProbP
    -
    \frac{\partial}{\partial \phi}
    \Bigg(
        \frac{H^2}{8\pi^2} \ProbP
    \Bigg) \;.
\end{equation}
\end{subequations}
If we are working in a larger state space,
for example in a scenario with multiple
scalar fields or
in a phase-space formulation,
$\Current$ should be promoted to a vector
in this larger space.
Letting $\alpha$ be an index in this space,
the probability flux across a boundary
$\partial B$
with unit normal $\hat{n}^\alpha$
and area element $\d A$
is given
by
$\int_{\partial B} \hat{n}^\alpha \Current_\alpha \, \d A$.
A specific example that will be important later
occurs when we work with the phase space formulation for a single
field. The state space is then labelled by the field and
momentum coordinates $(\phi, \pi)$,
and the boundary $\partial B$ corresponding to an end-of-inflation
surface at $\phi = \phiend$
is an infinite line with normal $\hat{n}^\alpha = (\pm 1, 0)$.
The sign is fixed by the orientation of the boundary.
The probability flux is
\begin{equation}
    \label{eq:flux-formula-marginalized}
    \text{flux}
    =
    \pm \int_{-\infty}^{\infty} \Current_\phi(\phiend, \pi) \, \d \pi ,
\end{equation}
corresponding to the
oriented $\phi$-component of $\Current_\alpha$
marginalized over the momentum $\pi$.

Eqs.~\eqref{eq:Starobinsky-Langevin}, \eqref{eq:kolmogorov-forwards}
and~\eqref{eq:forward-komogorov-transport}--\eqref{eq:probability-current}
constitute the basic apparatus of the stochastic approach to
inflation.
In particular, $\ProbP(\phi, N \divider \phiinit, \Ninit)$
provides a statistical
resolution to
the question of how
a sequence of fluctuations co-add
over an extended inflationary interval.

\subsubsection{Stochastic $\delta N$ formalism}
\label{sec:Est_tail}
To build observables,
the relevant quantity is not the
transition probability
but rather $\ProbP(\zeta)$.
When Eqs.~\eqref{eq:Starobinsky-Langevin}--\eqref{eq:kolmogorov-forwards}
are repurposed to estimate
this distribution,
Vennin \& Starobinsky~\cite{Vennin:2015hra}
introduced the term
``stochastic $\delta N$ formalism''.
We now briefly summarize how this can be
achieved.
The procedure was described in
Refs.~\cite{Vennin:2015hra,Pattison:2017mbe,Ezquiaga:2019ftu}.

Consider any patch of scale $L = 2\pi / k$,
and
write the smoothed fields interior to this patch as $\phi^\alpha$.
In general, the curvature perturbation $\zeta$
associated with this patch
is equal to
the perturbation $\delta N$ in the number of e-folds
that elapse 
(relative to the mean in a larger volume)
as the
fields
evolve from
a prescribed initial
configuration
$\phiinit^\alpha$,
up to a final
configuration of fixed energy density~\cite{Wands:2000dp,Lyth:2004gb}.
Depending on the balance between the noise and drift
contributions in Eq.~\eqref{eq:Starobinsky-Langevin},
during some parts of this evolution the fields may be dominated
by noiseless rolling, and in others they may be dominated by
quantum fluctuations.
In the remainder of this paper we focus on a single field
scenario for simplicity,
but
where appropriate we explain
how the analysis would
change in a multiple field scenario.
We will usually
assume that
the final configuration is determined by another fixed
field value $\phiend$.
In general this will be an oversimplification, but
this choice is convenient for analytic purposes.
If the final configuration corresponds
to a fixed energy density, it
will receive both kinetic and potential contributions.
Depending on the balance between them,
there might not be a unique termination
condition
but rather a range of possibilities.
Of course, if slow-roll applies at the final time then,
to a fair approximation,
the potential energy can be taken to dominate.

In conclusion,
we wish to determine the distribution of
the number of e-folds $\Nstar$
between $\phi = \phiinit$
and the time of first arrival
at some final
field configuration $\phiend$.%
    \footnote{We write this elapsed number of
    e-folds as $\Nstar$
    in order to indicate that it is distinct from $N$,
    which is simply the time coordinate in the problem.}
This distribution is described
as the
\emph{first-passage distribution}
and written $\ProbQ(\Nstar)$.
It is a density with respect to $\Nstar$,
but not the final field configuration.
The curvature perturbation in each patch
will be given by
$\zeta = \Nstar - \langle \Nstar \rangle$,
with $\langle \cdots \rangle$
denoting an expectation value in the larger volume.
Therefore,
neglecting coupling between patches,
the required distribution $\ProbP(\zeta)$
satisfies
\begin{equation}
    \label{eq:prob-zeta-estimate}
    \ProbP(\zeta)
    \,
    \d \zeta
    \approx
    \ProbQ(\Nstar = \zeta + \langle \Nstar \rangle)
    \,
    \d \zeta
    \;
    .
\end{equation}
It follows that
extreme values of $\Nstar$
will produce extreme values of $\zeta$.
Moreover,
the tail of $\ProbP(\zeta)$ is controlled
by the tail of $\ProbQ(\Nstar)$.

In the absence of noise,
almost all
combinations of $\phiinit$, $\phiend$
and $\Nstar$
will fail to describe
an allowed transition.
Instead,
the deterministic
evolution will usually select
a single
transition time $\NstarDet$
for which the field arrives precisely at $\phi = \phiend$.
Transitions of different durations are not allowed.
In contrast,
after inclusion of noise,
the transition from $\phiinit$ to $\phiend$
becomes possible for a much wider range of $\Nstar$.
We describe such transitions as ``noise supported.''
Those that take place in a time close
to the deterministic value
$\NstarDet$
require fairly typical realizations of the
stochastic process $\d\xi$,
and are relatively probable.
Transitions that occur in a time very different
to $\NstarDet$
involve rare or unusual realizations of the noise,
and are relatively improbable.
In this paper we will assume
that, to the accuracy required in Eq.~\eqref{eq:prob-zeta-estimate},
the expected number
of elapsed e-folds
in the enclosing region is
$\langle \Nstar \rangle \approx \NstarDet$.

Notice that, depending what we intend to compute,
the correct target field value $\phiend$
may not correspond to the end of inflation.
For example,
if (as described above)
we wish to evaluate the curvature perturbation
interior to some patch of scale $L$,
then we explain in~\S\ref{sec:decaying-noise}
that the noise amplitude decays rapidly outside the horizon,
so that most of the superhorizon evolution is dominated
by deterministic drift.
In this case, $\phiend$ should correspond to the
energy density at which the scale $k = 2\pi/L$
is roughly equal
to the horizon scale, or the coarse-graining
scale of~\eqref{eq:Starobinsky-Langevin}.
In the remainder of this paper, except in~\S\ref{sec:decaying-noise},
we leave
$\phiend$ arbitrary, with the understanding
that it should be chosen appropriately
for the observable under discussion.

Finally, we comment on
the use of $\ProbQ(\Nstar)$
as the required distribution on $\Nstar$.
In this paper, following Refs.~\cite{Vennin:2015hra,Pattison:2017mbe},
we generally make this identification.
If we are computing the distribution of $\Nstar$
up to the end of inflation,
this corresponds to disallowing ``backflow''
events where a coarse-grained patch
passes the terminal boundary
$\phi = \phiend$,
but later rejoins the inflating region.
This is reasonable because, once inflation ends,
fluctuations are no longer being generated,
preventing a reverse passage through $\phi = \phiend$.
Eq.~\eqref{eq:prob-zeta-estimate}
then gives the correct distribution for $\ProbP(\zeta)$.

On the other hand, if $\phiend$
corresponds to horizon exit for a particular scale,
as described above,
then perturbations continue to be generated and it
is no longer clear that we should exclude backflow.
In this case, we may need to identify $\ProbP(\zeta)$
with a different distribution
that includes these events.
This is the scenario considered by Tomberg in 
Ref.~\cite{Tomberg:2023kli}, and
discussed in \S\ref{sec:tomberg-langevin-method}.
Intuitively, we can regard scenarios where the tails
of $\ProbP$ and $\ProbQ$ (or $\ProbPre$) differ
as corresponding to cases where backflow events are
not strongly suppressed.
In this paper we generally assume that $\ProbQ(\Nstar)$
still gives the relevant distribution.
However,
it is necessary to select $\phiend$ with care.
It should be correspond to a time soon enough after horizon
exit that we do not incorrectly estimate the variance of
the smoothed fields in the patch,
but late enough that backflow events are strongly suppressed.
See also the discussion in~\S\ref{sec:decaying-noise}.

\subsubsection{Survival probability and first passage probability}
Now consider how to compute the first-passage distribution.
To do so,
notice that
the transition probability
$\ProbP( \phi, N \divider \phiinit, \Ninit)$
also satisfies the backward Kolmogorov equation
(or ``adjoint'' Fokker--Planck equation),
\begin{equation}
    \label{eq:kolmogorov-backwards}
    - \frac{\d \ProbP}{\d \Ninit}
    =
    - \frac{V'}{3H} \frac{\partial \ProbP}{\partial \phiinit}
    + \frac{H^2}{8\pi^2} \frac{\partial^2 \ProbP}{\partial \phiinit^2}
    =
    \LFPadj \ProbP
    .
\end{equation}
For further discussion of the backward equation, see
the appendices of Ref.~\cite{Assadullahi:2016gkk}.

We have introduced another
second-order differential operator $\LFPadj$.
If the drift and noise are independent of field-space
position, the forward and backward Kolmogorov equations
are equivalent
and $\LFP = \LFPadj$.
In other cases, the notation
reflects that they are adjoints of each other
in the inner product
$(f, g) = \int \d \phi \, f^\ast(\phi) g(\phi)$,
where `$\ast$' denotes a complex conjugate.

\para{Survival probability}
We now consider how
the transition probability and first passage
probability are related.
Define $\Survival(N \divider \phiinit, \Ninit)$
to be the \emph{survival probability}
given a starting location $\phi = \phiinit$ at time $\Ninit$.
This is the probability that,
at time $N$, the field has not yet
passed the terminal field value $\phiend$.
Assuming that $\phi$ rolls towards $\phiend$ from larger values, we have
\begin{equation}
    \label{eq:survival-probability-def}
    \Survival(N \divider \phiinit, \Ninit) =
    \int_{\phiend}^{\infty}
    \ProbPre(\phi, N \divider \phiinit, \Ninit) \, \d \phi \;.
\end{equation}
In this expression,
the transition probability
$\ProbPre$
should only count
paths that do not violate the first passage condition.
This is indicated by the prime $'$.
Kumar (1985)~\cite{kumar1985quantum}
described $\ProbPre$ as the \emph{restricted}
transition probability.

Since $\ProbPre$ is a density with respect to $\d \phi$,
Eq.~\eqref{eq:survival-probability-def}
makes $\Survival$ an honest probability,
not a density.
If the slow-roll approximation were dropped, the
transition probability $\ProbPre$
would also depend on the velocity $\d \phi / \d N$. In this case, one should
define a marginalized survival probability by
also integrating over the momentum
as in~\eqref{eq:flux-formula-marginalized}.

In any time interval, $\Survival$ decreases
because some
trajectories newly arrive at $\phi = \phiend$.
Such trajectories terminate
and no longer contribute to $\ProbPre$.
Hence, the $\ProbPre$ appearing in~\eqref{eq:survival-probability-def}
does not conserve probability;
in particular, it is \emph{not} normalized
to unity at all times.
It follows that the probability for a trajectory
originating at $\phi = \phiinit$
at time $\Ninit$
to
arrive for the first time at $\phiend$
at time $N$,
which we write $\ProbQ(\phiend, N \divider \phiinit, \Ninit)$,
must satisfy
\begin{equation}
    \label{eq:first-passage-probability-def}
    \ProbQ(\phiend, N \divider \phiinit, \Ninit)
    =
    - \frac{\partial}{\partial N}
    \Survival( N \divider \phiinit, \Ninit) \;.
\end{equation}
This argument was first given by
Sch\"{o}dinger~\cite{schrodinger1915theorie}.
A similar discussion in the context
of stochastic inflation was given recently
by Rigopoulos \& Wilkins~\cite{Rigopoulos:2022gso} 
(see Eqs.~(5.3)--(5.4) of that reference) and 
by Tomberg \& Dimopoulos~\cite{Tomberg:2025fku}.

Eq.~\eqref{eq:first-passage-probability-def}
makes $\ProbQ$ a probability density with respect to $N$.
Often, the absolute initial and final times of the transition are not important,
and we are interested only in the transition duration $\Nstar = N - \Ninit$.
It follows that $\partial / \partial N = \partial / \partial \Nstar$
at fixed $\Ninit$,
and $\partial / \partial \Ninit = - \partial / \partial \Nstar$
at fixed $N$,
and we can replace the $N$ derivative by an $\Nstar$
derivative in~\eqref{eq:first-passage-probability-def}.
We
abbreviate $\ProbQ = \ProbQ( \Nstar, \phiend \divider \phiinit)$,
which
can be regarded as a density with respect to $\Nstar$.

\para{Computation of $\Survival$}
Eq.~\eqref{eq:first-passage-probability-def}
enables us to evaluate $\ProbQ$
if $\Survival$ is known.
There are two main strategies.
One option is to solve for $\Survival$
directly.
It
satisfies the same backward
Kolmogorov equation as $\ProbP$,
\begin{equation}
    \label{eq:survival-kolmogorov-equation}
    -
    \frac{\d \Survival}{\d \Ninit}
    =
    -
    \frac{V'}{3H}
    \frac{\partial \Survival}{\partial \phiinit}
    +
    \frac{H^2}{8\pi^2}
    \frac{\partial^2 \Survival}{\partial \phiinit^2}
    =
    \LFPadj \Survival \;.
\end{equation}
There is no analogue of the forward equation for
$\Survival$ because the final configuration is integrated out.

We must specify suitable boundary conditions.
Clearly the survival probability at
the terminal point $\phiend$ is zero at all times.
Therefore we require the boundary condition
\begin{subequations}
\begin{equation}
    \label{eq:survival-phiend-bc}
    \Survival(N \divider \phiinit=\phiend, \Ninit) = 0
    \quad
    \text{for all $N$ and $\Ninit$}
    .
\end{equation}
In principle, we should set an analogous boundary condition
at $\phiinit = \infty$.
In practice, as noted by Ezquiaga {\etal}, this is inconvenient
if we intend to solve~\eqref{eq:survival-kolmogorov-equation}
numerically.
In that case,
it may be preferable to set a boundary condition
at some (arbitrary) finite position $\phiuv \gg \phiend$.
This boundary condition is not physical, but if
$\phiuv$ is taken sufficiently large
it will not
influence the result.%
    \footnote{In this discussion,
    we are continuing to assume that $\phi$ rolls towards $\phiend$
    from larger values.
    
    If we are setting boundary conditions on $\ProbPre$
    rather than $\Survival$ directly, we could choose
    a boundary condition such that $\Current = 0$
    at $\phiuv$,
    where $\Current$ is the Kolmogorov probability
    current defined in~\eqref{eq:probability-current}.
    There would then be no outflow of probability at the upper boundary.
    This has the desirable
    outcome that
    changes in $\Survival$ could be due only to trajectories
    arriving at $\phiend$.
    However, note that this is a boundary condition on the
    final field configuration in $\ProbPre$, rather than the initial
    one, so does not translate directly to $\Survival$.}
Finally,
we still require an initial condition.
When $N = \Ninit$,
no trajectories
beginning at $\phiinit \neq \phiend$
have yet been removed from the system. Therefore,
\begin{equation}
    \label{eq:survival-ic}
    \Survival(\Ninit \divider \phiinit, \Ninit) = 1
    \quad
    \text{for all $\phiinit > \phiend$} .
\end{equation}
\end{subequations}

\para{Flux formula, Feynman--Kac representation}
Direct solution of Eq.~\eqref{eq:survival-kolmogorov-equation}
is adequate to study first passage at a single boundary
in one dimension.
With multiple boundaries
(not actually needed for our applications),
or in higher dimensions,
this definition is not always convenient.

An alternative strategy
is to solve for the
restricted transition
probability $\ProbPre$,
and obtain $\ProbQ$
from~\eqref{eq:first-passage-probability-def}.
To do so,
note that the derivative in~\eqref{eq:first-passage-probability-def}
acts on the final time $N$.
The forward Kolmogorov equation~\eqref{eq:forward-komogorov-transport}
then implies
\begin{equation}
    \label{eq:first-passage-probability-current}
    \ProbQ(\phiend, N \divider \phiinit, \Ninit)
    =
    - \int_{\phiend}^{\infty} \frac{\d \ProbPre}{\d N} \, \d \phi
    =
    \int_{\phiend}^{\infty} \frac{\d \CurrentRe}{\d \phi} \, \d \phi
    = - \CurrentRe \big|_{\phi = \phiend} \;.
\end{equation}
We describe this as the \emph{flux formula} for $\ProbQ$.
It shows that, as we expect,
the first passage probability
only requires information about
transitions to the neighbourhood 
of the boundary.

In multiple dimensions, Eq.~\eqref{eq:first-passage-probability-current}
should be replaced by a probability
flux computed
by integrating the current $\Current_\alpha$ over a suitable
boundary, cf. Eq.~\eqref{eq:flux-formula-marginalized}.
It follows that
\begin{equation}
    \label{eq:first-passage-probability-current-higher-dim}
    \ProbQ(\partial B, N \divider \phi_0, N_0)
    =
    \int_{\partial B} \hat{n}^\alpha \CurrentRe_{\alpha} \, \d A .
\end{equation}
This is the primary definition of
$\ProbQ(\Nstar)$ used in this paper.
Note that the
direction of the flux
in Eqs.~\eqref{eq:first-passage-probability-current},
reflected by the overall sign,
depends on the orientation of boundary
for which we are computing first-passage
events.
In Eq.~\eqref{eq:first-passage-probability-current-higher-dim}
this sign is inherited from the orientation of
$\hat{n}^\alpha$.

It is critical that the
current $\CurrentRe$
appearing in Eqs.~\eqref{eq:first-passage-probability-current}--%
\eqref{eq:first-passage-probability-current-higher-dim}
is
constructed
from the \emph{restricted}
transition probability $\ProbPre$,
and not its unrestricted counterpart $\ProbP$.
Transport of probability is a local property,
and so $\ProbPre$
obeys the same forward and backward Kolmogorov
equations as $\ProbP$, Eqs.~\eqref{eq:kolmogorov-forwards}
and~\eqref{eq:kolmogorov-backwards}.
The global restriction
on paths
appears only at the level of a boundary condition.
Since paths terminate on arrival at $\phi = \phiend$
and are removed,
no probability accumulates there.
Hence, we must
have $\ProbPre(\phiend, \Nstop \divider \phiinit, \Ninit) = 0$,
sometimes described as an ``absorbing'' boundary condition.
This has the consequence
that the current $\Current$,
and hence the boundary flux $\ProbQ$,
receives contributions only from diffusion,
not deterministic drift.

There is no general relation between
$\ProbP$ and $\ProbPre$,
although
explicit formulas
are known for a limited number of cases.
These are usually based either on the method
of images
or renewal theory.
The method of images is limited
to scenarios with spatial homogeneity
(here meaning that the drift velocity and noise are
independent of position in field space),
or where there is a symmetry axis 
co-located with the boundary.
We describe renewal theory in
Appendix~\ref{sec:App_ren_eq}.
It is mostly useful only in one dimension.
Moreover,
it requires evaluation of an inverse Laplace
transform,
which is possible analytically only in
very limited cases.
Beyond these examples,
there appear to be very few analytic
tools.
Some attempts have been made to
provide a prescription for $\ProbPre$ in more
general scenarios; see, e.g., Ref.~\cite{nyberg2016simple},
but these require inputs
beyond $\ProbP$ alone.

In this paper we will build
path integral representations of
$\ProbP$ and $\ProbPre$,
although we will certainly not solve the problem
of their inter-relation.
To do so we reformulate
$\ProbP$ and $\ProbPre$ without
direct reference to the
Kolmogorov equations.
The unrestricted
transition probability
$\ProbP$ can be represented
by the Feynman--Kac formula,
\begin{equation}
    \label{eq:Feynman-Kac-transition}
    \ProbP( \phistop, \Nstop \divider \phiinit, \Ninit )
    =
    \Expect
    \Big\{
        \delta
        \big[
            \phi(\Nstop) - \phistop
        \big]
    \Big\}
\end{equation}
where the expectation $\Expect$ should be taken
over paths that begin
at $\phi = \phiinit$ at time $\Ninit$,
but are otherwise unconstrained.
The restricted transition probability
$\ProbPre$ can likewise be obtained
by replacing $\Expect$
with an expectation
only over paths
with $\phi(N) > \phiend$.
We denote this restricted
expectation value by $\ExpectRe$.
This restriction means that the
expectation value does not intersect the support
of the $\delta$-function,
reproducing the boundary condition
$\ProbPre(\phiend, \Nstop \divider \phiinit, \Ninit) = 0$.

We can give an analogous ``Feynman--Kac''-like
formula by using $\ExpectRe$
to evaluate the probability current
in Eq.~\eqref{eq:first-passage-probability-current-higher-dim},
\begin{equation}
    \label{eq:Feynman-Kac-Q}
    \ProbQ( \partial B, \Nstop \divider \phiinit, \Ninit)
    =
    \ExpectRe
    \bigg\{
        \bigg(
            \int_{\partial B}
            \hat{n}^\alpha
            v_\alpha(N_1)
        \bigg)
        \delta
        \big[
            \phi(\Nstop) - \phistop
        \big]
    \bigg\}
    ,
\end{equation}
where $v^\alpha = \d \phi^\alpha / \d N$
is the field-space (or phase-space) velocity.
The final boundary condition
$\phi(\Nstop) = \phistop$ should be understood
via a limiting procedure
from the allowed region $\phi(N) > \phistop$.
This result is the key step
needed
to obtaining a path integral representation
for $\ProbQ$.
The interpretation of $\hat{n}^\alpha v_\alpha \delta[ \phi(\Nstop) - \phistop ]$
as a microscopic probability current
is formally parallel to the
microscopic current operator $\vect{j} = q \vect{v} \delta[ \vect{x}(t) - \vect{x}_1 ]$
in electrodynamics.
In~\S\ref{sec:path-integral-flux-formula}
we discuss some subtle technical
details that arise
when verifying that Eq.~\eqref{eq:Feynman-Kac-Q}
reproduces the flux
formula~\eqref{eq:first-passage-probability-current}.

\subsection{Ezquiaga {\etal} spectral formalism}
\label{sec:ezquiaga-spectral-method}
\label{sec:spectral-solution}
In this section and the next, we review
the approaches of
Ezquiaga {\etal}~\cite{Ezquiaga:2019ftu}
and Tomberg~\cite{Tomberg:2023kli}.
These both aim to estimate the tail of the
probability distribution $\ProbP(\zeta)$
via Eq.~\eqref{eq:prob-zeta-estimate}.

In this section we present the method
of Ezquiaga {\etal},
which
is based on a spectral solution
of the
backward Kolmogorov equation.
Ezquiaga {\etal} framed their discussion in terms
of the restricted transition probability
$\ProbPre$.
Strictly, one should then translate
to $\ProbQ$
using the flux formula~\eqref{eq:first-passage-probability-current}.
In our framework,
Eq.~\eqref{eq:first-passage-probability-def}
enables us to give a slightly cleaner
discussion in terms of the survival probability
$\Survival$.

Using $\partial / \partial \Ninit = - \partial / \partial \Nstar$
at fixed $N$,
the Kolmogorov equation~\eqref{eq:survival-kolmogorov-equation} for
$\Survival$
has a formal solution
(matching that used by Ezquiaga {\etal})
\begin{equation}
    \label{eq:formal-Kolmogorov-soln}
    \Survival(\Nstar \divider \phiinit)
    =
    \exp( \Nstar \LFPadj )
    \Survival(\Nstar = 0 \divider \phiinit )\,,
\end{equation}
where the operator exponential should be understood to be defined
by its power series.
This solution is valid only if
$\LFPadj$ is time-independent.
That will not be true in general, but may be
approximately valid up to slow-roll
corrections.

$\LFPadj$ is a well-defined operator of Sturm--Liouville type.
A detailed discussion
of the spectral properties of such operators
is given in the standard reference by
Morse \& Feshbach~\cite{morse1946methods};
see also the recent summary in Ref.~\cite{Tomberg:2025fku}.
It was shown in Ref.~\cite{Ezquiaga:2019ftu}
that $\LFPadj$
possesses an infinite spectrum of real eigenvalues,
and that the corresponding eigenfunctions
form a complete orthogonal set.
Any suitably regular function
may therefore be
expanded as
a series in the $\Psi_n$,
analogous to a Fourier series.
Such
spectral methods were introduced for
the Fokker--Planck
equation by
Tomita {\etal}~\cite{tomita1976eigenvalue}
and
van Kampen~\cite{van1977soluble}.
They were applied to stochastic inflation
by Starobinsky \& Yokoyama~\cite{Starobinsky:1994bd}. 

To apply this to $\Survival(\Nstar \divider \phiinit)$ we must
select boundary conditions to the $\Psi_n$
that are compatible with $\Survival$.
Notice that the eigenvalues depend on the boundary conditions
chosen,
which therefore
have physical significance
and are not just a matter of convention.
Eq.~\eqref{eq:survival-phiend-bc}
requires $\Survival$ to vanish at $\phiend$,
so we should
choose $\Psi_n(\phiend) = 0$.
There is an arbitrary, unphysical boundary condition
at $\phiuv$.
As explained above, the eigenvalues $\Lambda_n$
are intended to
be insensitive to this unphysical boundary condition.

We now expand $\Survival(\Nstar=0 \divider \phiinit)$
in terms of the $\Psi_n$,
\begin{equation}
    \label{eq:survival-eigenfunction-expansion}
    \Survival(\Nstar=0 \divider \phiinit) = \sum_n a_n \Psi_n(\phiinit) \,,
\end{equation}
where the $a_n$ are to be determined by the
initial condition~\eqref{eq:survival-ic}.
To do so, we must
extend~\eqref{eq:survival-eigenfunction-expansion}
to $\Nstar > 0$ using
the formal solution~\eqref{eq:formal-Kolmogorov-soln},
\begin{equation}
    \label{eq:survival-time-dep}
    \Survival(\Nstar) = \sum_n a_n \Psi_n(\phiinit) \e{-\Lambda_n \Nstar} \;.
\end{equation}
A nearly equivalent analysis,
applied to the transition probability rather than
the survival probability,
was given by Caroli, Caroli
\& Roulet (1979)~\cite{caroli1979diffusion}.
They combined it with a WKB procedure
to estimate the $\Lambda_n$,
which enabled them to compute
a range of characteristic timescales
associated with stochastic tunnelling between minima.

Whatever initial condition is imposed on $\Survival$,
we can expect it to be sufficiently generic that the
low-lying eigenfunctions $\Psi_0$, $\Psi_1$, $\Psi_2$, \ldots,
will be excited
in Eq.~\eqref{eq:survival-eigenfunction-expansion}.
It follows
from Eq.~\eqref{eq:survival-time-dep}
that (again, generically)
only the lowest eigenvalue survives
in the
asymptotically rare limit,
\begin{equation}
    \Survival(\Nstar) \sim \e{-\Lambda_0 \Nstar}
    \quad
    \text{for $\Nstar \gtrsim \NKramers = \Lambda_0^{-1}$} .
\end{equation}
Caroli {\etal} described this as the \emph{Kramers regime},
and the timescale $\NKramers = \Lambda_0^{-1}$ as the
\emph{Kramers time}~\cite{caroli1979diffusion}.%
    \footnote{The name arises because this is the timescale
    controlling the escape time in the Kramers problem
    of escape from a narrow potential well.}
If $\Nstar \gg 1$ is large enough to be rare, but not yet
in its asymptotic regime,
it may be necessary to retain several of the smaller eigenvalues
$\Lambda_1$, $\Lambda_2$, etc.
Caroli {\etal} described this as the
\emph{intermediate regime}~\cite{caroli1979diffusion}.
In either regime,
combining
Eqs.~\eqref{eq:survival-time-dep}
and~\eqref{eq:first-passage-probability-def},
using $\partial / \partial N = \partial / \partial \Nstar$
at constant $\Ninit$,
shows that the same tail estimate will apply for the first-passage
distribution.
Hence, in the Kramers regime,
\begin{equation}
    \label{eq:eigenvalue-tail}
    \ProbQ( \Nstar, \phiend \divider \phiinit)
    \sim
    \e{-\Lambda_0 \Nstar} .
\end{equation}

We conclude that, when Eq.~\eqref{eq:formal-Kolmogorov-soln}
applies,
computation of the tail probability
reduces to
computation of the eigenvalue spectrum
of $\LFPadj$.
In simple examples this can be done explicitly.
In more complex cases it may be necessary to use an approximate
scheme such as a Rayleigh--Ritz procedure.
For the example of a linear potential,
Ezquiaga {\etal}~\cite{Ezquiaga:2019ftu} found,
in our notation,
\begin{equation}
    \label{eq:slow-roll-exp-tail}
    \ProbQ(\Nstar) \sim \exp\bigg(
        {-
        \frac{\Nstar}{2 \dimP_\zeta}}
    \bigg) 
    \,,
\end{equation}
where $\dimP_\zeta$ is the dimensionless power spectrum of $\zeta$,
\begin{equation}
    \dimP(k) = \frac{k^3}{2\pi^2} P_\zeta(k) \,,
\end{equation}
and $P_\zeta(k)$
is the ordinary power spectrum,
\begin{equation}
    \langle \zeta(\vect{k}_1) \zeta(\vect{k}_2) \rangle
    =
    (2\pi)^3
    \delta(\vect{k}_1 + \vect{k}_2)
    P_\zeta(k_1) \;.
\end{equation}

An interesting feature of this procedure is
that we can access $\ProbQ(\Nstar)$,
and hence $\ProbP(\zeta)$,
without ever having to specify the relationship
between $\zeta$ and $\phi$.
As explained above, this is important because
extreme fluctuations can probe the self-interactions
of $\phi$
at arbitrarily high order.
By comparison, for example, the Jacobian method
used
in Refs.~\cite{Tomberg:2023kli,Ballesteros:2024pwn}
requires an explicit formula for $\zeta$ in terms of $\phi$.
Emergence of the exponential tail depends on the
relationship between $\zeta$
and $\phi$ being exactly logarithmic.
This means that the relationship must be known
(and trustable) to all orders.
Here,
remarkably, computation of the eigenvalues
can be done purely in terms of the theory for $\phi$.
We will solve this model using our instanton
framework in~\S\ref{sec:slow-roll-instanton} below,
and show that it shares the same property.

The tail estimate~\eqref{eq:slow-roll-exp-tail}
scales more slowly than
the tail from the Celoria {\etal} instanton.
We infer that,
for sufficiently large $\Nstar$,
it is
less unlikely to build a
large fluctuation by an incremental
series of small fluctuations,
that is to nucleate it in a single event,
at least where this nucleation is dominated by the
$\dot{\zeta}^4$ interaction studied by Ref.~\cite{Celoria:2021vjw}.
A similar argument
(but more generally applicable)
has been suggested by Cohen {\etal}~\cite{Cohen:2022clv}
in the context of the large deviation theory of
Freidlin \& Wentzell.
The small fluctuations may or may not
be individually unlikely themselves,
but
the total \emph{sequence}
of such fluctuations must be exponentially
rare.
We will see further examples of this phenomenon
below.

\subsection{Tomberg's Langevin formalism}
\label{sec:tomberg-langevin-method}
We now consider an alternative formalism
due to Tomberg~\cite{Tomberg:2023kli}.
In Ref.~\cite{Tomberg:2023kli}
this was applied this to a model
of ``constant roll'' inflation,
to be described below.
We analyse this model using the instanton method in~\S\ref{sec:CR}.

To define constant roll models,
we introduce the conventional slow roll parameter
$\epsilon \equiv - \d \ln H / \d N$,
and also a second $\eta$-like parameter,
\begin{equation}
    \label{eq:epsilon2-def}
    \epsilon_2 \equiv \frac{\d \ln \epsilon}{\d N}\; .
\end{equation}
A model is said to be of constant roll type if $\epsilon_2$ is time-independent.
In these scenarios the scalar
field evolution is no longer overdamped,
and the Starobinsky equation~\eqref{eq:Starobinsky-Langevin}
must be supplemented by a second Langevin equation
for $\pi = \d \phi / \d N$.
In Appendix~\ref{sec:Lang_app} we show
that the required equations are
\begin{equation}
    \label{eq:constant-roll-evolution}
    \frac{\d \phi}{\d N} =
    \pi
    +
    \xi_\phi \,,
    \quad
    \frac{\d \pi}{\d N} = 
    -
    (3 - \epsilon)
    \bigg(
        \pi + \Mp^2 \frac{V'}{V}
    \bigg)
    + \xi_\pi\;.
\end{equation}
The noises $\xi_\phi$ and $\xi_\pi$
are correlated
according to
Eqs.~\eqref{eq:M21} and \eqref{eq:PS_matrix}.%
    \footnote{See also Eqs.~\eqref{eq:inst_cr_1} and \eqref{eq:inst_cr_2} below.
    
    In Appendix~\ref{sec:Inf_func}, the noise terms in the
    Langevin equations are written as $\xi_\phi/H^2$ and $\xi_\pi/H^2$.
    Here, we keep the simpler notation $\xi_\phi$ and $\xi_\pi$,
    understanding them simply as rescaled versions of those in the Appendix.}

\para{Noiseless evolution}
This gives the unperturbed background, with $\xi_\phi = \xi_\pi = 0$.
We adopt the notation of Ref.~\cite{Tomberg:2023kli}
and write $\epsilon_2 \equiv 2 \sigma$.
The solution can be written
\begin{equation}
    \label{eq:noiseless-constant-roll}
    \phi(N) - \Phi = \frac{(2\epsilon_0)^{1/2}}{\sigma} \Mp \e{\sigma N} \,,
    \quad
    \pi(N) = (2\epsilon_0)^{1/2} \Mp \e{\sigma N} \,,
\end{equation}
where
$\epsilon_0$ is the value of
$\epsilon(N)$,
and $\Phi$ is an integration constant
that can be used to fix the corresponding initial value of $\phi$.
Following Ref.~\cite{Tomberg:2023kli}
we redefine $\phi$ so that $\Phi = 0$,
in which case
we obtain
\begin{equation}
    \label{eq:unperturbed-constant-roll}
    \epsilon(N)
    =
    \epsilon_0
    \e{2 \sigma N}
    =
    \frac{\sigma^2}{2\Mp^2}
    \phi(N)^2
    \quad
    \text{and}
    \quad
    \pi(N) = \sigma \phi(N)\;.
\end{equation}

\para{Stochastic evolution}
We now reintroduce the noise terms.
In Refs.~\cite{Tomberg:2022mkt,Tomberg:2023kli}
it was shown
that microscopic correlations
require that $\xi_\phi$ and $\xi_\pi$ combine
to a single stochastic
kick along the unperturbed trajectory given
in~\eqref{eq:unperturbed-constant-roll}.
Therefore the linear relation $\pi = \sigma \phi$ is preserved.
Moreover, the noise terms depend only on our position on this trajectory.
Under these conditions we can rewrite the first equation
in~\eqref{eq:constant-roll-evolution}
as
\begin{equation}
    \label{eq:constant-roll-Langevin}
    \d \phi
    =
    \pi \, \d N
    +
    \dimP_\phi^{1/2}(\phi, \pi) \, \d \xi
    ,
\end{equation}
where $\dimP_\phi$ is a known function of $\phi$ and $\pi$ only,
and $\d \xi$ is a stochastic process with $\Expect(\d \xi^2) = \d N$.
The noise amplitude $\dimP_{\phi} = \dimP_\phi( \phi, \pi )$
corresponds to $\dimP_{11}$
of Eq.~\eqref{eq:PS_matrix}.
Together with the relation $\phi = \sigma \pi$,
this single Langevin equation is sufficient to determine the evolution
of $\phi$.

The solution to Eq.~\eqref{eq:constant-roll-Langevin}
with initial condition $\epsilon = \epsilon_0$ at $N=0$
can be written
\begin{equation}
    \label{eq:stochastic-constant-roll}
    \phi(N) =
    \tilde{\phi}(N)
    \bigg(
        1
        +
        \frac{\epsilon_2}{2}
        \Gamma(N)
    \bigg)\,
    ,
\end{equation}
where $\tilde{\phi}(N)$ is the unperturbed
evolution~\eqref{eq:unperturbed-constant-roll}
for this initial condition,
and $\Gamma(N)$ is the noise integrated with respect to the
stochastic process $\d \xi$,
\begin{equation}
    \label{eq:constant-roll-Gamma-def}
    \Gamma(N) = \int \dimP_\zeta^{1/2} \, \d \xi
    .
\end{equation}
We can regard $\Gamma(N)$
as the sum of all stochastic kicks in the noise
realization $\d \xi$.
It is
an aggregate fluctuation
of the kind required for the `Jacobian'
approach described in~\S\ref{sec:introduction}.
For typical realizations of the noise $\d \xi$ we expect
$\Gamma(N) \sim 0$.
Values of $\Gamma(N)$ significantly different from zero
represent extreme realizations.

Note that $\dimP_{\phi}$ in~\eqref{eq:constant-roll-Langevin}
becomes $\dimP_{\zeta} = \dimP_\phi / (2 \epsilon \Mp^2)$
in~\eqref{eq:constant-roll-Gamma-def}.
The stochastic integral in~\eqref{eq:constant-roll-Gamma-def}
should
be understood via a discretization, as discussed in Ref.~\cite{Tomberg:2023kli}.

\para{Transition duration}
We wish to use~\eqref{eq:stochastic-constant-roll}
to evaluate the probability of a transition to some final field
value $\phiend$ with total duration $\Nstar$ e-folds.
As above,
we write
the deterministic transition time as $\NstarDet$.
When $\Nstar$ is very different to $\NstarDet$ we require a
highly unlikely realization of the integrated noise $\Gamma(\Nstar)$,
in order that
$\phi(\Nstar)$
can be very different from its deterministic value.

The field arrives at the final surface
after exactly $\Nstar$ e-folds, and therefore
$\phi(\Nstar) = \phiend = \tilde{\phi}(\NstarDet)$.
Meanwhile, the deterministic solution
$\tilde{\phi}(\Nstar)$, defined in Eq.~\eqref{eq:stochastic-constant-roll},
will assume a very different value.
The difference must be compensated by an extremely
specific value
for the integrated noise,
\begin{equation}
    \label{eq:GDN_transf}
    \Gamma(\Nstar)
    \equiv
    \GammaStar
    =
    -
    \frac{2}{\epsilon_2}
    \bigg(
        1
        -
        \e{-\sigma \Delta N}
    \bigg) ,
\end{equation}
where $\Delta N \equiv \Nstar - \NstarDet$.
When $\Delta N \sim 0$,
we find $\GammaStar \sim 0$, corresponding to a typical
noise realization.
When $\Delta N \gg 0$
we have $\GammaStar \sim -2/\epsilon_2$,
although the exact value is exponentially
sensitive to the required $\Delta N$.

This approach does not determine the time history of the noise
required to mediate the transition,
which here would be represented by knowledge of the function $\Gamma(N)$.
Instead, we only determine the aggregate value $\Gamma(\Nstar)$
after precisely $\Nstar$ e-folds.
There may be many noise profiles that can match this
boundary condition.

How unlikely is it that $\Gamma(N)$ arrives at the required value
\eqref{eq:GDN_transf}?
Since $\GammaStar$ is a sum of Gaussian variables,
Tomberg suggested that its probability distribution could be identified
as
\begin{equation}
    \label{eq:PGamma}
    \ProbP( \GammaStar )
    =
    \frac{1}{\sqrt{2\pi} \sstar}
    \exp
    \bigg(
        {-
        \frac{\GammaStar^2}{2 \sstar^2}}
    \bigg)
    \quad
    \text{where}
    \quad
    \sstar^2
    =
    \int
    \dimP_\zeta
    \Big(
        \phi(N),
        \pi(N)
    \Big)
    \,
    \d N \;
    .
\end{equation}
Eq.~\eqref{eq:PGamma}
should be interpreted as a transition probability: that is,
the probability for the transition from $\phiinit$
to $\phiend$ to occur in $\Nstar$ e-folds.
With this interpretation, we will find
in~\S\ref{sec:CR}
that
Eq.~\eqref{eq:PGamma}
has very good support within the instanton framework---%
to the extent that we will derive
an equivalent formula,
although understood as a density with respect to the field configuration
rather than the integrated noise.

In this computation there are no restrictions
on the time history $\Gamma(N)$,
and therefore
$\ProbP(\GammaStar)$
represents an unrestricted
transition probability.
It
includes the probability of
realizations that
cause $\phi$ to cross $\phi = \phiend$,
before later fluctuating back
to produce the required $\GammaStar$.
In Ref.~\cite{Tomberg:2023kli}
this was regarded as the probability
that a smooth patch of fixed
comoving scale
arrives at horizon exit,
here represented by the fixed boundary
$\phi = \phiend$,
after $\Nstar$ e-folds.
It therefore properly
includes backflow events as described in~\S\ref{sec:Est_tail}.
The tails of
$\ProbP$ and $\ProbPre$
sometimes agree up to exponential accuracy,
in which case
we would obtain a similar
result if $\ProbP$ was exchanged for $\ProbPre$
or $\ProbQ$.%
    \footnote{It is not always true that the tails
    agree, even asymptotically.
    In~\S\ref{sec:CR}, we will see that the constant roll
    model
    for $\sigma < 0$
    provides an example where this is not the case.}
    
In the framework of this section,
Eq.~\eqref{eq:PGamma}
should be understood as an Ansatz rather than
an exact formula.
In particular,
the integral for the variance $\sstar^2$ should be evaluated
along a trajectory $\phi(N)$, $\pi(N)$
corresponding to the noise realization.
However, as we have explained,
this is not determined
by the procedure described above.
Indeed, there need not be a unique
realization, in which case the
meaning of~\eqref{eq:PGamma}
is undecided.
Therefore,
for Eq.~\eqref{eq:PGamma}
to be useful,
it must be supplemented
by a prescription
to assign some value to $\sstar^2$.
We explain below how Tomberg's method can be extended
to obtain an estimate for the time history
of the noise,
which provides one possible prescription.
Later, in~\S\S\ref{sec:slow-roll-instanton}--\ref{sec:instanton-phase},
we explain how the instanton formalism
is able to supply this information.

In Eq.~\eqref{eq:PGamma},
$\ProbP$ represents the probability for a transition between
two field values at fixed $\Nstar$.
In principle, we wish to determine a related question:
given that a smoothed patch has arrived at
the boundary $\phi = \phiend$,
what is the distribution on the number of e-folds it has experienced?

Tomberg interpreted~\eqref{eq:PGamma}
as the required distribution
by expressing it
as a density with respect to $\Nstar$
using
a suitable Jacobian factor.
This is an example of the `Jacobian' approach described
in~\S\ref{sec:introduction}.
In our view, the justification for this step
is not completely clear.
Proceeding in this way,
however,
the required
factor can be obtained
using~\eqref{eq:GDN_transf}
to relate $\GammaStar$ and $\Delta N$,
and then $\Delta N = \Nstar - \NstarDet$
to relate $\Delta N$ and $\Nstar$.
Hence,
\begin{equation}
    \label{eq:prob-nstar-jacobian-tomberg}
    \ProbP(\Nstar)
    =
    \ProbP(\GammaStar)
    \left| \frac{\d \GammaStar}{\d \Delta N} \frac{\d \Delta N}{\d \Nstar} \right|
    =
    \ProbP(\GammaStar)
    \left| \frac{\d \GammaStar}{\d \Delta N} \right| \;.
\end{equation}
Using~\eqref{eq:prob-nstar-jacobian-tomberg}
with $\GammaStar$ determined by~\eqref{eq:GDN_transf},
we reproduce
the result reported by Tomberg,
\begin{equation}
    \label{eq:Tomberg-constant-roll-P}
    \ln \ProbP(\Nstar)
    \sim
    - \frac{2}{\epsilon_2^2 \sstar^2}
    \Big(
        1 - \e{-\epsilon_2 (\Nstar - \NstarDet) / 2}
    \Big)^2
    -
    \frac{\epsilon_2}{2}
    (\Nstar - \NstarDet) \;.
\end{equation}
The final term $\epsilon_2 (\Nstar - \NstarDet) / 2$ comes from the Jacobian factor.

This procedure depends on being able to isolate the aggregate fluctuation
$\GammaStar$ needed to bring the field to $\phiend$ after
exactly $\Nstar$
e-folds.
In the scenario considered here,
this is possible because we have an explicit
solution~\eqref{eq:stochastic-constant-roll}.
Further, this solution separates into a part that depends only on
deterministic evolution from the initial conditions,
and a part that depends only on the noise.
In turn, this follows from the linearity of the effective
Langevin equation~\eqref{eq:constant-roll-Langevin}
and the relation $\pi = \sigma \phi$.
These conditions are satisfied only in a limited set of models.
Where they do not apply, it may still be possible
to find solutions for $\Gamma(\Nstar)$,
but the procedure would be more involved.
The instanton trajectories
discussed in~\S\S\ref{sec:slow-roll-instanton}--\ref{sec:instanton-phase}
can be regarded as a way to find
a \emph{single}, least unlikely
noise realization that does this.

\para{Least unlikely noise realization supporting the transition}
As explained above, this analysis only determines
$\GammaStar$ and not the entire time history
$\Gamma(N)$.
However, Tomberg suggested an approximate procedure
to estimate it.
To do so, write the solution
for
$\phi(N)$, including stochastic effects, as
\begin{equation}
    \phi(N)
    =
    \frac{(2 \epsilon_0)^{1/2}}{\sigma}
    \Mp
    \e{\sigma \tilde{N}(N)}
    ,
\end{equation}
where $\tilde{N}(N)$ can be regarded as an
\emph{effective} e-folding number
that accounts for stochastic corrections.
It need not increase monotonically.
At the final surface, $\tilde{N}$ will equal
$\NstarDet$, but for noise-supported
transitions $N = \Nstar$ will typically be much larger.
In terms of $\tilde{N}(N)$,
Eq.~\eqref{eq:constant-roll-Langevin}
can be written
\begin{equation}
    \d \tilde{N}
    =
    \bigg(
        1
        +
        \frac{\epsilon_2}{2}
        \frac{\dimP_\zeta(\tilde{N})}{2}
    \bigg)
    \d N
    -
    \dimP_\zeta^{1/2}(\tilde{N})
    \, \d \xi ,
\end{equation}
where
the power spectrum $\dimP_\zeta$ should be
regarded as a function
of the effective e-folding number $\tilde{N}$.
The term proportional to $\epsilon_2$
arises from the quadratic
part of the It\^{o} chain rule.
If $\dimP_\zeta \ll 1$ this first term
is typically small.
Accordingly, we drop it in what follows.

Since $\d \xi$ is a stochastic process with unit variance,
the probability for any given realization satisfies
\begin{equation}
    \label{eq:tomberg-onsager-machlup}
    \ProbP(\xi)
    \sim
    \exp
    \bigg(
        -
        \frac{1}{2}
        \int_{\Ninit}^{\Nstar}
        (\xi')^2
        \, \d N
    \bigg)
    \approx
    \exp
    \bigg(
        -
        \frac{1}{2}
        \int_{\Ninit}^{\Nstar}
        \frac{( \tilde{N}' - 1 )^2}{\dimP_\zeta(\tilde{N})}
        \, \d N
    \bigg)
    .
\end{equation}
We are using a prime $'$ to denote
the derivative
of a function with respect to its argument.
Eq.~\eqref{eq:tomberg-onsager-machlup}
is the Onsager--Machlup functional
for the process $\tilde{N}(N)$~\cite{PhysRev.91.1505}.
The least unlikely realization
supporting the transition
can be found by minimizing $\ln \ProbP(\xi)$.
This yields the constraint
\begin{equation}
    \label{eq:tomberg-deltaN-minimizer}
    \frac{\tilde{N}''}{(\tilde{N}')^2 - 1}
    =
    \frac{1}{2}
    \frac{\dimP_\zeta'(\tilde{N})}{\dimP_\zeta(\tilde{N})}
    .
\end{equation}
$\tilde{N}(N)$ in this expression should no
longer be regarded as a stochastic quantity,
but rather a deterministic function
obtained by solving~\eqref{eq:tomberg-deltaN-minimizer}
subject
to boundary conditions
corresponding to the transition in question.
It follows that we can formally
estimate $\Gamma(N)$
on this
realization, viz.,
\begin{equation}
    \Gamma(\Nstop)
    =
    \int \dimP_\zeta^{1/2} \, \d \xi
    \approx
    -
    \int_{\Ninit}^{\Nstop}
    \Big(
        \tilde{N}'(N)-1
    \Big)
    \,
    \d N
    .
\end{equation}
Notice that information about the most likely
noise realization is not easily available
in the Ezquiaga {\etal} method~\cite{Ezquiaga:2019ftu}.

\para{Discussion}
We close this section with some observations.
Eq.~\eqref{eq:Tomberg-constant-roll-P}
is interesting because it shows that
the transition probability $\ProbP$
can depend on $\Nstar$ in a complicated way,
not just as a sum of simple exponentials,
as the spectral method would suggest.
Even so,
in the formal limit $\Delta N \rightarrow \infty$,
the heaviest part of the tail
for~\eqref{eq:Tomberg-constant-roll-P}
would come from the linear $\epsilon_2 \Nstar/2$ piece.
(However, we shall see in~\S\ref{sec:CR}
that there are obstructions to taking this limit for 
first-passage statistics.)

More generally, one might wonder 
whether the spectral method can
encode contributions
to $\ln \ProbP(\Nstar)$
that are not simply linear in $\Nstar$.
This would include cases where the tail distribution is lighter
than exponential,
such as a Gaussian,
or heavier,
such as a power-law.
It is not difficult to achieve this for small
$\Nstar$, near the centre of the distribution and
far from the Kramers regime.
In this region,
all eigenvalues may contribute to the $\Nstar$ dependence.
The result may be a very complicated function of $\Nstar$,
and no clear statement can be made.
To have such contributions survive for rare $\Nstar$,
the formal solution~\eqref{eq:formal-Kolmogorov-soln}
must apparently be invalidated.
One way this could occur is if the adjoint Fokker--Planck
operator $\LFPadj$
has explicit time dependence.
We will see an example where this produces a Gaussian tail
in~\S\ref{sec:decaying-noise}.%
    \footnote{In Ref.~\cite{Ezquiaga:2019ftu},
    an alternative justification was given for the eigenfunction
    expansion, based on the calculus of residues.
    It was assumed that the Fourier transform
    of the characteristic function
    for $\ProbP(\Nstar)$ was meromorphic.
    In this case, the decay properties of the probability
    distribution at infinity are encoded in the pole
    structure of its Fourier transform.
    This relation is made precise by the Paley--Wiener
    theorem and its generalizations.

    If $\ProbP(\Nstar)$ decays exponentially
    at infinity, but not faster, then its
    Fourier transform will be meromorphic, as assumed
    in Ref.~\cite{Ezquiaga:2019ftu}.
    However, if it decays faster than exponentially
    the Fourier transform may be entire.
    In that case, there are no simple poles at finite
    locations in the complex plane;
    instead,
    information about the decay is encoded in
    the analytic structure at infinity.
    An example is provided by the Gaussian distribution,
    which is entire in the complex plane.
    It is well-known that the Fourier transform of a Gaussian is another Gaussian,
    so the characteristic function is also entire,
    but has an essential singularity at infinity.}

\section{A noise-supported instanton for the slow-roll first passage problem}
\label{sec:slow-roll-instanton}

In this section we
begin our presentation
of instanton methods
for the computation of tail probabilities.
We formulate the Starobinsky--Langevin
equation~\eqref{eq:Starobinsky-Langevin}
using a path integral method,
and
use this representation to compute
the tail of $\ProbQ(\Nstar)$
for the single-field, slow-roll
model studied by Ezquiaga {\etal}~\cite{Ezquiaga:2019ftu}
(\S\ref{sec:ezquiaga-spectral-method}).
We
demonstrate that the instanton
approach is able to reproduce
their tail estimate exactly.

In~\S\ref{sec:instanton-phase}
we repeat this analysis from the opposite
direction, ``top down'' rather than ``bottom up'',
in the sense that we work down from the
Schwinger--Keldysh path integral,
rather than up from the phenomenological Starobinsky equation.
This provides a systematic way to
drop the slow-roll approximation,
and also to
incorporate different levels of
microscopic detail into the tail estimate.
In~\S\ref{sec:Apps}
we report a series of case studies
showing how the instanton method can be used in practice,
including the constant roll model
studied by Tomberg~\cite{Tomberg:2023kli}
(\S\ref{sec:tomberg-langevin-method}).

\subsection{Path integral formulation of transition probabilities}
\label{sec:path-integral-transitions}

A number of
path integral representations
for the probability distribution function
have been discussed in
the literature.
Ref.~\cite{Seery:2006wk}
expressed the distribution function
as a functional Fourier
transform of a corresponding characteristic function.
By evaluating the path integral explicitly,
it was possible to reproduce the (perturbative)
Edgeworth expansion.
The same technique was used by Maggiore \&
Riotto~\cite{maggiore2009pathintegralapproachnonmarkovian}
for applications to collapsed structures.

As explained in~\S\ref{sec:introduction},
an approach of this type
works well near the centre of the
probability distribution, where the
characteristic function can be approximated by a few low-order
correlation functions.
However, to accurately evaluate tail probabilities
we require a representation that is valid even
for rare excursions.

\subsubsection{The Martin--Siggia--Rose action}
We now construct such a representation.
To compute expectation values for observables
associated with the
stochastic process $\phi(N)$,
we should average over all possible realizations
of $\phi(N)$ with an appropriate
probability weight.
In turn, this implies integration
over all realizations of the noise $\xi(N)$,
subject to the requirement that $\phi(N)$
responds to $\xi(N)$
as described by
the Starobinksy--Langevin equation~\eqref{eq:Starobinsky-Langevin}.
Assuming $\xi(N)$ to be a standard Brownian motion,
it follows that the required averaging
prescription is
\begin{equation}
    \label{eq:primitive-MSR-1}
    \ProbZ
    =
    \Normalization
    \int [ \d \phi \, \d \xi ]
    \;
    J
    \;
    \delta
    \bigg[
        \frac{\d \phi}{\d N}
        +
        \frac{V'}{3H^2}
        -
        \frac{H}{2\pi} \xi
    \bigg]
    \exp
    \bigg(
        {-\frac{1}{2}} \int \d N \, \xi^2
    \bigg)
    .
\end{equation}
$\Normalization$
is a normalization constant, to be
discussed below.
We write $\ProbZ$ to indicate that
this path integral should not yet be identified
with $\ProbP$, $\ProbPre$ or $\ProbQ$.
These identifications depend on the boundary conditions
we apply to the $\phi$ integral,
and also on further insertions
as described below.
In~\S\ref{sec:instanton-phase}
we will see that
the status of $\ProbZ$ is comparable to
the Schwinger--Keldysh partition function.

If $\xi$ has a more complicated
probability measure, it should replace the
Gaussian measure
in Eq.~\eqref{eq:primitive-MSR-1}.
The limits of the $\d N$ integral
match the time interval
of the transition,
and
the integral over $\xi$ is unrestricted.
The interpretation of
the $\delta$-functional
$\delta[ \cdots ]$
depends on the discretization of the
differential operator $\d / \d N$.
Once a discretization has been chosen,
$J$ is the corresponding Jacobian 
determinant.
In It\^{o}
discretization, which we are using,
it can be shown
that $J=1$.

Interpreted as an averaging
prescription,
$\ProbZ$ can be combined wth the
Feynman--Kac formulae~\eqref{eq:Feynman-Kac-transition}
for $\ProbP$ (or $\ProbPre$)
and~\eqref{eq:Feynman-Kac-Q}
for $\ProbQ$.
These yield the boundary
conditions to be satisfied by $\phi$.
To compute
an unrestricted transition
probability
$\ProbP( \phistop, \Nstop \divider \phiinit, \Ninit )$,
we should apply~\eqref{eq:Feynman-Kac-transition},
\begin{equation}
    \ProbP( \phistop, \Nstop \divider \phiinit, \Ninit )
    =
    \Normalization
    \int
    [\d \phi \, \d \xi]
    \;
    \delta
    \bigg[
        \frac{\d \phi}{\d N}
        +
        \frac{V'}{3H^2}
        -
        \frac{H}{2\pi} \xi
    \bigg]
    \exp
    \bigg(
        {-\frac{1}{2}} \int \d N \, \xi^2
    \bigg)
    .
\end{equation}
The integral is to be taken over all
$\phi(N)$
with $\phi(\Ninit) = \phiinit$
and $\phi(\Nstop) = \phistop$,
but without other constraints.
However, we caution that these
boundary conditions need to be applied
with care.
In particular, if we intend to use
$\ProbP$ to compute the average of a
noisy operator
evaluated at the final time,
it may disturb the boundary
condition required to produce the intended
transition.
For an explicit calculation,
see~\S\ref{sec:path-integral-flux-formula} below.
To obtain accurate results,
it is necessary to handle these disturbances correctly.

Alternatively, if we wish
to compute the restricted
transition probability
$\ProbPre$,
we should limit the integral to paths $\phi(N)$
that do not violate the first-passage condition
$\phi(N) > \phiend$~\cite{kumar1985quantum}.
We indicate this by attaching a prime to the integral
sign, viz. $\primedint$.
Because $\primedint$ does not include paths crossing
$\phi = \phiend$,
we
reproduce the boundary condition
$\ProbPre = 0$
at $\phistop = \phiend$.
Finally, we can compute the survival probability
$\Survival$
by integrating over paths satisfying the
initial boundary condition $\phi(\Ninit) = \phiinit$,
and which do not violate the first-passage condition,
but with $\phi(\Nstop)$ unrestricted except that it should remain
in the region $\phi > \phiend$.
The ability to represent all of $\ProbP$, $\ProbPre$
and $\Survival$ using the same integrand,
but with different choices for the integration domain,
makes the path integral an extremely flexible tool.
For either $\ProbP$ or $\ProbPre$, we can compute the corresponding
expectation value of any observable $\mathcal{O}$
by inserting it under the path integral.

Now consider the normalization $\Normalization$.
In principle, this
emerges from a careful consideration
of the path integral measure.
In practice it can be determined by demanding that
$\ProbP$ is properly normalized
to unity.
The $\Normalization$ obtained in this way can
depend on parameters describing the transition,
such as the
initial time $\Ninit$
and the final time $\Nstop$.
For $\ProbPre$
the normalization is
more complicated because of the limitation on paths.
An accurate evaluation may require Monte Carlo
methods.

We now
formulate a path integral for $\ProbQ$.
Using the
Feynman--Kac representation~\eqref{eq:Feynman-Kac-Q}
we obtain
\begin{multline}
    \label{eq:primitive-MSR-1-Q}
    \ProbQ( \phistop, \Nstop \divider \phiinit, \Ninit )
    =
    -
    \NormalizationRe(\Ninit, \Nstop)
    \primedint
    [ \d \phi \, \d \xi ]
    \;
    \left.\frac{\d \phi}{\d N}\right|_{N = \Nstop}
    \\
    \times
    \delta
    \bigg[
        \frac{\d \phi}{\d N}
        +
        \frac{V'}{3H^2}
        -
        \frac{H}{2\pi} \xi
    \bigg]
    \exp
    \bigg(
        {-\frac{1}{2}} \int \d N \, \xi^2
    \bigg)
    ,
\end{multline}
where the boundary conditions
on $\phi(N)$
are again $\phi(\Nstart) = \phistart$
and $\phi(\Nstop) = \Nstop$.
The same issues apply regarding interaction of the
boundary conditions with noise-sensitive
insertions under the path integral.
Also, 
$\phi(N)$
should satisfy the first-passage
constraint
$\phi > \phistop$ 
at intermediate times.
We have explicitly
noted that the
the normalization
$\NormalizationRe(\Ninit, \Nstop)$
may depend on
on $\Ninit$ and $\Nstop$,
or usually just $\Nstar$.
The prime $'$ indicates
that this normalization should be
inherited from the normalization of
$\ProbPre$.
This factor is awkward to handle,
because
(as explained above)
it can be difficult to determine.
In this paper,
as we explain in~\S\ref{sec:SP_restr_trans}
below, we will generally sidestep
this problem
by working
with $\ProbP$ rather than $\ProbPre$.

The remainder of the construction
follows the standard procedure
of Martin, Siggia \& Rose~\cite{Martin:1973zz}.
The path
integral formulation used here was proposed
by Janssen~\cite{janssen1976lagrangean},
de Dominicis~\cite{dominicis_1976}
and Phythian~\cite{R_Phythian_1977}.
The formalism has recently been used in a cosmological
context by Wilkins {\etal}~\cite{Wilkins:2020bsz,Wilkins:2021wde,Rigopoulos:2022gso}.%
    \footnote{Note that Wilkins {\etal} retain the Jacobian $J$
    in Eq.~\eqref{eq:primitive-MSR-1}, which they
    represent using a pair of Fadeev--Popov-like fields.
    This corresponds to a choice of discretization.
    Something similar was done in Ref.~\cite{Seery:2009hs}, which
    worked with an ``advanced'' anti-It\^{o} discretization
    and hence an opposite operator ordering convention.}
In what follows we give explicit formulae for $\ProbQ$,
with the understanding that
expressions for $\ProbP$, $\ProbPre$ and $\Survival$
can be obtained by suitable modifications.
The changes needed to represent
$\ProbQ$ in a multiple field model were discussed below
Eq.~\eqref{eq:Feynman-Kac-Q}.

To proceed,
note that
the $\delta$-functional
can be expressed
as a Fourier integral
using an auxiliary field $\Xmom$.
We use the symbol
``$\approx$''
to indicate that
we do not attempt to keep track of constant
prefactors,
which can be fixed by normalization of the final
probability distribution.
With this understanding, we have
\begin{multline}
    \label{eq:primitive-MSR-2}
    \ProbQ( \phistop, \Nstop \divider \phistart, \Nstart )
    \approx
    -
    \NormalizationRe(\Ninit, \Nstop)
    \primedint [ \d \phi \, \d \xi \, \d \Xmom ]
    \left.\frac{\d \phi}{\d N}\right|_{N = \Nstop}
    \\
    \times
    \exp
    \Bigg(
        \im \int_{\Nstart}^{\Nstop} \d N \;
        \Bigg[
            \Xmom
            \bigg(
                \frac{\d \phi}{\d N}
                +
                \frac{V'}{3H^2}
                -
                \frac{H}{2\pi} \xi
            \bigg)
            +
            \frac{\im}{2}
            \xi^2
        \Bigg]
    \Bigg) \;.
\end{multline}
The $\Xmom$ integral is
unconstrained.
The remaining $\xi$ integral is Gaussian
and can be performed explicitly.
There is a complication because
the insertion $\d \phi / \d N$ evaluated
at $N = \Nstop$
typically depends polynomially on
$\xi(\Nstop)$.
The effect of the Gaussian
integral over $\xi(\Nstop)$ is to replace this by
a (different) polynomial dependence on $X(\Nstop)$.
We continue to write $\d \phi / \d N$,
but from this point it should be regarded as a function of
$X(\Nstop)$,
rather than $\xi(\Nstop)$,
obtained by following this procedure.
If we wish only to compute $\ProbP$ or $\ProbPre$,
the insertion $\d \phi / \d N$ is not present and this
complication does not arise.
We conclude
\begin{multline}
    \label{eq:MSR-path-integral}
    \ProbQ( \phistop, \Nstop \divider \phistart, \Nstart )
    \approx
    -
    \NormalizationRe(\Ninit, \Nstop)
    \primedint [ \d \phi \, \d \Xmom ]
    \left.\frac{\d \phi}{\d N}\right|_{N = \Nstop}
    \\
    \times
    \exp
    \Bigg(
        \im
        \int_{\Nstart}^{\Nstop} \d N \;
        \Bigg[
            \Xmom
            \bigg(
                \frac{\d \phi}{\d N}
                +
                \frac{V'}{3 H^2}
            \bigg)
            +
            \frac{\im}{2}
            \bigg(
                \frac{H}{2\pi}
            \bigg)^2
            \Xmom^2 
        \Bigg]
    \Bigg)
    \\
    \equiv
    -
    \NormalizationRe(\Ninit, \Nstop)
    \int [ \d \phi \, \d \Xmom ]
    \left.\frac{\d \phi}{\d N}\right|_{N = \Nstop}
    \exp
    \Big(
        \im
        \SMSR[\phi, \Xmom]
    \Big) \;.
\end{multline}
Eq.~\eqref{eq:MSR-path-integral} is
the MSR~\cite{Martin:1973zz}
path integral for $\ProbQ$,
corresponding to the Langevin
equation~\eqref{eq:Starobinsky-Langevin}.
The MSR action
$\SMSR$ is defined in the last step.

This formulation of the
restricted transition probability
$\ProbPre$
in terms of
a path integral
seems to have been used first by Kumar (1985)~\cite{kumar1985quantum}.
An equivalent discussion was later
given,
in the context of reliability analysis
for a mechanical system,
by Iourtchenko {\etal} (2008)~\cite{10.1115/1.2967896},
and again in a cosmological context by Maggiore \&
Riotto (2009)~\cite{maggiore2009pathintegralapproachnonmarkovian}.

In the ``top down''
formulation of~\S\ref{sec:instanton-phase},
we will see that Eq.~\eqref{eq:MSR-path-integral}
can be obtained
from the Schwinger--Keldysh path integral
on a closed time contour.
In this framework, $\Xmom$
arises from the ``difference''
or ``quantum'' field
of the Keldysh basis,
$\qu{\phi} \equiv \phiplus - \phiminus$,
whereas
$\phi$
arises from the ``average''
or ``classical'' field $\cl{\phi} \equiv (\phiplus + \phiminus)/2$.
The $X^2$ term encoding the
statistics of the noise
would correspond to a term $\sim \qu{\phi}^2$
in the closed time path action,
which is ordinarily forbidden.
Terms of this kind, which mix the histories on the $+$
and $-$ contours,
are characteristic of interaction with an environment.
The Schwinger--Keldysh
path integral provides a systematic tool to
compute these effective vertices---%
and hence the statistics of the
noise---%
direct from microphysical
data, rather than impose it by hand as we have
done here.
For details, see~\S\ref{sec:instanton-phase}
and Appendix~\ref{sec:Inf_func}.

\para{Onsager--Machlup functional}
A related representation for the
transition probability density $\ProbP$
was introduced by
Onsager \& Machlup (1953)~\cite{PhysRev.91.1505}.
They
computed it using a functional
(the ``Onsager--Machlup functional''),
equivalent to Eq.~\eqref{eq:tomberg-onsager-machlup}
in Tomberg's formulation.
This functional
can be obtained from the MSR path integral
by integrating out the
Fokker--Planck momentum $\Xmom$
in
Eq.~\eqref{eq:MSR-path-integral}.
This approach was followed in Appendix B of Ref.~\cite{Seery:2009hs}.
It is less general than the
MSR path integral~\eqref{eq:MSR-path-integral}
because one cannot study the noise realization
and the stochastic response $\phi(N)$ separately.

\subsubsection{Fokker--Planck Hamiltonian and Kolmogorov equations}\label{sec:FPH}
We now temporarily revert
to the unrestricted transition probability $\ProbP$.
We wish to show that the MSR path integral
for $\ProbP$
is equivalent to the
forward and backward Kolmogorov
equations~\eqref{eq:kolmogorov-forwards}
and~\eqref{eq:kolmogorov-backwards}.

To see the equivalence, notice that
the MSR action
$\SMSR$
in Eq.~\eqref{eq:MSR-path-integral}
has a Hamiltonian structure, in the sense
that
$\SMSR \sim \Xmom \dot{\phi} - \Hamiltonian$,
where $\dot{\phi} = \d \phi / \d N$
and
$\Xmom$ plays the role of the
momentum conjugate to $\phi$.
This is only a formal interpretation
based on the structure of the integrand,
and does not mean that $\Xmom$ really \emph{is}
the canonical momentum, which here would be $\d \phi / \d N$.
In reality, we will see that $\Xmom$ can be regarded as
encoding a particular realization of the noise.

We take $\ProbP$ to be
defined by~\eqref{eq:MSR-path-integral},
without a first-passage constraint on the paths $\phi(N)$,
and with the insertion $\d \phi / \d N$ removed.
It should be
regarded as a functional of the initial and final
field configurations, $\phistart$ and $\phistop$, respectively.
First, take the initial time $\Nstart$ to be fixed.
From any given final time $\Nstop$, we can evolve
to a later time
using the Schr\"{o}dinger equation
obtained from its Hamiltonian structure.
One option is to calculate the time derivative
explicitly by
cutting open the path
integral at time $\Nstop$ and considering the
evolution to $\Nstop + \d \Nstop$.
This approach was described by Feynman \& Hibbs~\cite{feynman1965path}.

Alternatively,
we may exploit the
formal Hamiltonian
structure associated with $\SMSR$.
Define a Hamiltonian
$\HFP(\Xmom, \phi)$
by the usual Legendre transform.
This yields
\begin{equation}
    \label{eq:Fokker-Planck-Hamiltonian}
    \HFP(\Xmom, \phi) 
    =
    -
    \Xmom \frac{V'}{3H^2}
    -
    \im \Xmom^2 \frac{H^2}{8\pi^2} \;.
\end{equation}
This Hamiltonian has a similar interpretation to the ``momentum'' $\Xmom$.
It is not the true Hamiltonian for $\phi$, but only a formal tool.
We describe it as the \emph{Fokker--Planck Hamiltonian}.
In It\^{o} regularization, time-ordering
means that the momenta $\Xmom$
should be positioned to the left of all
fields $\phi$.
In the usual coordinate representation we have
\begin{equation}
    \Xmom
    \rightarrow \hat{\Xmom}
    =
    -
    \im \frac{\partial}{\partial \phi} ,
    \quad
    \phi
    \rightarrow \hat{\phi}
    =
    \phi \;.
\end{equation}
The resulting
time-dependent Schr\"{o}dinger equation is the
forward Kolmogorov equation.
After cancelling a common factor of $\im$ we
reproduce Eq.~\eqref{eq:kolmogorov-forwards},
\begin{equation}
    \label{eq:Euclidean-Schrodinger}
    \frac{\d \ProbP}{\d \Nstop}
    = \HFP
    \big(
        \hat{\Xmom},
        \hat{\phi}
    \big)
    \ProbP
    =
    \frac{\partial}{\partial \phi}
    \bigg(
        \frac{V'}{3H^2}
        \ProbP
    \bigg)
    +
    \frac{\partial^2}{\partial \phi^2}
    \bigg(
        \frac{H^2}{8\pi^2}
        \ProbP
    \bigg) \;.
\end{equation}
In the same way, $\ProbP$ satisfies the backward
Kolmogorov equation~\eqref{eq:kolmogorov-backwards},
considered as a function of the 
initial time $\Nstart$ and field value $\phistart$.

The same analysis,
for both the forward and backward
equations, applies to the restricted transition
probability $\ProbPre$.
At the level of the Kolmogorov equations, the difference
between $\ProbP$ and $\ProbPre$
amounts to a boundary condition.

\subsubsection{Flux formula from the path integral}
\label{sec:path-integral-flux-formula}
As a final step before discussing rare events,
we validate the path integral
expression for $\ProbQ$,
Eq.~\eqref{eq:MSR-path-integral},
by showing that it reproduces the flux formula~\eqref{eq:first-passage-probability-current}
obtained from the
Kolmogorov probability current $\Current$.
This calculation is related to the regularization
of operator products, and will influence our
interpretation of the instanton approximation.

The Feynman--Kac formula~\eqref{eq:Feynman-Kac-Q}
shows that the first-passage distribution can be evaluated
from insertion of
the microscopic current
operator
$\d \phi / \d N |_{\Nstop} \delta [ \phi(\Nstop) - \phiend ]$
under the path integral.
As we now describe,
because the operator $\d \phi / \d N$
is sensitive to the noise, it can interact
with the final boundary condition enforced by the $\delta$-function.

The Langevin equation for $\d \phi / \d N$ is
Eq.~\eqref{eq:Starobinsky-Langevin}.
It was
explained below Eq.~\eqref{eq:primitive-MSR-2}
that
$\xi$ should be replaced
during the Hubbard--Stratonovich transformation to $\Xmom$.
We should therefore regard $\d \phi / \d N$ as the operator
\begin{equation}
    \label{eq:HS-Langevin-equation}
    \frac{\d \phi}{\d N}
    =
    -
    \frac{V'}{3H^2}
    -
    \im
    \left(
        \frac{H}{2\pi}
    \right)^2
    \Xmom ,
\end{equation}
in which $\Xmom$ encodes the properties of the noise.
This equation is equivalent to the later instanton equation~\eqref{eq:SR-instanton-phi}.
The insertion required to compute $\ProbQ$
therefore contains the
composite operator
$\Xmom(\Nstop) \delta[ \phi(\Nstop) - \phiend ]$,
which involves
a product of operators defined at the same time
and must be regularized.
Heuristically, one can regard the noise field
$\Xmom(\Nstop)$
as slightly displacing the
value of the field at the final time $\Nstop$.
To study the intended transition,
we need to change the final boundary condition
we impose on the
$[\d \phi]$ integration
so that we no longer integrate over paths
that arrive exactly at $\phiend$.
Instead, we should have $\phi(\Nstop) = \phiend$
only after accounting for the disturbance caused
by $\Xmom(\Nstop)$.

Consider
an expectation value such as
$\langle \Xmom(\Nstop) F[ \phi(\Nstop) ] \rangle$,
where $F$ is any differentiable function.
To regularize it,
we define the operator product
by symmetric point splitting,
\begin{equation}
    \label{eq:noise-correlation-point-splitting}
    \langle \Xmom(\Nstop) F[ \phi(\Nstop) ] \rangle
    \overset{\text{def}}{=}
    \lim_{\epsilon \rightarrow 0}
    \bigg\{
        \frac{1}{2}
        \Big\langle \Xmom(\Nstop) F[ \phi(\Nstop + \epsilon) ] \Big\rangle
        +
        \frac{1}{2}
        \Big\langle \Xmom(\Nstop) F[ \phi(\Nstop - \epsilon) ] \Big\rangle
    \bigg\}
    .
\end{equation}
For simplicity, we assume the drift velocity
$V'/3H^2$ and noise amplitude $(H/2\pi)^2$
are constant.
This parallels the approximation we will use in~\S\ref{sec:instanton-linear-model}.
The correlation functions obtained from Eq.~\eqref{eq:MSR-path-integral}
are
\begin{subequations}
\begin{equation}
    \label{eq:noise-field-correlation}
    \Big\langle
        \phi(N) \Xmom(N')
    \Big\rangle
    =
    \Bigg\{
        \begin{array}{l@{\hspace{3ex}}l}
            - \im & N > N' \\
            0 & N < N'
        \end{array}
    ,
\end{equation}
and
\begin{equation}
    \label{eq:noise-noise-correlation}
    \Big\langle
        \Xmom(N) \Xmom(N')
    \Big\rangle
    =
    \left( \frac{H}{2\pi} \right)^{-2} \delta(N - N') .
\end{equation}
\end{subequations}
In Eq.~\eqref{eq:noise-field-correlation}
we have imposed retarded boundary conditions.
This corresponds to the causality requirement that there
is no correlation between the field $\phi(N)$
and a noise event $\Xmom(N')$
at time $N'$,
until the noise has been absorbed into $\phi$.
This absorption
is described by the Langevin equation~\eqref{eq:HS-Langevin-equation}.
A small time interval $\epsilon$ after the noise acts, we find
\begin{equation}
    \phi(N + \epsilon)
    \approx
    \phi(N)
    -
    \frac{V'}{3H^2} \epsilon
    -
    \im
    \left( \frac{H}{2\pi} \right)^2
    \int_{N}^{N+\epsilon} \, \d N' \, \Xmom(N')
    .
\end{equation}
It follows that, up to terms of order $\Or(\epsilon^2)$,
\begin{multline}
    \Big\langle
        \Xmom(N_1) F[\phi(N_1 + \epsilon)]
    \Big\rangle
    \approx
    \Big\langle
        \Xmom(N_1) F[\phi(N_1-0)]
    \Big\rangle
    \\
    +
    \bigg\langle
        \Xmom(N_1)
        \bigg(
            \frac{\partial}{\partial \phi} F[\phi(N_1-0)]
        \bigg)
        \bigg(
            -
            \frac{V'}{3H^2} \epsilon
            -
            \im \Big( \frac{H}{2\pi} \Big)^2
            \int_{N_1-0}^{N+\epsilon} \d N' \, \Xmom(N')
        \bigg)
    \bigg\rangle
    +
    \Or(\epsilon^2)
\end{multline}
The notation $N_1 - 0$ indicates that, for the purpose
of computing correlations, we regard $N_1 - 0$ as being
infinitesimally earlier than $N_1$.
This ensures the correct causality properties.
Using Eqs.~\eqref{eq:noise-field-correlation}--\eqref{eq:noise-noise-correlation}
we obtain
\begin{equation}
    \lim_{\epsilon \rightarrow 0}
    \frac{1}{2}
    \Big\langle
        \Xmom(N_1) F[\phi(N_1 + \epsilon)]
    \Big\rangle
    =
    -
    \frac{\im}{2}
    \bigg \langle
        \frac{\partial}{\partial \phi}
        F[\phi(N_1)]
    \bigg \rangle
    .
\end{equation}
A similar calculation shows that the
other
expectation value in~\eqref{eq:noise-correlation-point-splitting}
is zero.
It follows that
\begin{equation}
    \ProbQ
    =
    - \NormalizationRe
    \primedint [ \d \phi \, \d \Xmom ]
    \bigg(
        - \frac{V'}{3H^2}
        - \frac{1}{2} \left( \frac{H}{2\pi} \right)^2
        \frac{\partial}{\partial \phiend}
    \bigg)
    \delta[ \phi(\Nstop) - \phiend ]
    \exp
    \Big(
        \im \SMSR
    \Big)
    ,
\end{equation}
where now the end-point of the $[\d \phi]$ integration is unrestricted. We conclude
\begin{equation}
    \ProbQ
    =
    \left(
        \frac{V'}{3 H^2} \ProbPre
        +
        \frac{H^2}{8\pi^2}
        \frac{\partial \ProbPre}{\partial \phi}
    \right)_{\phi = \phiend}
    =
    - \Current \big|_{\phi = \phiend} ,
\end{equation}
in agreement with Eq.~\eqref{eq:first-passage-probability-current}.

Note that the diffusion term would appear with an incorrect
coefficient
if we did not impose the point-splitting regularization,
Eq.~\eqref{eq:noise-correlation-point-splitting}.
It is well known that regularization effects
can change the equations of motion satisfied by composite operators.
In principle, it would have been possible impose
a different regularization by weighting the
point-splitting formula~\eqref{eq:noise-correlation-point-splitting}
asymmetrically.
In this case,
we should
regard the symmetric
splitting
as the required regularization
needed to produce a conserved current.

A similar issue arises in any path integral where we
attempt to compute an expectation value by insertion
of a microscopic operator that produces a noise disturbance
in the final state. The lesson of the calculation in this section is
that, although it is safe to exchange the $\delta$-function
$\delta[\phi(\Nstop) - \Nstop]$
for a final boundary condition on the fields \emph{in the transition
probability}, this is \emph{not} safe in general.
Where an inserted operator disturbs the final state, we should instead
retain the $\delta$-function explicitly
and regularize the resulting operator product.
Regularization
produces a perturbed $\delta$-function that imposes
the correct boundary condition for the
intended transition, in the presence of the disturbance
in the final state.
It will be important to bear this in mind when constructing
instanton approximations to path integrals in~\S\ref{sec:instanton-solution}
and~\S\ref{sec:instanton-phase}
below,
because the boundary conditions on the instanton
are inherited from boundary conditions on the fields
entering the path integral.

\subsection{Rare events and instantons}
\label{sec:instanton-solution}
The conclusion of~\S\ref{sec:path-integral-transitions}
is that the
MSR path integral
is equivalent to knowledge of the forward and
backward Kolmogorov equations,
or (equivalently)
the underlying Langevin equation.
To analyse a typical transition it is
a matter of convenience which we use.
Neither the Kolmogorov equations,
the Langevin equation,
or the
path integral are easy to solve explicitly,
so in changing representation
we only exchange one set of difficulties for another.

However, our interest is in \emph{rare}
transitions.
For rare events the path integral has
significant advantages,
because it provides the option
of a saddle point approximation.
It is possible to obtain
similar approximations in other ways;
see, e.g.,
Ach\'{u}carro {\etal}~\cite{Achucarro:2021pdh}
who combined a saddle point method with the
approach of Ezquiaga {\etal}~\cite{Ezquiaga:2019ftu}.
However, the path integral formulation makes
such approximations especially easy to obtain.
Further, we shall see that in the MSR
framework, the saddle point trajectory has
a clear physical interpretation
involving the least unlikely noise realization
that is able to support the transition.

\para{Noiseless transitions and stationary phase}
First, consider
a typical transition,
for which $\phiinit$, $\phistop$ and $\Nstar \equiv \Nstop - \Ninit$
are
close to an allowed noiseless transition.
We continue to focus on $\ProbP$,
with the changes needed for $\ProbPre$
and $\ProbQ$
to be discussed below.

The noiseless limit corresponds to switching off the
$\im \Xmom^2$ term in $\SMSR$,
so for such transitions
we expect that
$\SMSR$ is almost real.
It follows that
the probability density
is localized near the noiseless trajectory.
This is because
most
trajectories generate
rapid phase variations in
$\exp( \im \SMSR )$,
producing destructing interference.
The contribution from such trajectories
is exponentially suppressed.
Destructive interference is absent only
near a critical point of $\SMSR$,
which
occurs for trajectories $\phi(N)$
that
satisfy the almost-noiseless evolution equation.
This is the
stationary phase (or ``semiclassical'')
approximation to the path integral.
This scenario was considered by
Wiegel (1967)~\cite{WIEGEL1967105}
and later Moreau (1978)~\cite{MOREAU1978410}.

\para{Noise-supported transitions}
Rare transitions correspond to combinations
of $\phiinit$, $\phistop$ and $\Nstar$
for which there is no noiseless trajectory,
and therefore no
stationary point of $\SMSR$
with $\Xmom \approx 0$.
Instead, we expect any critical point
to be noise-supported with $\Xmom \neq 0$.

Critical points may occur for real
or complex values
of $\phi(N)$
and $\Xmom(N)$.
Although the original contour of integration
does not pass through any complex critical points,
they may still control the
behaviour of the path integral.
We can regard $\SMSR$
as a holomorphic
function of complexified
fields
$\phi(N)$ and $\Xmom(N)$,
and the original integral as a contour integral.
It follows from complex analyticity that any critical
point of $\SMSR$ must be a saddle.
If it is possible to deform the contour of integration
to pass through the saddle, then
contributions from its neighbourhood will be
exponentially dominant.
Therefore,
a simple approximation to
the tail probability can be obtained
immediately by evaluation of the MSR action
at the saddle-point solution.%
    \footnote{Clearly this strategy is similar to
    estimation of the tail probability by saddle-point
    evaluation of Eq.~\eqref{eq:pdf-from-fourier-characteristic}.
    The difference is that an action is being used
    to obtain the location of the saddle
    \emph{nonperturbatively},
    unlike the perturbative
    series expansion of $\ln \chi(t_k)$
    entailed by computing correlation functions as an intermediate step.}
It is this feature that makes the path integral
and saddle point approximation
so powerful for rare events.
The trajectory corresponding to the saddle point is called an \emph{instanton},
or sometimes an \emph{optimal path},
by analogy with the instanton method in quantum field 
theory~\cite{Belavin:1975fg,tHooft:1976snw,Coleman:1978ae}.
(For a textbook treatment, see Ref.~\cite{Paranjape:2017fsy}.)

\subsubsection{The instanton equations}
The equations for a critical point of $\SMSR$
are
\begin{subequations}
\begin{align}
    \label{eq:SR-instanton-phi}
    \frac{\d \phi}{\d N} + \frac{V'(\phi)}{3H^2} + \im \left( \frac{H}{2\pi} \right)^2 \Xmom
    &
    =
    0 \,, \\
    \label{eq:SR-instanton-X}
    -\frac{\d \Xmom}{\d N} + \frac{V''(\phi)}{3H^2} \Xmom
    &
    =
    0 \,,
\end{align}
\end{subequations}
The boundary conditions for
$\phi(N)$
are inherited from the
path integral.
If this describes a transition between
$\phiinit$ and $\phistop$,
then we must set
$\phi(\Ninit) = \phiinit$
and $\phi(\Nstop) = \phistop$
(subject to the discussion of~\S\ref{sec:path-integral-flux-formula}).
There are no further boundary conditions for $\Xmom(N)$.
Together, the initial and final conditions on $\phi(N)$ fix the
expected two constants of integration from
Eqs.~\eqref{eq:SR-instanton-phi}--\eqref{eq:SR-instanton-X}.

Eqs.~\eqref{eq:SR-instanton-phi}--\eqref{eq:SR-instanton-X}
have a clear physical interpretation, as follows.
The first instanton equation,
Eq.~\eqref{eq:SR-instanton-phi},
is the Starobinsky--Langevin
equation~\eqref{eq:Starobinsky-Langevin} for
a very specific realization of the noise,
corresponding to $\xi(N) = - \im (H/2\pi) \Xmom(N)$.
In order to generate a real $\xi(N)$,
we must take $\Xmom(N)$ to be imaginary.
Therefore, we set $\Xmom(N) = \im \XEmom(N)$.
The time dependence of $\XEmom(N)$
is determined
by the second instanton equation,
Eq.~\eqref{eq:SR-instanton-X},
which yields a self-consistent real solution.
We write this saddle-point solution
as $\phiSP(N)$,
$\XEmomSP(N)$.
It is located
at a real value of $\phi$,
but an imaginary value of $\Xmom$.
This is reasonable,
in order that $\phi(N)$ can still be given
an unambiguous interpretation as a physical trajectory
connecting $\phiinit$ and $\phistop$.

Under which circumstances is it possible to deform
the contour of integration to pass through
the saddle point?
To do so,
define the path integral by time slicing,
meaning that
we discretize the time axis
into a mesh of points $N_1$, $N_2$, \ldots, $N_K$
spaced evenly between the initial and final times.
We define
$\phi_j = \phi(N_j)$
and
$\Xmom_j = \Xmom(N_j)$
for $1 \leq j \leq K$,
and likewise
$\phiSP_j = \phiSP(N_j)$
and 
$\XEmomSP_j = \XEmomSP(N_j)$.
For $\ProbP$, both the
$\phi(N)$
and $\Xmom(N)$ integrals are unrestricted.
Therefore we should define the path integral $\int [\d \phi \, \d X]$
as
\begin{equation}
    \label{eq:path-integral-time-slice-measure}
    \int [\d \phi \, \d X]
    \sim
    \int_{-\infty}^{+\infty} \d \phi_1
    \int_{-\infty}^{+\infty} \d \phi_2
    \cdots
    \int_{-\infty}^{+\infty} \d \phi_K
    \int_{-\infty}^{+\infty} \d X_1
    \int_{-\infty}^{+\infty} \d X_2
    \dots
    \int_{-\infty}^{+\infty} \d X_K ,
\end{equation}
up to a normalization that we do not write explicitly,
with the understanding that we are to take the
limit $K \rightarrow \infty$
in which the mesh becomes dense.
Other interpretations of
the path integral measure
are possible~\cite{kleinert2006path},
but
we expect
that these will yield equivalent results
when the
contour deformation is well-defined.
The saddle point at time $N_j$
occurs at $(\phi_j, \Xmom_j) = (\phiSP_j, \im \XEmomSP_j)$.
Since each $\Xmom_j$ integral
extends to $\pm \infty$,
the
$\im \Xmom_j^2$ term in $\SMSR$
causes exponential decay for $|\Xmom_j| \rightarrow \infty$
provided the imaginary part of $\Xmom_j$ remains bounded.
Therefore we should displace the $\Xmom_j$ integral
to run along the contour
$\Xmom_j = x_j + \im \XEmomSP_j$, where $-\infty < x_j < +\infty$.
There is no contribution
from the small contours at infinity needed to connect
this path to the original contour.
Finally, expanding the phase function
on the displaced contour to quadratic
order around
$(\phiSP_j, \im \XEmom_j)$
yields a convergent Gaussian
integral
centred on the saddle.
Repeating this procedure for each $N_j$
yields
the
required approximation.

\subsubsection{Saddle point approximation for restricted transitions}
\label{sec:SP_restr_trans}
Now consider the changes needed to apply this argument to the
restricted path integral $\primedint [\d \phi \, \d \Xmom]$
used in $\ProbPre$ and $\ProbQ$.

\para{Approximation for $\ProbPre$}
As explained in~\S\ref{sec:stochastic-inflation},
there is no known model-independent relation between
$\ProbP$ and $\ProbPre$.
However,
in scenarios where such a relation exists,
it can be used to
translate an instanton approximation for $\ProbP$ to
one for $\ProbPre$.
For example, for pure Brownian motion there is a
famous reflection
formula~\cite{kumar1985quantum,Balakrishnan2021}
\begin{equation}
    \ProbPre(\phistop, \Nstop \divider \phiinit, \Ninit)
    \sim
    \ProbP(\phistop, \Nstop \divider \phiinit, \Ninit)
    -
    \ProbP(2 \phiend - \phistop, \Nstop \divider \phiinit, \Ninit)
    .
    \label{eq:method-of-images}
\end{equation}
This satisfies the absorbing boundary condition
$\ProbPre(\phiend, \Nstop \divider \phiinit, \Ninit) = 0$.
If $\phistop$ is far from the boundary, we expect the
second ``image'' term to be exponentially
suppressed relative the first term.
However, where $\phistop$ is close to the boundary,
the subtraction of the image term is important.

To relate this to the saddle point approximation for
$\ProbPre$,
note that the restricted path integral
$\primedint$
enforces the constraint $\phi(N) > \phiend$.
Therefore
the 
$\Xmom_j$ integrals in Eq.~\eqref{eq:path-integral-time-slice-measure}
are unchanged,
but
each $\phi_j$ integral in~\eqref{eq:path-integral-time-slice-measure}
now runs only over the interval $\phiend < \phi_j < +\infty$.
If the saddle point $\phiSP_j$ is far from the boundary $\phi = \phiend$
at all times, then we may extend the $\phi_j$ integral
to the entire real line at the cost of an exponentially small error.
This erases knowledge of the boundary
and corresponds to dropping the exponentially small image term
in~\eqref{eq:method-of-images}.
On the other hand,
if $\phiSP_j$ is close to $\phiend$
then we cannot extend the $\phi_j$ contour in this way.
If a saddle occurs
at the endpoint of the range of integration,
it is still possible to use the saddle point
approximation but the details are different.
After expansion of the phase function,
the resulting Gaussian integral over the contour covers only a half-line.
It would be interesting to see how ``method of images''
formulae such as
Eq.~\eqref{eq:method-of-images}
emerge from such a careful saddle point analysis,
but we do not pursue this idea here.

The main conclusion is that it is not straightforward to
obtain a direct
estimate of $\ProbPre$ from the path integral
(at least near the boundary),
even in the instanton approximation.
We would also have the problem of obtaining an
accurate
normalization $\NormalizationRe$.
In this paper, when we need $\ProbPre$ explicitly,
we prefer to build an instanton
approximation for the unrestricted transition
probability $\ProbP$,
and then use one of the standard methods
to relate it to $\ProbPre$.
As discussed below Eq.~\eqref{eq:primitive-MSR-1-Q},
this has the advantage that we avoid the
need to determine $\NormalizationRe$.

\para{Approximation for $\ProbQ$}
A similar discussion applies for $\ProbQ$.
In particular, because $\ProbQ$ is built from the same
restricted path integral $\primedint$
we must account for the presence of the absorbing
boundary at $\phi = \phiend$.
For $\ProbQ$ we inevitably encounter the problem
of interaction with the boundary, because
the path integral is based on the restricted
transition probability to arrive there.

However,
for $\ProbQ$ there is a further complication because the
discussion of~\S\ref{sec:path-integral-flux-formula}
shows that we must be careful to account
for interactions
between the insertion
$\d \phi / \d N$
and the final boundary condition.
It is not clear how to do this in the instanton approximation,
where we would normally
evaluate such a prefactor on the instanton solution,
yielding $\d \phiSP / \d N$.
We have checked in explicit examples that doing so
produces a diffusion contribution that is too large by a factor of 2,
exactly as the analysis of~\S\ref{sec:path-integral-flux-formula}
would suggest.
Fortunately,
it would appear that
this discrepancy
typically affects only the subexponential prefactor,
not the exponential rate estimate.

In the language of~\S\ref{sec:path-integral-flux-formula},
use of
$\d \phiSP / \d N$ in the instanton approximation
corresponds to working in the purely
``advanced'' regularization where the noise field
$\Xmom(\Nstop)$ is taken to occur
earlier than the boundary
constraint $\delta[ \phi(\Nstop) - \phiend ]$.
Therefore, at least if wish to
accurately evaluate the subexponential
prefactor in $\ProbQ$,
the conclusion is apparently that we must
use
the flux formula~\eqref{eq:first-passage-probability-current},
which we know to be
correctly reproduced by the regularized path integral.
Clearly this technical situation
deserves further attention, which we defer to the future.

We therefore arrive at the following
prescription to estimate $\ProbQ$.

First, if an explicit relation can be found between
$\ProbPre$ to $\ProbP$,
then the instanton approximation for $\ProbP$
can be converted into an instanton approximation for $\ProbQ$
by use of Eq.~\eqref{eq:first-passage-probability-current}.
This is the most favourable situation.
In this scenario, we can accurately obtain subexponential
prefactors that correct the dominant exponential
estimate.
These prefactors would come from the fluctuation determinant
produced by the Gaussian integrals over the displaced
contour (see below), combined with the correct normalization of
$\ProbP$, and factors from the flux formula~\eqref{eq:first-passage-probability-current}.

Second, if no explicit relation
can be found between $\ProbPre$ and $\ProbP$,
it may still be possible to estimate $\ProbQ$
to exponential accuracy
by using the flux formula~\eqref{eq:first-passage-probability-current}
to make an estimate,
essentially in the form
\begin{equation}
    \ProbQ(\phiend) = - \CurrentRe(\phiend)
    =
    \bigg(
        v \ProbPre
        +
        D \frac{\partial \ProbPre}{\partial \phistop}
    \bigg)
    \bigg|_{\phistop = \phiend}
    \approx
    D
    \frac{\partial \ProbP}{\partial \phistop}
    \bigg|_{\phistop = \phiend}
    .
\end{equation}
Here, $v$ schematically denotes the drift velocity
and $D$ the diffusion constant,
and we have used the absorbing boundary condition
$\ProbPre(\phistop = \phiend) = 0$.
In the last step we have simply
made the crude estimate that the tail of
$\partial \ProbPre / \partial \phi$
is
equivalent to the tail of 
$\partial \ProbP / \partial \phi$.
Clearly, we cannot expect to obtain
a sensible estimate of the prefactor
in this way.
For many of the models considered in this paper
this procedure will work, but not for all.
In particular,
it
fails for
the constant roll model considered in~\S\ref{sec:CR}.

\para{Further considerations}
There is no guarantee that an instanton exists
for all possible combinations
of $\phiinit$, $\phiend$, and $\Nstar$.
If there is no instanton,
we have no clear prediction for the
probability of the transition but it is presumably
very strongly suppressed---more strongly than the exponential
suppression for rare transitions that \emph{do} have an
instanton description.
It is also possible that there is more than one saddle point.
If so, one saddle may be exponentially less
suppressed
than the others, in which case it dominates
the probability distribution.
Alternatively,
if several saddles contribute equally, to exponential accuracy,
then we must usually sum over their contributions.
Where several saddles exist, it is possible to obtain
Stokes-like phenomena
as the dominant contribution switches between them.

This approach to rare events,
based on saddle points obtained by
analytic continuation,
was initiated by
Zittarz \& Langer (1966)~\cite{zittartz1966theory,LANGER1967108,LANGER1969258}.
Any path integral formulation makes these methods especially easy to apply.
However, with more effort, the same results can be
recovered from a WKB approximation to the
Kolmogorov equations (forward or backward).
This is analogous to computation of quantum tunelling probabilities
using the Schr\"{o}dinger
equation.%
    \footnote{To obtain an MSR-like formulation with response fields,
    it is possible to use an eikonal approximation.
    See, e.g., §7.10 of Weinberg~\cite{weinberg2015lectures}.}
Many early results were obtained in this way.
Calculation of transition probabilities for rare noise-supported events
appears to have been considered first by Caroli, Caroli \& Rouet
(1979)~\cite{caroli1979diffusion,caroli1981diffusion}.
They worked at the level of the Fokker--Planck equation, using a WKB
approximation
to find an instanton describing transitions between multiple minima.
A similar analysis was given in the context of
chemical kinetics by Dykman {\etal} (1994)~\cite{dykman1994large}.
The method of Caroli {\etal}
is very similar to the spectral analysis of
Starobinsky's Fokker--Planck equation,
introduced by Starobinsky \& Yokoyama (1994)~\cite{Starobinsky:1994bd},
and refined in Refs.~\cite{Pattison:2017mbe,Ezquiaga:2019ftu}.
Indeed, Starobinsky \& Yokoyama already noted
that the Fokker--Planck equation could describe certain
instanton-like solutions.

Application of saddle-point
methods specifically
to the MSR path integral
was initiated by Falkovitch {\etal} (1995)~\cite{Falkovich:1995fa}.
Gurarie \& Migdal (1995)~\cite{Gurarie:1995qc}
applied their formalism
to determine the rare tail for events
in a turbulent flow described by Burgers' equation.
These authors worked in an older stochastic formalism
due to Wyld (1961)~\cite{WYLD1961143}.
However,
this is equivalent to the MSR method,
at least to the order used here~\cite{Berera_2013}.

\subsection{Linear potential model of Ezquiaga {\etal}}
\label{sec:instanton-linear-model}
As an example,
we apply this formalism to
the single-field, slow-roll model
with linear potential.
Ezquiaga {\etal}~\cite{Ezquiaga:2019ftu}
obtained a tail estimate for this model,
described in~\S\ref{sec:ezquiaga-spectral-method}.

In the linear model we take $V'/3H^2$ to be constant.
This is equivalent to the stochastic model of biased
diffusion. 
Dropping the slow-roll suppressed term
$V''/3H^2$ in~\eqref{eq:SR-instanton-X}
shows that $\XEmom(N) = \text{constant}$.
The boundary condition
$\phi(\Nstar) = \phiend$
serves to fix the amplitude of $\XEmom$.
If we assume for convenience that $\Ninit = 0$,
one can verify that the required solution is
\begin{subequations}
\begin{align}
    \label{eq:SR-instanton-soln-phi}
    \phiSP(N)
    &
    =
    \phiinit
    +
    \frac{N}{\Nstar} (\phiend - \phiinit) \,,
    \\
    \label{eq:SR-instanton-soln-P}
    \XEmomSP(N)
    &
    =
    \left( \frac{H}{2 \pi} \right)^{-2}
    \left(
        \frac{V'}{3 H^2}
        +
        \frac{\Delta\phi}{\Nstar}
    \right) \,,
\end{align}
\end{subequations}
where $\Delta\phi = \phiend - \phiinit$.
Eq.~\eqref{eq:SR-instanton-soln-phi}
describes steady, incremental progress
from $\phiinit$ to $\phiend$,
without large jumps.
This is consistent with the conclusions
of Cohen {\etal}~\cite{Cohen:2022clv},
who argued that inclusion of large fluctuations would
make the noise realization exponentially more unlikely.
Notice that the solution
$\phiSP(N)$ is independent of all details
of the model, except the initial and final field values.
Model-dependent details
appear only in the noise realization $\XEmomSP(N)$.

In particular,
Eq.~\eqref{eq:SR-instanton-X}
shows that the noise amplitude is adjusted to
\emph{almost}
precisely cancel the deterministic drift term $V'/3H^2$.
The small imbalance, proportional to $\Delta\phi / \Nstar$,
is tuned to
bring $\phi(N)$ to the end-point $\phiend$
in exactly $\Nstar$
e-folds.
Clearly, this is closely related to the interpretation of the
solution~\eqref{eq:stochastic-constant-roll}--\eqref{eq:GDN_transf}
in Tomberg's formalism.
The primary difference is that the instanton equations
select a single, specific noise realization $\xi(N)$.
Access to this realization is extremely useful.
It is a special feature of the MSR formalism
that has no parallel
in the frameworks of Ezquiaga {\etal} or Tomberg, both of which
are expressed in terms of quantities averaged over noise realizations.

\para{Evaluation of the tail probability}
The saddle-point approximation to the
path integral
for $\ProbP$
is
\begin{equation}
    \label{eq:instanton-approximation}
    \ProbP \approx
    \Normalization
    \exp
    \Big(
        \im \SMSR[ \phiSP, \XEmomSP ]
    \Big)
    \left\{
        \det
        \bigg(
            - \im \frac{\delta^2 \SMSR}{\delta \varphi_i \delta \varphi_j}
        \bigg)
    \right\}^{1/2}
    +
    \text{subleading}
    ,
\end{equation}
where the symbol ``$\approx$''
has the same meaning explained above,
and to write the functional determinant
we have combined $\phi$ and $\Xmom$
into a 2-component field vector $\varphi = (\phi, \Xmom)$.
There are formally subleading corrections that we have not
written explicitly.
In Eq.~\eqref{eq:instanton-approximation},
all occurrences of $\phi$ and $\Xmom$ should be evaluated
on the solutions
$\phiSP$, $\XEmomSP$
to the instanton
equations~\eqref{eq:SR-instanton-phi}--\eqref{eq:SR-instanton-X}.
We describe~\eqref{eq:instanton-approximation}
as
the \emph{instanton approximation}.

The functional determinant
represents the effect of fluctuations around the
instanton trajectory~\eqref{eq:SR-instanton-soln-phi}--\eqref{eq:SR-instanton-soln-P}.
If it is independent of the field configurations
$\phiinit$ and $\phiend$, then it can
be absorbed into the overall normalization $\Normalization$
and need not be computed.
This is the situation for the Gaussian examples
considered in this paper,
but will not be true more generally.
Evaluation of the determinant can be done
exactly in the Gaussian case, but
otherwise is not expected to be easy.
Some general methods are available for determinants
of second order operators,
such as $\zeta$-function regularization
and the Gelfand--Yaglom formula.
However, in our case, the fluctuation matrix
$\delta^2 \SMSR / \delta \varphi_i \delta \varphi_j$
is a first order operator,
for which these general methods do not apply.

In this paper we are always able to drop
the fluctuation determinant.
The MSR action evaluated at the saddle point is
\begin{equation}
    \label{eq:SR-instanton-SMSR}
    \im \SMSR[\phiSP, \XEmomSP]
    =
    -
    \frac{1}{2} \int_{0}^{\Nstar}
    \d N \;
    \left(
        \frac{H}{2\pi}
    \right)^2
    \XEmomSP(N)^2
    =
    -
    \frac{1}{2}
    \left(
        \frac{H}{2\pi}
    \right)^2
    (\XEmomSP)^2
    \Nstar \;.
\end{equation}
It follows
from~\eqref{eq:SR-instanton-SMSR}
that the probability associated with the instanton
is
determined by the probability of the noise realization,
as proposed by Tomberg~\cite{Tomberg:2023kli};
compare Eq.~\eqref{eq:tomberg-onsager-machlup}.
Indeed, we have already noted
that it is the noise realization that carries
almost all physical
information about the transition.
The property~\eqref{eq:SR-instanton-SMSR}
follows from partial cancellation
between the two terms in~\eqref{eq:MSR-path-integral},
and the instanton
equation~\eqref{eq:SR-instanton-phi} for $\phi$.
It is generic for Gaussian models, but
more generally it need not always apply.
In the final step we have used that $\XEmomSP$ is constant in
this model.
One can interpret Eq.~\eqref{eq:SR-instanton-SMSR}
to mean that an exponential tail forms because of
the constant unlikeliness
``cost'' per e-fold associated with this noise
realization.

The normalization $\Normalization$ can be determined
using~\eqref{eq:SR-instanton-SMSR}.
It yields
a rational function of $\Nstar$,
and therefore it does not contribute to
exponential accuracy.
It follows that the tail estimate
is determined only by the saddle-point action,
\begin{equation}
    \label{eq:Prob-instanton-approx}
    \ln
    \ProbP
    \approx
    \im \SMSR[\phi, \XEmom]
    .
\end{equation}
Eq.~\eqref{eq:Prob-instanton-approx}
should be valid to exponential
accuracy
whenever $\Nstar$ is large enough
that the
instanton~\eqref{eq:SR-instanton-phi}--\eqref{eq:SR-instanton-X}
exponentially dominates any neighbouring trajectories.
In this model,
it is possible to
find an explicit relation between
$\ProbP$ and $\ProbPre$ by the method of images
(see Eq.~\eqref{eq:method-images-linear-potential}).
Further, the coefficients needed for the flux
formula~\eqref{eq:first-passage-probability-current}
are known.
Neither of these relations introduces exponential
contributions.
Therefore,
to leading order in $\Nstar$,
Eqs.~\eqref{eq:SR-instanton-SMSR}
and~\eqref{eq:Prob-instanton-approx}
precisely
reproduce the exponential tail estimate~\eqref{eq:slow-roll-exp-tail},
\begin{equation}
    \label{eq:instanton-slow-roll-exp-tail}
    \ln \ProbQ(\Nstar)
    \approx
    - \frac{1}{2}
    \frac{\Nstar}{\dimP_\zeta} 
    +
    \Or(1)
    \quad
    \text{for $\Nstar \gg 1$.}
\end{equation}
The notation
$+ \Or(1)$
indicates that the leading correction
to~\eqref{eq:instanton-slow-roll-exp-tail}
is a constant
(in the limit of large $\Nstar$),
plus a tower
of subleading corrections from powers of $1/\Nstar$.
In principle, these corrections can be obtained from
the instanton.
However, they are very small for $\Nstar \gg 1$,
and
it is not clear that they are numerically
more important than the
subexponential prefactor
that we have not computed.
This issue deserves further investigation.
For now, we omit these corrections to avoid
giving a misleading
result.

Evidently
the instanton procedure gives a prediction for
$\ProbQ(\Nstar)$, and hence
$\ProbP(\zeta)$ via~\eqref{eq:prob-zeta-estimate},
without ever needing to specify the relation between the
field $\phi$ and the curvature perturbation $\zeta$.
We have already observed that this is an important
convenience (\S\ref{sec:ezquiaga-spectral-method}),
since we need only solve the instanton
equations~\eqref{eq:SR-instanton-phi}--\eqref{eq:SR-instanton-X}
and do not need details of the gauge transformation
to $\zeta$.
We also see how the instanton method avoids the
problem of specifying a suitable probability
measure for the noise realization, as in
Eq.~\eqref{eq:PGamma}.
The instanton equations pick out a specific
realization whose probability can be evaluated
unambiguously.

The structure of Eq.~\eqref{eq:instanton-slow-roll-exp-tail}
reproduces expected
behaviour from Freidlin--Wentzell theory~\cite{2021JPhA...54q5001A}.
The main result
is that, under certain hypotheses
and if $\epsilon$ represents the noise
experienced by a stochastic process, then
in the limit of weak noise,
\begin{equation}
    \label{eq:freidlin-wentzell}
    - \lim_{\epsilon \rightarrow 0}
    \epsilon \ln \ProbP
    =
    S_T
    ,
\end{equation}
where $S_T$ is the Freidlin--Wentzell action,
sometimes described as the \emph{rate function}.
Note that weak noise at fixed rareness, and
extreme rareness at fixed noise, are usually interchangeable.
For our applications, we usually consider the latter
with $\dimP_\zeta$ as a proxy for the noise amplitude.
Eq.~\eqref{eq:freidlin-wentzell}
shows that the rate function
is essentially equivalent to the Onsager--Machlup
functional,
or the MSR action at the saddle point,
and hence reproduces~\eqref{eq:SR-instanton-SMSR}.
In Friedlin--Wentzell theory,
a distribution $\ProbP$
satisfying~\eqref{eq:freidlin-wentzell}
is said to have
a \emph{large deviation principle}.
It has
already been discussed
in the present context by
Cohen {\etal}~\cite{Cohen:2022clv}.

\para{Likeliness of noise realization}
The noise realization~\eqref{eq:SR-instanton-soln-P}
should be regarded as the least unlikely configuration
that allows the transition to proceed.
Nevertheless, it is still exceptionally unlikely.
In particular, its amplitude is constant, and
it acts in the
same direction during each time interval.
Both properties are highly atypical.
In order to cancel the deterministic motion,
as explained above,
the amplitude of the noise
in each time step
must be fairly large,
relative to the typical fluctuation amplitude
$\sim H/2\pi$.
Eq.~\eqref{eq:SR-instanton-soln-P}
yields the estimate $|\xi| \sim \dimP_\zeta^{-1/2}$.
Since the amplitude of a typical fluctuation is
$|\xi| \sim 1$, these are individually unlikely events,
at least during a slow-roll epoch
where $\dimP_\zeta \ll 1$.

To exponential accuracy, the probability is hardly altered
if the noise~\eqref{eq:SR-instanton-soln-P}
is slightly modified, perhaps by allowing a few small-amplitude steps
(or steps in the wrong direction) compensated by larger
steps elsewhere.
It is
trajectories of this kind,
obtained from minor perturbations around the instanton,
that are accounted for by the fluctuation
determinant in Eq.~\eqref{eq:instanton-approximation}.

\section{Schwinger--Keldysh description of the instanton}
\label{sec:instanton-phase}

\subsection{Non-equilibrium field theory on a closed time contour}
We now extend the analysis beyond the
overdamped slow-roll regime,
in which (in the absence of noise)
$\d \phi / \d N$ is a function of $\phi$.
To do so, we re-interpret
the
stochastic instanton within a more fundamental
framework, that of the
Schwinger--Keldysh path integral.
This replaces the effective Langevin equation~\eqref{eq:Starobinsky-Langevin}
with a more detailed microscopic description.
We then systematically integrate out
short-scale modes.
The result is
a path integral
describing the evolution of
large-scale modes
under the influence
of a bath
of fluctuating short-scale modes.
The part of the action describing interactions
between the large-scale modes and the bath
is called the \emph{influence functional}.
Such functionals were introduced by
Feynman \& Vernon~\cite{feynman1963theory}.
Their primary characteristic is mixing of the Keldysh $+$ and $-$
vertices, which is usually forbidden.
This mixing is typical of interactions with an
external environment.

The exact choice of short-scale modes
over which we integrate
depends on the observable to be computed.
To make contact with the Langevin equation~\eqref{eq:Starobinsky-Langevin}
we should integrate out all modes shorter than the
smoothing scale.
In Eq.~\eqref{eq:Starobinsky-Langevin}
this was not specified exactly, but taken to be a fixed multiple
of the horizon.
Influence functions appropriate for this choice have been
constructed by a number of authors,
including Moss \& Rigopoulos~\cite{Moss:2016uix}, 
Pinol \etal~\cite{Pinol:2020cdp}, 
and Andersen \etal~\cite{Andersen:2021lii}.
Alternatively,
if we wish to make predictions for the value of $\phi$
associated with a fixed comoving scale $\vect{k}$
after horizon exit,
we would instead integrate out all modes
with magnitude greater than $k$.
We comment briefly on this scenario in~\S\ref{sec:decaying-noise} below.

The combination of
a Schwinger--Keldysh path integral
coupled to an influence function
is very general.
In addition to the fluctuations
from the short scale modes,
which
are already captured by the Starobinsky--Langevin
equation~\eqref{eq:Starobinsky-Langevin},
the influence functional may describe renormalization
of the couplings and dissipative
effects.
It encodes correlation times
and lengths associated
with the short-scale fluctuations,
memory effects, and non-Markovian
behaviour.
In summary,
it provides a means to import
first-principles microscopic information into the
computation of tail probabilities.
These properties are
interesting because, by construction,
the tail is populated by rare events.
As we have seen in~\S\ref{sec:instanton-solution},
such events are mediated by highly unlikely
realizations.
The relative unlikeliness of such realizations
could plausibly depend on small changes in
the statistical characteristics of the noise.

\para{Phase space path integral}
We begin by setting up a description of the
background model in the absence of noise
from short-scale fluctuations.
In order to account for interactions with the
bath of short-scale modes,
it is convenient to work in a
Hamiltonian phase-space description.%
    \footnote{In the Langevin description this is
    equivalent to writing \emph{two} Langevin equations,
    one analogous to Eq.~\eqref{eq:Starobinsky-Langevin},
    and the other for $\mom$ (or $\dot{\phi}$).
    Each Langevin equation may have its own sources
    of noise, which may be correlated.}
We begin with the action
\begin{equation}
    S[\phi]
    =
    \int
    \d^4 x \, \sqrt{-g} \,
    \bigg(
        -\frac{1}{2}
        g^{\mu\nu}
        \partial_{\mu}\phi
        \partial_{\nu}\phi
        -
        V(\phi)
    \bigg)
    ,
\end{equation}
As in~{\S}\ref{sec:ezquiaga-spectral-method},
we take the e-folding number $\d N = H \d t$ as our time
variable~\cite{SalopekBond1991,Vennin:2015hra}.  
To move to phase space, we introduce
$\mom \equiv \delta S / \delta \dot{\phi}$
as the momentum canonically conjugate to $\phi$,
where $\dot{\phi} = \d \phi / \d N$.
The
Hamiltonian density $\Hamiltonian$
is
\begin{equation}
    \label{eq:phase-space-legendre}
    \mom
    \equiv
    -\sqrt{-g}\,
    g^{0\mu}
    \partial_{\mu} \phi
    =
    a^3 H \frac{\d \phi}{\d N}
    \,,
    \qquad 
    \Hamiltonian \equiv
    \frac{1}{2}
    \frac{\mom^2}{a^3 H^2}
    -
    \frac{a}{2H}
    (\nabla \phi)^2
    +
    \frac{a^3}{H}
    V(\phi)
    \;.
\end{equation}
The Schwinger--Keldysh
path integral computes
the transition
probability between a field configuration $\phiinit$
at some early time $\Ninit$,
and a final field configuration $\phistop$
at a later time $\Nstop$.
By analogy with the discussion of~\S\ref{sec:slow-roll-instanton}
we write this transition probability
as $\ProbP$,%
    \footnote{\label{footnote:Keldysh-generating-functional}%
    More generally, one can work with a generating functional
    rather than the transition probability.
    The full generating functional, allowing for an arbitrary density matrix
    $\rho$ characterizing the initial state, takes the form
    \begin{multline*}
        Z_\rho[J_\pm]
        =
        \Normalization
        \int_{-\infty}^{+\infty} \d \phistop
        \int_{-\infty}^{+\infty} \d \phiinit^\pm
        \,
        \rho(\phiinit^\pm)
        \!
        \int_{\phi_\pm(\Ninit)=\phiinit^\pm}^{\phi_\pm(\Nstop) = \phistop}
        \!
        [\d \phi_\pm]
        \\
        \times
        \int [\d \mom_\pm]
        \exp\left(
            \im
            \int_{\Ninit}^{\Nstop} \d^4 x
            \sum_{\alpha = \pm} \alpha
            \left[
                \mom_\alpha \frac{\d\phi_\alpha}{\d N}
                -
                \Hamiltonian(\mom_\alpha, \phi_\alpha)
                +
                J_\alpha \phi_\alpha
            \right]
        \right)
        \;.
    \end{multline*}
    The notation $\int_{-\infty}^{+\infty} \d \phiinit^\pm$
    indicates a double integral over $\phiinit^+$ and $\phiinit^-$ separately
    with the specified limits, i.e.,
    $\int_{-\infty}^{+\infty} \d \phiinit^+ \int_{-\infty}^{+\infty} \d \phiinit^-$.
    The path integral $\int [\d \phiplusminus]$
    should be interpreted in the same way.
    The initial conditions are
    now part of the specification of the density matrix $\rho$
    rather than being applied as boundary conditions
    on the $\phiplusminus$ integral.
    }
\begin{equation}
    \label{eq:Schwinger-Keldysh-P}
    \ProbP(
        \phistop, \Nstop
        \divider
        \phiinit^\pm, \Ninit
    )
    =
    \Normalization
    \int
    [\d\phiplusminus \, \d\momplusminus] 
    \exp
    \bigg(
        \im
        \int_{\Ninit}^{\Nstop}
        \d^4 x
        \sum_{\alpha \in \pm}
        \alpha
        \Big[
            \mom_\alpha \frac{\d \phi_\alpha}{\d N}
            -
            \Hamiltonian(\mom_\alpha, \phi_\alpha)
        \Big]
    \bigg)
    ,
\end{equation}
where $\alpha = \pm$ is an index
labelling the forward and backward branches of the
closed-time-path (``CTP'') contour.
We must again allow for a normalization
factor $\Normalization$
that will typically depend on the parameters of the transition.
The integral should be taken over field configurations
$\phiplusminus$
matching specified spatial configurations
$\phiinit^\pm(\vect{x})$
at the initial time $\Ninit$,
and a common final configurations
$\phistop(\vect{x})$
at the final time $\Nstop$.
The $+$ and $-$ $\phi$ fields must agree
at the final time $\Nstop$,
but
the integrals over $\momplusminus$
are unrestricted.
In the language of~\S\ref{sec:slow-roll-instanton},
Eq.~\eqref{eq:Schwinger-Keldysh-P}
computes an unrestricted transition
probability, because we do not impose
any first-passage conditions on the fields
$\phiplusminus$
at intermediate times.
In this form,
$\ProbP$ is a density
with respect to the initial and final spatial configurations
$\phiinit^\pm(\vect{x})$
and
$\phistop(\vect{x})$.

In Eqs.~\eqref{eq:phase-space-legendre}
and~\eqref{eq:Schwinger-Keldysh-P},
$\mom$ is the ``true'' canonical momentum associated
with $\phi$,
not the ``fake'' Fokker--Planck momentum $\Xmom$ that appeared
in the Martin--Siggia--Rose construction.
We will see momentarily that the $+$, $-$ fields can be reorganized
to give \emph{both} $\phi$ and $\mom$ their own ``fake''
Fokker--Planck momentum.

\para{Keldysh basis}
At this stage,
if we were to integrate
over the $\mom_\alpha$,
we would obtain 
the standard Lagrangian path integral
$\int [\d\phiplusminus] \exp\big( \im S[\phiplus] - \im S[\phiminus] \big)$.
In what follows, we
make two minor abuses of notation.
First, we refer to the effective CTP action
$S_+ - S_-$
as simply $S$.
Second, we use the same term ``action''
to refer to
its phase space counterpart.
We now transition to the
Keldysh basis,
defined by
\begin{equation*}
    \cl{\phi} \equiv \frac{\phiplus + \phiminus}{2} \,,
    \quad
    \qu{\phi} \equiv \phiplus - \phiminus \,,
\end{equation*}
and likewise
\begin{equation*}
    \cl{\mom} \equiv \frac{\momplus + \momminus}{2} \,,
    \quad
    \qu{\mom} \equiv \momplus - \momminus \;.
\end{equation*}
The labels ``cl'' (for \emph{classical})
and ``q'' (for \emph{quantum}) are conventional.
However, despite their names, it
should be noted that the ``q''
fields are needed to describe classical processes.
Further, both ``cl'' and ``q'' fields are subject
to quantum fluctuations.
The action becomes%
    \footnote{The symplectic
    structure of the Keldysh variables couples the ``cl'' and
    ``q'' components in such a way that $\cl{\mom}$ is the
    conjugate momentum to $\qu{\phi}$, and vice-versa.
    This may appear surprising.
    The reason is that the corresponding equations of
    motion enforce the expected identification of
    $\cl{\mom}$ with $\d \cl{\phi} / \d N$,
    and likewise for $\qu{\mom}$ and
    $\d \qu{\phi} / \d N$.}
\begin{equation}
    \label{eq:S-Keldysh-basis}
    S[\phi, \mom] = \int \d^4 x
    \bigg[ 
        - \qu{\phi} \frac{\d \cl{\mom}}{\d N}
        + \qu{\mom} \frac{\d \cl{\phi}}{\d N}
        - \frac{\qu{\mom}\cl{\mom}}{a^3 H} 
        - \frac{a}{H} \nabla \cl{\phi} \cdot \nabla \qu{\phi}
        - \frac{a^3}{H} V(\cl{\phi} + \frac{\qu{\phi}}{2}) 
        + \frac{a^3}{H} V(\cl{\phi} - \frac{\qu{\phi}}{2})
    \big]\; .
\end{equation}
For future convenience we have integrated
by parts in the first term,
in order that the ``q'' fields
all appear undifferentiated.
Later,
this will assist us
in re-interpreting~\eqref{eq:S-Keldysh-basis}
in terms of a ``fake'' Fokker--Planck phase space.
After expanding the potentials
$V$
in powers of $\qu{\phi}$,
each term in~\eqref{eq:S-Keldysh-basis}
contains at least one ``q''
field.
This is a consequence of the
requirement
that $S$ vanish when all ``q'' fields
are zero, which
follows from the CTP structure
and
forbids terms involving only two ``cl'' fields.
Also,
in~\eqref{eq:S-Keldysh-basis},
there are no terms involving only two ``q'' fields.
These are not forbidden, but
cannot be produced
(without environmental interactions)
by the CTP structure of $S$.

We now define 
$\pi = \d\phi/\d N$
and use $\pi$ as the momentum variable in favour of $\mom$.
Eq.~\eqref{eq:phase-space-legendre}
shows that $\pi$ and $\mom$
are not the same, but they are proportional,
up to a factor involving only the background fields.
This choice makes the description as similar as possible
to the MSR framework of~\S\ref{sec:slow-roll-instanton}.
Both $\pi$ and $\phi$ are to be treated as independent variables
of integration
in the path integral.
It follows that deviations from
the ``on-shell'' relation
$\pi = \d\phi/\d N$
will arise
naturally in a
stochastic setting.
In terms of $\phi$ and $\pi$, the
Hamiltonian form of action becomes
\begin{equation}
    \label{eq:actqc}
    S[\phi,\pi]
    =
    \int \d^4 x \, \sqrt{-g} \,  H^2
    \bigg[
        -\qu{\phi}
        \bigg(
            \frac{\d \cl{\pi}}{\d N}
            +
            (3 - \epsilon)\cl{\pi}
            +
            \frac{V'(\cl{\phi})}{H^2}
        \bigg)
        +
        \qu{\pi}
        \bigg(
            \frac{\d \cl{\phi}}{\d N}
            -
            \cl{\pi}
        \bigg)
        -
        \frac{1}{(aH)^{2}} \nabla \cl{\phi} \cdot \nabla \qu{\phi}
    \bigg] \,
    .
\end{equation}
The potential
terms in~\eqref{eq:S-Keldysh-basis}
have been expanded to linear order in $\qu{\phi}$,
which
will replicate the MSR formalism
of~\S\ref{sec:slow-roll-instanton}.
Terms at higher order in $\qu{\phi}$
could be retained if desired.
We will see below that these
lead to a more complex description of the noise.
(In particular, they would invalidate the result~\eqref{eq:SR-instanton-SMSR},
that the transition probability derives
only from the probability of the noise realization.)
The first such correction is the cubic term
$V'''(\cl{\phi})\qu{\phi}^3$; see Eq.~\eqref{eq:Vcq}.

Let us now consider
boundary conditions on the ``cl'' and ``q'' fields.
In Eq.~\eqref{eq:Schwinger-Keldysh-P}
the integrals over
$\momplus$ and $\momminus$
were unrestricted
at the initial and final times,
and therefore the integrals over
$\cl{\pi}$ and $\qu{\pi}$
should be likewise unrestricted.

The integrals over
$\phiplus$ and $\phiminus$
satisfy the CTP condition
$\phiplus = \phiminus$
at the final time,
and therefore we must have
$\qu{\phi} = 0$ there.
Meanwhile, $\cl{\phi}$ should match
the specified final spatial configuration $\phistop(\vect{x})$.
In Eq.~\eqref{eq:Schwinger-Keldysh-P},
$\phiplus$ and $\phiminus$
were regarded as having fixed initial
configurations
$\phiinit^+(\vect{x})$
and $\phiinit^-(\vect{x})$.
Under these conditions we would conclude that
both $\cl{\phi}$ and $\qu{\phi}$
should be likewise fixed at the initial time.
However, we are going to identify the ``q'' fields
with the MSR response field $\Xmom$.
We therefore impose that
$\cl{\phi}$ has a specified initial configuration
$\phiinit(\vect{x})$,
but integrate unrestrictedly over the initial value
of $\qu{\phi}$.
The result of these choices
is that the path integral over ``cl'' and ``q'' fields
should be understood as
a density with respect to the initial and final configurations
of $\cl{\phi}$, with all other boundary
data being integrated over.
In this sense, the transition probability
incorporates information from more than one initial state.

\subsection{Stochastic dynamics in phase space}
We are now able to obtain
a stochastic description 
from~\eqref{eq:actqc}
by coarse-graining the fields,
and integrating out short-wavelength (UV) modes.
The result is an effective theory for the
remaining
long-wavelength (IR) modes.
It is described by the effective action
\begin{equation}
    \exp(\im \Seff [\phi, \pi])
    =
    \exp(\im \SIR [\phi, \pi])
    \,
    \mathcal{F}[\phi, \pi]\,,
\end{equation}
where $\SIR[\phi, \pi]$ is the IR
phase-space action obtained
by restricting Eq.~\eqref{eq:actqc} to
the long-wavelength fields.
$\mathcal{F}[\phi, \pi]$ is the Feynman--Vernon influence functional.
It encodes the effect of the (UV) environment on the (IR) system.
Up to quadratic order it can be written%
    \footnote{As shown in Appendix~\ref{sec:Inf_func},
    the influence functional also has terms
    associated with dissipation.
    However, these are vanishingly small at leading-order.}
\begin{align}
    \FeynmanVernon[\phi,\pi]
    =
    C
    \exp
    \Bigg[
        -
        \frac{1}{2}
        \int \d^4 x \, \big({-g(x)}\big)^{1/2}
        \int \d^4 x' \, \big({-g(x')}\big)^{1/2}
        \bigg(\begin{matrix}
            \qu{\pi} (x) \\ -\qu{\phi}(x)
        \end{matrix}\bigg)^\transpose
        {\cal M}(x,x')
        \bigg(\begin{matrix}
            \qu{\pi} (x') \\ -\qu{\phi}(x')
        \end{matrix}\bigg)
    \Bigg]
    ,
\end{align}
where $C$ is a normalization constant,
whose explicit form is not required.
$\mathcal{M}(x,x')$ is the noise correlation matrix,
originally calculated by
Sasaki, Nambu \& Nakao~\cite{Sasaki:1987gy,Nakao:1988yi};
see also Ref.~\cite{Stewart:1991dy}.
Its general form is
given in Eq.~\eqref{eq:M2},
which depends on the window function $W_k$
used to separate the long- and short-wavelength
fields.
In Eq.~\eqref{eq:M21} we evaluate it
for a step-function window $W_k = \Theta(k - \mu a H)$.
In this case there is a sharp
division between short and long modes,
with the short modes being those with $k/aH > \mu$.
This prescription is chosen for compatibility with
the Starobinsky equation~\eqref{eq:Starobinsky-Langevin}.
The corresponding noise correlation matrix is
\begin{equation}\label{eq:M20}
    \mathcal{M}_{ij}(x,x')
    =
    \Big(
        1 - \epsilon(N)
    \Big)
    \,
    H(N)^2 H(N')^2
    \, 
    \sinc(k_\mu |\vt{x} - \vt{x}'|)
    \,
    \dimP_{ij}(k_{\mu})
    \,
    \delta(N - N')
    ,
\end{equation}
where $N$ and $N'$ are the time coordinates
associated wth $x$ and $x'$, respectively.
Here, $\sinc(x) \equiv \sin x / x$,
$k_\mu \equiv \mu aH$,
and $\dimP_{ij}$
denotes the dimensionless power spectra
of the phase-space variables
($i,j \in \{\phi, \pi\}$),
as defined in Eq.~\eqref{eq:PS_matrix}.
The $\delta$-function
$\delta(N - N')$
shows that the noise has zero correlation
time~\cite{Nakao:1988yi,Starobinsky:1994bd}.
It is possible this property could be relaxed
if higher-order effects were retained in
the computation of $\FeynmanVernon$.

In principle,
the correlation~\eqref{eq:M20}
could be used to study the stochastic
evolution of an ensemble of nearby inflating patches,
experiencing spatially correlated noise.
However, for comparison with the MSR
formalism of~\S\ref{sec:slow-roll-instanton}
we wish to describe a single,
isolated spatial patch.
To do so we select a sufficiently large coarse-graining
scale by choosing $\mu$ so that
$\sinc(k_\mu |\vect{x} - \vect{x}'|) \approx 1$~\cite{Nakao:1988yi}.
This corresponds to choice of an appropriate
smoothing scale in the Starobinsky equation~\eqref{eq:Starobinsky-Langevin}.
We also integrate out the temporal delta function and drop
terms of $\Or(\epsilon)$
in the Feynman--Vernon sector.
However, we retain these terms in the noiseless part
of the action for the
long-wavelength fields.
Finally, we define a diffusion matrix $D_{ij}$
by
$D_{ij} \equiv \dimP_{ij} / 2$,
and refer to the full matrix compactly as $D$.
The effective action for the long-wavelength field
values
interior to a single
patch becomes
\begin{multline}
    \label{eq:Seff1}
    S_{\rm eff}[\phi,\pi]
    =
    \int \d N
    \Bigg[
        \int \d ^3x \;
        a^3 H
        \bigg\{
            - \qu{\phi}
            \bigg(
                \frac{\d \cl{\pi}}{\d N}
                +
                (3-\epsilon)\cl{\pi}
                +
                \frac{1}{H^2} V'(\cl{\phi})
            \bigg)
            + \qu{\pi}
            \bigg(
                \frac{\d \cl{\phi}}{\d N}
                -
                \cl{\pi}
            \bigg)
        \bigg\}
        \\
        + \im \int \d^3 x \; \d^3 x' \;
        a^3 H
        \bigg(\begin{matrix}
                \qu{\pi} (x) \\ -\qu{\phi}(x)
        \end{matrix}\bigg)^\transpose
        \D(x,x')
        \bigg(\begin{matrix}
                \qu{\pi} (x') \\ -\qu{\phi}(x')
        \end{matrix}\bigg)
    \Bigg]
    ,
\end{multline}
We have dropped spatial gradients
because
the fields are taken to be approximately homogeneous.
The spatial integrals in~\eqref{eq:Seff1}
should extend only over
a single such patch.

In this approximation,
the novel feature of the Feynman--Vernon functional
is that it introduces effective vertices
that couple two ``q'' fields alone.
It was noted above that these cannot be produced by the closed time
path action before coarse-graining,
because they amount to explicit coupling between the $+$ and $-$
vertices.
These terms are the analogue of the $\im \Xmom^2$ noise term
that appears in the MSR action
of~\S\ref{sec:slow-roll-instanton}.

\subsection{Martin--Siggia--Rose from Schwinger--Keldysh}
\label{sec:MSR-phase-space}
We now interpret
Eq.~\eqref{eq:Seff1}
as an MSR-like action for
the classical ``cl'' fields.
At tree level, these
fields
obey the classical equations
of motion when the ``q'' fields are set to zero.
Therefore, they
are the correct degrees of freedom
to associate with the observables $\phi$ and $\pi$.
Meanwhile,
inspection of~\eqref{eq:Seff1}
shows that the ``q'' fields
appear in the same way as the $\Xmom$ field
of the MSR formalism,
including the occurrence of
quadratic vertices
such as $\im \qu{\phi}^2$, as noted above.
The interpretation of these fields follows from the
corresponding saddle point equations,
to be discussed below.
These are structurally the same as the MSR
saddle point equations,
and
show that the ``q'' fields can be interpreted as
encoding a specific realization of the noise.
This relation between the Schwinger--Keldysh
and MSR actions was first noticed
by Zhou, Su, Hao \& Yu (1980)~\cite{PhysRevB.22.3385}.

It can be shown that the
construction of the Schwinger--Keldysh path integral
translates to It\^{o} discretization when interpreted
as an MSR action.
For details, see the book by Kamenev~\cite{Kamenev_2011}.
This is why we have adopted It\^{o} discretization
from the outset.
For example, this is why the overdamped limit
of Eq.~\eqref{eq:Seff1}
reproduces the It\^{o}-discretized
action obtained in~\S\ref{sec:slow-roll-instanton}
without any extra Jacobian factor.
If a different discretization is required,
it could be introduced
at the cost of introducing
such a Jacobian.

At the level of the current discussion, the main benefit
of working down
from the Schwinger--Keldysh path integral
is to provide a systematic tool
to extend the Starobinsky equation~\eqref{eq:Starobinsky-Langevin}
beyond the slow-roll regime.
However, in principle, it is much more general.
First, Eq.~\eqref{eq:actqc}
shows that we can expect corrections of the form
$\qu{\phi}^3$, $\qu{\phi}^5$, \ldots, and so on,
which appear in combination with higher derivatives of
the potential $V(\cl{\phi})$.
These provide corrections to the interaction
of the noise
with the observable field.
When such terms are included, we lose the simplest
interpretation of
$\qu{\phi}$ as precisely the noise realization
in the Langevin equation.
However, the broader picture that $\qu{\phi}$
encodes this realization remains valid.
Similar terms,
occurring at higher order in the ``q''
fields,
will arise if we
compute the Feynman--Vernon influence
functional $\FeynmanVernon$
beyond quadratic level
in the long-wavelength fields.

Second, we have already observed that,
beyond lowest order,
the Feynman--Vernon effective vertices will be
nonlocal, encoding the appearance of nonzero
correlation times and lengths,
or non-Markovian effects.
Such effects are very difficult to include in the
Langevin description.
Remarkably,
they
make almost no conceptual
difference to the calculation of the instanton
solution: they occur simply as new, nonlocal
source terms in the instanton equation.
Of course, we expect that inclusion of such nonlocal
sources would be numerically challenging in practice.

Third, the Schwinger--Keldysh
description includes all quantum effects.
Therefore we can consider building a
full quantum effective
action, accounting
for ultraviolet contributions from loops.
These will correct the instanton equations.
Note that the instanton solution itself has an interpretation
as a partial resummation of diagrams, as described
by Celoria {\etal}~\cite{Celoria:2021vjw}.
The effect of including loop corrections to the instanton equations
would therefore be something like resummation of skeleton
diagrams in the 2PI formalism.

Finally, we can incorporate the effect
of a nontrivial initial state
by specifying a suitable density matrix.
For details, see footnote~\ref{footnote:Keldysh-generating-functional}
on p.~\pageref{footnote:Keldysh-generating-functional}.
One possible application is to combine
a Celoria {\etal}-type instanton,
describing production of a rare fluctuation
at horizon exit,
with subsequent noisy evolution,
described by interactions of the Keldysh ``cl'' and ``q'' fields.
The final state of the instanton can be encoded
by a suitable density matrix.
Loosely, we can regard the Wigner function associated with the
density matrix as giving a quasiprobability
distribution of the field fluctuation at horizon exit.

(In fact, this construction applies whether or not the horizon exit
fluctuation is sufficiently rare to be described by an instanton.
This shows how to include the effect of even ordinary
initial fluctuations in our framework. However,
up to~\S\ref{sec:decaying-noise},
we
continue to ignore such fluctuations,
and take the initial state to correspond to a fixed
value for $\phiinit$.)

\para{Fokker--Planck phase space}
We are now ready to identify the coordinates of the Fokker--Planck
phase space.
As explained above, the primary
field-like degrees of freedom are
$\cl{\phi}$ and $\cl{\pi}$.
In order to achieve a unified notation we define
\begin{equation}
    \qa 
    \equiv
    \cl{\phi}
    ,
    \qquad
    \qb
    \equiv
    \cl{\pi}
    .
\end{equation}
(Note that this use of $\varphi$ differs from
that in~\S\ref{sec:slow-roll-instanton}.)
The effective action~\eqref{eq:Seff1}
has a ``true''
Hamiltonian structure, in which
$\qu{\pi}$ is the
canonical momentum for
$\cl{\phi}$
and
$\cl{\pi}$
is the canonical momentum for
$\qu{\phi}$.
However, inspection of~\eqref{eq:Seff1}
(in which we have integrated by parts)
shows that
there is also a formal, ``fake''
structure in which we can regard the ``q''
fields as momenta for the ``cl'' fields.
This identification pairs
$\qu{\pi}$ as a momentum for $\cl{\phi}$,
and $-\qu{\phi}$ as a momentum for $\cl{\pi}$.
The Schr\"{o}dinger equation for
this second, ``fake''
Hamiltonian structure yields the Fokker--Planck
equation.

For rare events we expect the saddle point to lie
at imaginary values of the ``q''
fields.
We therefore define
\begin{equation}
    \im \pa \equiv \int \d^3x\ a^3 H  \qu{\pi} ,
    \qquad 
    \im \pb \equiv \int \d^3x\ a^3 H (-\qu{\phi}) ,
\end{equation}
in which we have absorbed the spatial integral over the
patch into the noise variables.
The relative minus sign
reflects the relative signs
of $\qu{\phi}$ and $\qu{\pi}$
in the influence functional.
Since the fields are taken to be homogeneous within each
patch, the spatial integrals evaluate to a fixed comoving
volume, which we write $\SpatialVolume$.
Hence, the noise fields reduce to
\begin{equation}
    \pa = - \im a^3 H \SpatialVolume \qu{\pi} ,
    \qquad 
    \pb = \im a^3 H \SpatialVolume \qu{\phi} .
\end{equation}
Therefore the amplitude of the
noise scales with the volume of
the region.
This agrees with the idea that
stochastic fluctuations should be
extensive over a spatial domain.

Finally, we define an effective MSR-like action
by identifying
$
    \im \Seff[\phi_{\rm cl}, \phi_{\rm q}, \pi_{\rm cl}, \pi_{\rm q}] 
    = \im \SMSR[\varphi, \XEmom]
$,
where
$\varphi = (\qa, \qb)$
and
$\XEmom = (\pa, \pb)$.
The MSR action can be written
\begin{align}\label{eq:Smps}
    \im \SMSR[\varphi,\XEmom]
    =
    -
    \int \d N
    \bigg\{
        \pb
        \bigg(
            \frac{\d \qb}{\d N}
            +
            (3-\epsilon) \qb
            +
            \frac{V'(\qa)}{H^2}
        \bigg)
        +
        \pa
        \bigg(
            \frac{\d \qa}{\d N}
            -
            \qb
        \bigg)
        -
        \XEmom_i
        \D_{ij}
        \XEmom_j
    \bigg\}
    ,
\end{align}
where
summation over $i,j = \{1,2\}$ is implied in the last term.
From this we can read off the corresponding
Fokker--Planck Hamiltonian,
\begin{equation}\label{eq:HFP}
    \HFP(\varphi,\XEmom)
    =
    \pa \qb
    -
    \pb
    \bigg(
        (3-\epsilon) \qb
        +
        \frac{V'(\qa)}{H^2}
    \bigg)
    +
    \XEmom_i
    \D_{ij}
    \XEmom_j
    .
\end{equation}
The CTP structure guarantees that
$\Seff$
vanishes when the ``q'' fields are set to zero.
At the level of the Fokker--Planck Hamiltonian
this requires
$\HFP(\varphi, 0) = 0$,
which guarantees that the corresponding
Fokker--Planck equation has the structure of
a continuity equation
(compare Eq.~\eqref{eq:forward-komogorov-transport}).
Therefore we continue to have a concept
of a Kolmogorov probability
current analogous
to Eq.~\eqref{eq:probability-current}~\cite{Kamenev_2011}.

For the Gaussian
action~\eqref{eq:Smps},
the volume $\mathcal{V}$
is absorbed into the normalization of the response fields
$\pa$ and $\pb$ and does not need to be specified
explicitly.
However, if we were to retain vertices from the
Schwinger--Keldysh action
at higher order in ``q'' fields, then this convenience would
be lost. In that case, we would have to specify the
exact volume $\mathcal{V}$.
As a result, we would acquire explicit dependence on the
coarse-graining scale.

\para{First passage distribution}
Our interest is in using $\SMSR$
to evaluate the rare tail of the first passage distribution.
Formally, we can introduce a restricted
transition probability
by translation
of~\eqref{eq:Schwinger-Keldysh-P}
to our current variables of integration,
\begin{equation}
    \label{eq:Schwinger-Keldysh-P-restricted}
    \ProbPre( \phistop, \Nstop \divider \phiinit, \Ninit )
    =
    \NormalizationRe
    \primedint
    [\d \qa \, \d \qb \, \d \pa \, \d \pb]
    \,
    \exp
    \Big(
        \im \SMSR[ \varphi, \XEmom ]
    \Big) .
\end{equation}
Here, as in~\S\ref{sec:slow-roll-instanton},
$\primedint$ should be understood to impose the
first passage condition $\qa(N) > \phiend$
at each intermediate time.
As above, this should be
understood as a density with respect to the boundary
configurations
$\phiinit$ and $\phistop$.

Note that, in these variables,
the path integral appears
unusual
because the primary fields $\varphi_i$
are integrated over
the real axis (or a subset of it for $\qa$), whereas
the response fields $\XEmom_i$
are integrated over the imaginary axis.
The field
$\qa$ satisfies
initial and final boundary conditions
set by $\phiinit$ and $\phistop$, respectively,
at times $N = \Ninit$ and $N = \Nstop$.
However, the boundary values
of $\qb$ are unrestricted.
The integrals over the response fields $\XEmom_i$
run over the entire imaginary axis,
except that
the CTP condition $\qu{\phi} = 0$
at the final time requires
$\pb$ to approach zero
there. Its initial value is unrestricted.
Finally,
the other response field $\pa$
is unrestricted in its initial and final values.

The condition that the noise variable $\pb$
approach zero
at the final time
is a new feature, inherited from the
Schwinger--Keldysh path integral.
It did not have an analogue in the ``bottom up''
MSR action constructed from the phenomenological
Starobinsky equation~\eqref{eq:Starobinsky-Langevin}.
Guarie \& Migdal
found it necessary to impose a similar condition by hand
when constructing an instanton
for turbulent flow described by Burgers' equation~\cite{Gurarie:1995qc},
in order to have convergence of the action
at late times.

As before, our main interest is in the first
passage distribution $\ProbQ$.
There is an interesting literature on arrival processes
in quantum mechanical systems;
see, e.g., Refs.~\cite{kumar1985quantum,Grot:1996xu,
PhysRevA.57.4130,2000quant.ph..9111M,dhar2015quantum},
although in this paper we do not make any use of
these results.
We will simply assume that the Feynman--Kac-like
formula~\eqref{eq:Feynman-Kac-Q}
for $\ProbQ$
can be promoted to a quantum-mechanical expectation
value computed using the Schwinger--Keldysh path integral.
It is an interesting question whether this formula
requires corrections
(beyond those discussed in~\S\ref{sec:path-integral-flux-formula})
when interpreted within a fully
quantum-mechanical setting.

With this assumption,
and using~\eqref{eq:flux-formula-marginalized},
it follows that $\ProbQ$
can formally be computed from the path integral representation
\begin{equation}
    \label{eq:Schwinger-Keldysh-Q}
    \ProbQ( \phistop, \Nstop \divider \phiinit, \Ninit )
    =
    -
    \NormalizationRe
    \primedint
    [\d \qa \, \d \qb \, \d \pa \, \d \pb]
    \,
    \left.\frac{\d \qa}{\d N}\right|_{N = \Nstop}
    \exp
    \Big(
        \im \SMSR[ \varphi, \XEmom ]
    \Big)
    .
\end{equation}
There is no need to explicitly marginalize over the velocity
if the $\qb$ integral is unrestricted at the final time.
The sign arises from assuming that $\qa$
approaches the end-of-inflation
boundary from the right;
see the discussion below Eq.~\eqref{eq:Feynman-Kac-Q}.
The generalization to multiple fields would
follow in the expected way
from~\eqref{eq:first-passage-probability-current-higher-dim}.

The fields entering the path integral
for $\ProbQ$ should satisfy the
same boundary conditions identified for
the restricted transition probability $\ProbPre$,
and the primed integral
$\primedint$ continues to indicate that
the field $\qa$
should satisfy the first passage constraint
$\qa > \phistop$ at intermediate times.
For application to the ``stochastic $\delta N$
formalism''
we identify $\phistop = \phiend$
and set $\Nstar = \Nstop - \Ninit$.

In practice,
this explicit path integral representation is difficult to use
for the same reasons identified in~\S\ref{sec:slow-roll-instanton}.
In particular, there is generally still an issue with the
regularization of the probability
current at the boundary,
at least in models where the $\qa$ equation
mixes with $\pa$.
Because the CTP condition forces $\pb = 0$ at the future boundary
there would appear to be no similar
regularization issue for mixing with $\pb$.
These issues all deserve further attention.
In conclusion,
as in the slow roll case, we will generally prefer to work with an
instanton approximation to the unrestricted transition
probability,
and translate this to an estimate for $\ProbQ$.

\para{Instanton equations}
We still expect rare events to be controlled by
saddle points of the MSR-like action.
In the rare limit,
the first-passage distribution can be
evaluated from an instanton approximation to
the path integral~\eqref{eq:Schwinger-Keldysh-Q}
based on these saddle points.
This approximation is exactly
analogous to Eq.~\eqref{eq:instanton-approximation}.
As in that case,
to obtain an answer to exponential
accuracy
it is possible to neglect the fluctuation determinant
for the Gaussian models considered here.

The instanton equations
for the saddle point
are the Hamiltonian equations for
$\HFP$, i.e.,
\begin{equation*}
    \frac{\d \varphi_i}{\d N}
    =
    \frac{\partial \HFP}{\partial \XEmom_i}
    ,
    \qquad 
    \frac{\d \XEmom_i}{\d N}
    =
    -
    \frac{\partial \HFP}{\partial \varphi_i}
    .
\end{equation*}
This yields
\begin{subequations}
\begin{align}
    \label{eq:inst_ps_2}
    \frac{\d \qa}{\d N}
    & =
    \qb
    +
    2 \D_{1j} \XEmom_j
    ,
    &
    \frac{\d \qb}{\d N}
    & =
    - (3 - \epsilon)\qb - \frac{V'(\qa)}{H^2}
    +
    2 \D_{2j} \XEmom_j
    ,
    \\
    \label{eq:inst_ps_1}
    \frac{\d \pa}{\d N}
    & =
    \pb
    \frac{V''(\qa)}{H^2}
    ,
    &
    \frac{\d \pb}{\d N}
    & =
    - \pa
    +
    (3 - \epsilon) \pb
    .
\end{align}
\end{subequations}
These equations reduce to the standard background dynamics
when $\XEmom = 0$.
They clearly illustrate how stochastic corrections
modify the classical trajectory via coupling to the
noise kernel $\D_{ij}$.
Notice the ``wrong sign''
contribution $(3-\epsilon) \pb$
in the evolution equation for $\pb$.
This is opposite to the usual
Hubble friction term $-(3-\epsilon) \qb$
appearing in the evolution equation for $\qb$,
and shows that the noise terms will typically
have exponentially \emph{growing}
solutions.
We comment on numerical issues
associated with this exponential
growth in~\S\ref{sec:conclusions}.

In writing Eq.~\eqref{eq:inst_ps_1}
for the time evolution of the $\XEmom_i$,
we have regarded the background quantities
$\epsilon(N)$ and $H(N)$
as fixed time-dependent quantities
and ignored their variation with
$\varphi_i$.
We have also dropped terms coming from derivatives
of the noise matrix $\D_{ij}$ with respect to
the $\varphi_i$.
In slow-roll models,
and perhaps others,
we expect that
these terms are typically small.
However,
this approximation is not required by the instanton
method,
and all these terms can be restored
in models where they are needed to accurately
compute the time dependence of the $\XEmom_i$.

The instanton
equations~\eqref{eq:inst_ps_2}--\eqref{eq:inst_ps_1}
have a close relationship
to Tomberg's method for estimating
the most likely transition
trajectory,
discussed in~\S\ref{sec:tomberg-langevin-method}.
Specifically, Tomberg's
Onsager--Machlup functional~\eqref{eq:tomberg-onsager-machlup}
could be obtained from the
MSR-like transition probability
by integrating out the noise fields $\XEmom_i$.
Further,
the
critical
trajectory~\eqref{eq:tomberg-deltaN-minimizer},
obtained by maximizing this functional,
has a similar status
to the saddle point equations obtained in this section.
The key difference is that in Tomberg's approach
(as also in the spectral method), the
noise variables have already been marginalized over.
In the instanton approach, we
obtain the full
time-dependence of the most likely noise realization
\emph{automatically}
from solving~\eqref{eq:inst_ps_1}.
We also obtain full time-dependent
information about the most likely transition
trajectory.
As explained in~\S\ref{sec:methodoogy-review},
both of these carry useful information that
is difficult to obtain
using either the spectral method
or Tomberg's formalism.
In practical terms, the
fact that the noise variables have been eliminated
makes the critical point
equation~\eqref{eq:tomberg-deltaN-minimizer}
more nonlinear than
Eqs.~\eqref{eq:inst_ps_2}--\eqref{eq:inst_ps_1}.

It is still necessary to check
that the contour of integration can be
safely deformed to pass through the saddle.
When expressed in terms of $\varphi$ and $\XEmom$,
it should be remembered that
the $\varphi_i$ are integrated over the real
axis,
but the $\XEmom_i$
are integrated over
the \emph{imaginary} axis.
Based on our results in~\S\ref{sec:slow-roll-instanton}
we expect the saddle point to be located
at real values of both $\varphi$
and $\XEmom$.
Provided the matrix $\D_{ij}$ is positive definite,
and
using the same argument as in~\S\ref{sec:slow-roll-instanton},
one can verify that it is possible the deform the
integration contour for the $\XEmom_i$
so that it is displaced
slightly away from the imaginary axis.
These allows the contour to pass through the saddle point
where it crosses the real axis.

\section{Applications}\label{sec:Apps}

In this section, we apply the framework
developed in~\S\S\ref{sec:slow-roll-instanton}--\ref{sec:instanton-phase}
to a number of idealized cases that
admit analytical solutions. These examples enable
us to benchmark the instanton approach
against results reported in the literature
using
one of the techniques
described in~\S\ref{sec:methodoogy-review}.
In some cases we are also able to compare
to results from the mathematical literature.

In this section, to simplify notation, we always
define time so that $\Ninit = 0$,
and therefore $\Nstop = \Nstar$.

\subsection{Slow-roll inflation}\label{sec:SR}

We begin by
re-analysing the slow-roll scenario
with a linear potential,
previously discussed
in~\S\ref{sec:instanton-linear-model}
within the
framework of the slow-roll MSR instanton.
In this section, we are able to drop the slow-roll
approximation.

First, we identify the necessary
noise terms.
In the exact de Sitter limit,
the only non-vanishing component of the diffusion matrix
(see Eq.~\eqref{eq:PS_matrix})
is $\D_{11} = H^2 / (8\pi^2)$.
The instanton equations~\eqref{eq:inst_ps_2}--\eqref{eq:inst_ps_1}
reduce to
\begin{subequations}
\begin{align}
    \label{eq:srins1}
    \frac{\d \pa}{\d N}
    &=
    0
    ,
    &
    \frac{\d \pb}{\d N}
    &=
    -
    \pa
    +
    3\pb
    ,
    \\
    \label{eq:srins2}
    \frac{\d \qa}{\d N}
    &=
    \qb
    +
    \frac{H^2}{4\pi^2}
    \pa
    ,
    &
    \frac{\d \qb}{\d N}
    &=
    -3 \qb
    -
    \frac{V'}{H^2}
    ,
\end{align}
\end{subequations}
where we have approximated
$V'(\qa)/H^2 \approx \text{const.}$,
and neglected $V''(\qa)$ accordingly.
To simplify the following expressions,
we define the drift velocity $v \equiv V'/(3H^2)$
and a diffusion constant
$D \equiv D_{11}$.

This system of equations is readily solved.
Eq.~\eqref{eq:srins1} shows that $\pa$ is constant,
and
the remaining equations can be integrated sequentially.
Before imposing the boundary conditions
to identify the instanton trajectory,
it is useful to look at the general solution.
This highlights the effect of the stochastic terms.
The solution is
\begin{subequations}
\begin{align}
    \label{eq:linear-phase-space-instanton-P}
    \pa
    &=
    \text{const.}
    , 
    &
    \pb(N)
    &=
    \frac{\pa}{3}
    +
    \kappa \e{3N}
    ,
    \\
    \label{eq:linear-phase-space-instanton-fields}
    \qa(N)
    &=
    \alpha
    -
    \frac{\beta}{3} \e{-3N} 
    +
    (2 D \pa - v) N
    ,
    &
    \qb(N)
    &=
    \beta \e{-3N}
    -
    v
    .
\end{align}
\end{subequations}
As
explained in~\S\ref{sec:MSR-phase-space}
above,
there is an
exponentially growing mode for $\pb$,
which is a counterpart to the
decaying mode that appears in both $\qa$ and $\qb$.

We determine the constants
$\alpha$, $\beta$, $\kappa$,
and $\pa$
by imposing the instanton
boundary conditions.
With our conventions in this section,
the field starts at $\qa(0) = \phiinit$
and reaches $\qa(\Nstar) = \phistop$
after $\Nstar$ e-folds,
assumed to be much larger than the time required under
deterministic evolution alone.
We will eventually be interested in the
case $\phistop = \phiend$.
We also assume that the system begins on the
slow-roll attractor, so that $\qb(0) = -v$.
This sets $\beta = 0$.
Finally, we must set $\pb(\Nstar) = 0$,
which fixes the constant $\kappa$.
This yields the instanton
trajectory
\begin{subequations}
\begin{align}
    \label{eq:sol_q_sr}
    \qa(N)
    &=
    \phiinit
    +
    \Delta\phi \frac{N}{\Nstar}
    ,
    &
    \qb(N)
    &=
    - v
    .
\end{align}
The noise realization is
\begin{align}
    \label{eq:sol_p_sr}
    \pa
    & =
    \frac{1}{2D}
    \left(
        v
        +
        \frac{\Delta\phi}{\Nstar}
    \right)
    ,
    &
    \pb(N)
    &
    =
    \frac{1}{6D}
    \left(
        v
        +
        \frac{\Delta\phi}{\Nstar}
    \right)
    \Big(
        1 - \e{3(N - \Nstar)}
    \Big)
    .
\end{align}
\end{subequations}
We see that, in fact,
$\kappa$ does not contribute to the instanton
approximation for the path integral.
The noise realization
$\pa$
matches that found in the overdamped limit,
cf.~Eq.~\eqref{eq:SR-instanton-soln-P}.

As in the overdamped analysis,
we see that the noise term precisely adjusts itself
to cancel the deterministic drift $v$.
This allows the field to traverse
the required field-space distance
$\Delta\phi$ in $\Nstar$ e-folds.
The key difference is that, here,
the noise is activated by $\qu{\pi}$
rather than $\qu{\phi}$.
Ultimately, however, the effect on the tail of the
distribution is the same, as we now confirm.

As a first step,
we compute the Fokker--Planck Hamiltonian,
Eq.~\eqref{eq:HFP}, on the instanton trajectory.
This produces
\begin{equation}
    \HFPtilde(\varphi,\XEmom)
    =
    D \pa^2
    -
    v \pa
    -
    3 \beta \kappa
    =
    -
    \frac{1}{4D}
    \bigg(
        v^2
        -
        \frac{(\Delta\phi)^2}{\Nstar^2}
    \bigg)
    ,
\end{equation}
where we have used $\beta = 0$.
At the level of approximation to which we are working,
this value is a constant along the
instanton trajectory.
We can interpret
$\HFPtilde$ as the energy required
to drive the field from $\phiinit$ to $\phistop$
in exactly $\Nstar$ e-folds.
In other applications, the focus is often on
zero-energy trajectories, such as the case of
transitions between two stationary points
driven by stochastic forces.
In our example, the emergence of the non-zero $\HFP$
reflects (at least partly)
the requirement that the transition completes
in a finite time interval.

Finally, we evaluate the MSR action at the saddle point.
Eqs.~\eqref{eq:Smps}
and~\eqref{eq:Schwinger-Keldysh-Q}
show that
\begin{multline}
    \im
    \SMSR[\varphi,\XEmom]
    =
    -
    \int_0^{\Nstar}
    \d N
    \left[
        \pa
        \frac{\d \qa}{\d N}
        +
        \pb
        \frac{\d \qb}{\d N}
        -
        \HFPtilde(\varphi,\XEmom)
    \right]
    \\
    =
    -
    D
    \pa^2
    \Nstar
    =
    -
    \frac{1}{4D}
    \bigg(
        v
        +
        \frac{\Delta\phi}{\Nstar}
    \bigg)^2
    \Nstar
    .
    \label{eq:SMSR_sr_5}
\end{multline}
Eq.~\eqref{eq:SMSR_sr_5}
can be used to build an instanton approximation
for $\ProbP$.
Including the correct
normalization, we find
\begin{equation}
    \label{eq:Prob_phie_SR}
    \ProbP(\phistop, \Nstar \divider \phiinit, 0)
    =
    \frac{1}{(4\pi D \Nstar)^{1/2}}
    \exp
    \bigg(
        {-
        \frac{1}{4D \Nstar}}
        \Big[
            \phistop
            -
            \big(
                \phiinit
                -
                v \Nstar
            \big)
        \Big]^2
    \bigg)
    .
\end{equation}
This is a Gaussian distribution for $\phistop$,
centred around the deterministic value
$\phi(\Nstar) = \phiinit - v \Nstar$. The width of this distribution is
modulated by the transition duration, $\Nstar$. 

Eq.~\eqref{eq:Prob_phie_SR}
shows that the normalization
is a power law in $\Nstar$
and therefore does not affect the result
to exponential accuracy.
Also, using the method of images,
we can relate
$\ProbP$ to $\ProbPre$.
The required relation is
\begin{equation}
    \label{eq:method-images-linear-potential}
    \ProbPre (\phistop, \Nstop \divider \phiinit)
    =
    \ProbP(\phistop, \Nstop \divider \phiinit)
    -
    \e{- v(\phiend - \phiinit)/D}
    \ProbP(\phistop, \Nstop \divider 2\phiend - \phiinit)
    .
\end{equation}
Notice that Eq.~\eqref{eq:method-images-linear-potential}
uses a slightly different formulation
of the method of images, where the ``image''
is in the initial value, not the final value.
In this formulation it is manifest that
$\ProbPre$ satisfies the forward Kolmogorov equation.
Still working
to exponential accuracy in $\Nstar$,
Eq.~\eqref{eq:method-images-linear-potential}
clearly introduces no changes
to our estimate.
Finally,
since $\ProbPre(\phistop = \phiend) = 0$
and $D$ is a constant,
it can be checked that
the mapping to $\CurrentRe(\phiend)$
also produces no change.
In conclusion, taking the $\Nstar \gg 1$ limit
of~\eqref{eq:Prob_phie_SR},
we can write the tail of the
$\ProbQ$ distribution in the form
\begin{equation}
     \label{eq:SMSR_sr_51}
     \ln \ProbQ(\Nstar)
     \sim
     -
     \frac{1}{2}
     \frac{\Nstar}{\dimP_{\zeta}}
     +
     \Or(1)
     ,
\end{equation}
with the same meaning for $+ \Or(1)$
as explained below Eq.~\eqref{eq:instanton-slow-roll-exp-tail}.
The leading part of the tail is independent of our precise choice
for the initial velocity.
If the constant $\beta$ had been retained, it would
enter only at the level of the first $\Or(1)$ correction.
See also the discussion in~\S\ref{sec:decaying-noise}.

This result is unchanged
compared
to the overdamped
estimate~\eqref{eq:instanton-slow-roll-exp-tail}.

\para{Survival probability and subexponential estimate}
Combining Eqs.~\eqref{eq:Prob_phie_SR}
and~\eqref{eq:method-images-linear-potential}
yields an explicit formula for the survival
probability~\eqref{eq:survival-probability-def},
\begin{equation}
    \Survival(\Nstar, \phiend \divider \phiinit)
    =
    \frac{1}{2}
    \left[
        \Erfc
        \left(
            \frac{\phiend - \phiinit + v \Nstar}{\sqrt{4\D \Nstar}}
        \right)
        -
        \e{- v (\phiend - \phiinit)/D}
        \Erfc
        \left(
            \frac{\phiinit - \phiend + v \Nstar}{\sqrt{4 \D \Nstar}}
        \right)
    \right]
    ,
\end{equation}
where $\Erfc(z)$ denotes the complementary error function.
Evaluating its decay rate gives an exact formula for the
first passage distribution, including the subexponential
prefactor,
\begin{equation}
    \label{eq:exact-Q-linear-potential}
    \ProbQ(\Nstar)
    =
    -
    \frac{
        \partial \Survival(\Nstar)
    }{
        \partial \Nstar
    }
    = 
    \frac{
        \phiinit - \phiend
    }{
        (4\pi \D \Nstar^3)^{1/2}
    }
    \exp
    \Big(
        {-
        \frac{1}{4\D \Nstar}}
        \Big[
            \phiend
            -
            \big(
                \phiinit - v \Nstar
            \big)
        \Big]^2
    \Big)
    .
\end{equation}
As promised, the exponential dependence of
this exact result
matches our estimate, Eq.~\eqref{eq:SMSR_sr_51}.
If the end-of-inflation boundary is an upper limit, so that
$\phiinit < \phiend$,
the same formula is valid up to a simple sign change.
It can be verified that we obtain the same result
from the flux formula
applied to $\ProbPre$.

Finally, notice that
if we apply the flux formula to $\ProbP$ rather than
$\ProbPre$,
we obtain
\begin{equation}
    -\Current (\Nstar)
    =
    \left.
        \left(
            v \ProbP
            +
            D 
            \frac{\partial \ProbP}{\partial \phi}
        \right)
    \right|_{\phi = \phiend}
    =
    \frac{
        \phiinit - \phiend + v \Nstar
    }{
        4(\pi D \Nstar^3)^{1/2}
    }
    \exp
    \Big(
        {-
        \frac{1}{4\D \Nstar}}
        \Big[
            \phiend
            -
            \big(
                \phiinit
                -
                v \Nstar
            \big)
        \Big]^2
    \big)
    .
\end{equation}
Up to exponential accuracy this agrees with the exact
result~\eqref{eq:exact-Q-linear-potential}.
However, as expected, it does not produce the correct
subexponential prefactor.

\subsubsection*{Validation through the renewal equation}

This short section lies outside our main line of
argument,
and can be omitted.

One of the useful features of the slow-roll model
with linear potential is its simplicity.
This allows the problem to be approached
in a number of different ways,
even to the subexponential accuracy required for~\eqref{eq:exact-Q-linear-potential}.
We can leverage this feature to validate
our estimate for $\ProbQ$
without using the method of images.

In Appendix~\ref{sec:App_ren_eq}
it is shown that
the transition probability $\ProbP$
and the first passage distribution $\ProbQ$
are related through a \emph{renewal equation}.
This can be written
\begin{equation}
    \label{eq:ren_eq_0}
    \ProbP( \phistop, \Nstop \divider \phiinit, \Ninit )
    =
    \int_{\Ninit}^{\Nstop}
    \ProbP( \phistop, \Nstop \divider \phi, N )
    \ProbQ( \phi, N \divider \phiinit, \Ninit )
    \,
    \d N 
    ,
\end{equation}
Eq.~\eqref{eq:ren_eq_0}
can be obtained by noticing that the system must
first cross the (arbitrary) intermediate
value $\phi$ at some time
$N$ between the initial and final times,
i.e., $\Ninit \leq N \leq \Nstop$.
These events are exclusive and exactly one of them must
apply. Eq.~\eqref{eq:ren_eq_0} then follows.

Assuming Markovian dynamics,
both $\ProbP$ and $\ProbQ$
depend only on time differences.
Hence, we may interpret the right-hand side of
Eq.~\eqref{eq:ren_eq_0} as a convolution.
Applying a Laplace transform
(denoted by a tilde)
yields the simple relation
\begin{equation}
    \label{eq:Q_LT0_SR}
    \tilde{\ProbQ}( s, \phi \divider \phiinit )
    =
    \frac{
        \tilde{\ProbP}(\phistop, s \divider \phiinit)
    }{
        \tilde{\ProbP}(\phistop, s \divider \phi)
    }
    .
\end{equation}
Further details are provided in Appendix~\ref{sec:App_ren_eq}.

We can now make use of the expressions derived above.
Eq.~\eqref{eq:Prob_phie_SR}
already gives the normalized,
unrestricted transition probability from $\phiinit$
to $\phistop$.
In what follows we temporarily change notation to write this
as a function of an arbitrary future time $\Nstop$.
Its Laplace transform is
\begin{align}
    \tilde{\ProbP}(\phistop,s \divider \phiinit)
    =
    \frac{1}{(4 D s + v^2)^{1/2}}
    \exp
    \left(
        -
        \frac{
            \phiinit - \phistop
        }{
            2\D
        }
        \left[
            \sqrt{4\D s + v^2 }
            -
            v
        \right]
    \right)
    .
\end{align}
Using this result in the Laplace-domain
renewal equation~\eqref{eq:Q_LT0_SR}
produces
\begin{equation}\label{eq:Q_LT1_SR}
    \tilde{\ProbQ}(s, \phi \divider \phiinit)
    =
    \exp
    \left(
        -
        \frac{\phiinit - \phi}{2D}
        \left[
            \sqrt{4\D s + v^2 }
            -
            v
        \right]
    \right)
    .
\end{equation}
As expected from the assumption
of Markovianity,
all dependence on $\phistop$ has cancelled.
Finally, Bromwich inversion of
Eq.~\eqref{eq:Q_LT1_SR} exactly
reproduces exactly
our existing estimate for
$\ProbQ$, Eq.~\eqref{eq:exact-Q-linear-potential}.

Unfortunately,
this procedure is rarely applicable in more general
scenarios,
due to the difficulty of obtaining analytic expressions
for the Laplace transforms and their inverses.

\subsection{Ultra-slow-roll inflation}
\label{sec:USR}

As a second example,
we consider an epoch of ultra-slow-roll (USR)
inflation.
This model is particularly interesting
because it produces an enhanced power spectrum,
which
may have important implications for the
formation of primordial black holes
or other early collapsed objects.
The enhancement of power increases the amplitude
of fluctuations, and therefore also increases the
importance of stochastic effects.
If the tail of the probability distribution decays
more slowly than a Gaussian, rare fluctuations
become more probable
than one would predict based on behaviour of
$\ProbP(\zeta)$ near its centre.
This significantly
boosts the likelihood of PBH production,
and can completely change the viability of
models where PBHs form a significant fraction of the
dark matter.

Here,
our focus is on an idealized model
that captures the essential features of
a USR epoch.
Specifically,
we consider a background evolution governed by
\begin{equation}
    \frac{\d^2 \phi}{\d N^2}
    +
    (3 - \epsilon) \frac{\d \phi}{\d N}
    \approx
    0 ,
\end{equation}
where $\epsilon \ll 1$ is taken to be vanishingly small.
Therefore, its contribution to the dynamics can be
neglected at leading order.
However, the slow-roll parameter
$\epsilon_2 \equiv \d \ln \epsilon / \d N$
remains significant.
In what follows we work in the limit
$\epsilon_2 = -6$,
which characterizes an exact USR phase.
This choice yields a power
spectrum
$\dimP_{\delta\phi}$
that matches the spectrum produced in a SR epoch.
This correspondence is an example of a \emph{Wands duality}~\cite{Wands:1998yp},
and is especially convenient
because the necessary elements of the noise matrix
$\D_{ij}$ have already been determined.

Under these assumptions,
the
deterministic inflaton trajectory is
\begin{equation}
    \phi(N)
    =
    \phiasm
    +
    (\phiinit - \phiasm)
    \e{-3N}
    \quad
    \implies
    \quad
    \frac{\d\phi}{\d N}
    =
    -3 (\phiinit - \phiasm) \e{-3N}
    ,
\end{equation}
where $\phiasm$ denotes the asymptotic value
approached by the field. Without loss of generality,
we shift the field so that $\phiasm  = 0$.

The structure of the instanton equations
mirrors the slow-roll case, but
the
deterministic drift velocity is absent.
Moreover, the noise structure remains the same,
because
only the $\D_{11}$ component survives at leading
order in the slow-roll expansion;
see Eqs.~\eqref{eq:PS_matrix} and~\eqref{eq:Hankel_index}.
As a result, the instanton trajectory satisfies
\begin{subequations}
\begin{align}
    \frac{\d \qa}{\d N}
    & =
    \qb
    +
    \frac{H^2}{4\pi^2} \pa
    ,
    &
    \frac{\d \pa}{\d N}
    & =
    0
    ,
    \label{eq:usrins1}
    \\
    \frac{\d \qb}{\d N}
    & =
    -3 \qb
    ,
    &
    \quad
    \frac{\d \pb}{\d N}
    & =
    -
    \pa
    +
    3\pb
    .
    \label{eq:usrins2}
\end{align}
\end{subequations}
The general solution is
\begin{subequations}
\begin{align}
    \qa(N)
    & =
    \alpha
    -
    \frac{\beta}{3} \e{-3N}
    +
    \frac{H^2}{4\pi^2} \pa N
    ,
    &
    \pa
    & =
    \text{const.}
    ,
    \\
    \qb(N)
    & =
    \beta \e{-3N}
    ,
    &
    \pb(N)
    & =
    \kappa \e{3N}
    +
    \frac{\pa}{3}
    .
\end{align}
\end{subequations}
Boundary conditions should be
imposed in a similar way to the
slow-roll case studied in~\S\ref{sec:SR} above.
We set
$\qa(0) = \phiinit$
and $\qa(\Nstar) = \phistop$.
Ultimately, we are again
interested in the case $\phistop = \phiend$.
We will also require that the system begins on a noiseless
trajectory, i.e., $\qb(0) = \phi'(0) = -3\phiinit$.
This yields the following solutions for the noise
fields
\begin{subequations}
\begin{align}
    \label{eq:sol_pa_usr}
    \pa
    & =
    \left(
        \frac{H}{2\pi}
    \right)^{-2}
    \frac{
        \phistop -\phiinit \e{-3\Nstar}
    }{
        \Nstar
    }
    ,
    \\
    \label{eq:sol_pb_usr}
    \pb(N)
    & =
    \frac{1}{3\Nstar}
    \left(
        \frac{H}{2\pi}
    \right)^{-2}
    \left(
        \phistop - \phiinit \e{-3\Nstar}
    \right)
    \left(
        1 - \e{3(N - \Nstar)}
    \right)
    .
\end{align}
Just as for the slow-roll case, these
noise realizations are sensitive to microphysical information.
Meanwhile, the primary fields
interpolate between $\phiinit$ and $\phistop$,
\begin{align}
    \label{eq:sol_qa_usr}
    \qa(N)
    &
    =
    \phiinit \e{-3N}
    +
    \left(
        \phistop - \phiinit \e{-3\Nstar}
    \right)
    \frac{N}{\Nstar}
    ,
    \\
    \label{eq:sol_qb_usr}
    \qb(N)
    & =
    -3\phiinit \e{-3N}
    .
\end{align}
\end{subequations}
The integration constant $\kappa$
appearing in $\pb$
is fixed by the CTP condition
$\qu{\phi}(\Nstop) = 0$.
As for the case of the slow-roll linear potential,
it has no effect on the
instanton approximation to the effective action.
However, it does contribute to the
Fokker--Planck
Hamiltonian $\HFPtilde$,
which can be seen in Eq.~\eqref{eq:HFP-USR}
below.

To obtain
$\HFPtilde$,
insert
Eqs.~\eqref{eq:sol_pa_usr}--\eqref{eq:sol_qb_usr}
into Eq.~\eqref{eq:HFP}.
This produces a constant
along the the instanton trajectory,
\begin{equation}
\label{eq:HFP-USR}
\begin{split}
    \HFPtilde
    & =
    \frac{1}{2}
    \left(
        \frac{H}{2\pi}
    \right)^{2}
    \pa^2
    -
    3 \beta \kappa
    \\
    & =
    \frac{1}{2}
    \left(
        \frac{H}{2\pi}
    \right)^{-2}
    \Big[
        -6 \phiinit \e{-3\Nstar}
        \frac{
            \phiend
            -
            \phiinit \e{-3\Nstar}
        }{
            \Nstar
        }
        +
        \Big(
            \frac{
                \phiend
                -
                \phiinit \e{-3\Nstar}
            }{
                \Nstar
            }
        \Big)^2
    \Big]
    .
\end{split}
\end{equation}
In the second line we have set $\phistop = \phiend$,
which makes explicit the ``energy'' required to reach the target.

The MSR action evaluated on the instanton
trajectory is
\begin{equation}
\label{eq:SMSR_USR}
\begin{split}
    \im
    \SMSR
    & =
    -
    \int_0^{\Nstar}
    \d N
    \left[
        \pa \frac{\d \qa}{\d N}
        +
        \pb \frac{\d \qb}{\d N}
        -
        \HFPtilde
    \right]
    =
    -
    \frac{1}{2}
    \left(
        \frac{H}{2\pi}
    \right)^{2}
    \pa^2 \Nstar
    \\
    & =
    -
    \frac{1}{2}
    \left(
        \frac{H}{2\pi}
    \right)^{-2}
    \frac{
        \big( \phistop - \phiinit \e{-3\Nstar} \big)^2
    }{
        \Nstar
    }
    . 
\end{split}
\end{equation}
We may
use Eq.~\eqref{eq:SMSR_USR}
to build an instanton approximation for
$\ProbP$.
Imposing correct normalization of the resulting
distribution
yields a normalization constant $\Normalization$
that is a power law in $\Nstar$.
Hence,
this
does not affect our estimate,
up
to exponential accuracy.

A distinctive feature of the USR model is that the instanton trajectory~\eqref{eq:sol_pa_usr}--\eqref{eq:sol_qb_usr}
may develop overshoot behaviour.
This occurs when the solution
overshoots $\phi = \phiend$
and
moves into the forbidden region $\phi < \phiend$
\emph{before} $\Nstar$, but then turns around
and returns to $\phiend$.
We describe such trajectories as
\emph{turnovers}.
The appearance of this
behaviour depends sensitively on the choice
of $\Nstar$,
and also the relative positions of
$\phiasm$, $\phiinit$, and $\phiend$.
Because they enter the forbidden region,
the
interpretation of
these trajectories
becomes more delicate.
We will return to this issue below.

\begin{figure}[ht]
    \centering
    \includegraphics[width=\textwidth]{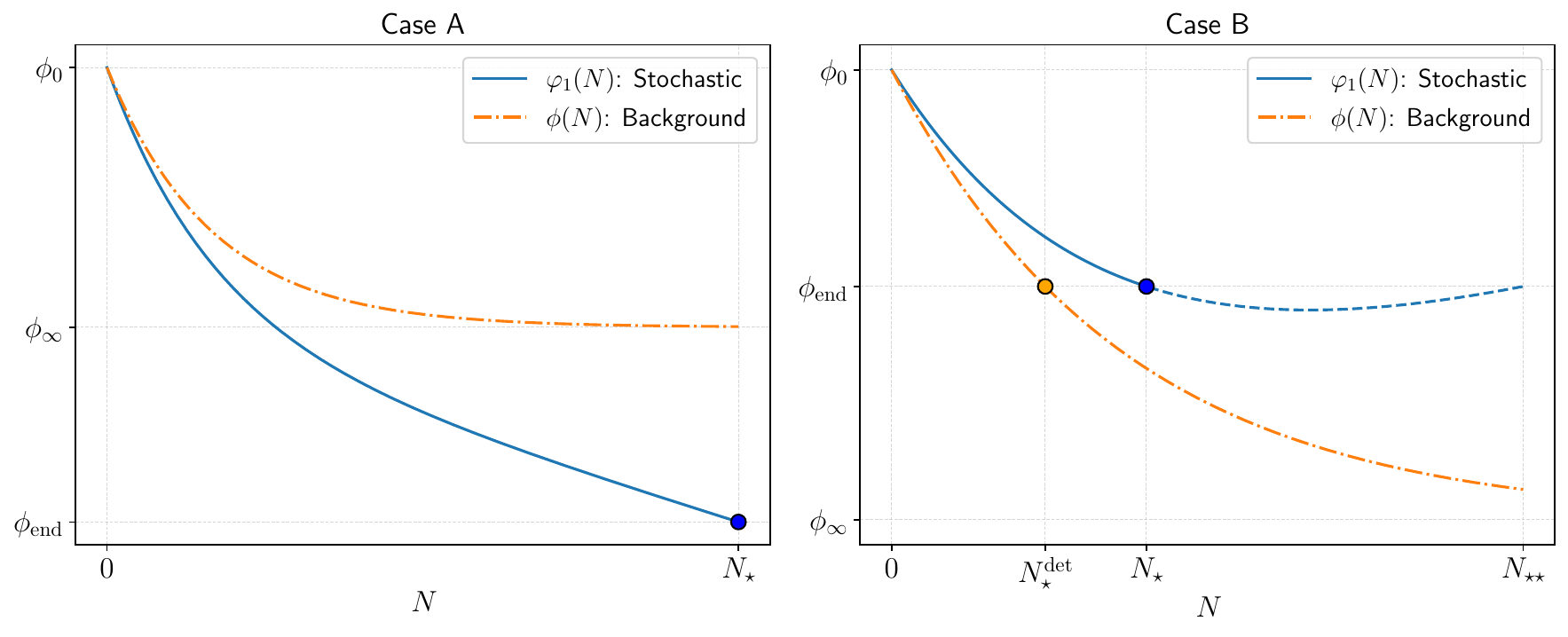}
    \caption{Background (noiseless) and stochastic trajectories for a field during a USR phase. \semibold{Left:} Case A ($\phiend < \phiasm < \phiinit$). The target value $\phiend$ is unreachable under noiseless evolution, and the stochastic trajectory approaches it monotonically. \semibold{Right:} Case B ($\phiasm < \phiend < \phiinit$). A single noise realization $\pa$ gives rise to two crossing times, $\Nstar$ and $N_{\star\star}$. } 
    \label{fig:USR_traj}
\end{figure}

\begin{figure}[ht]
    \centering
    \includegraphics[width=\textwidth]{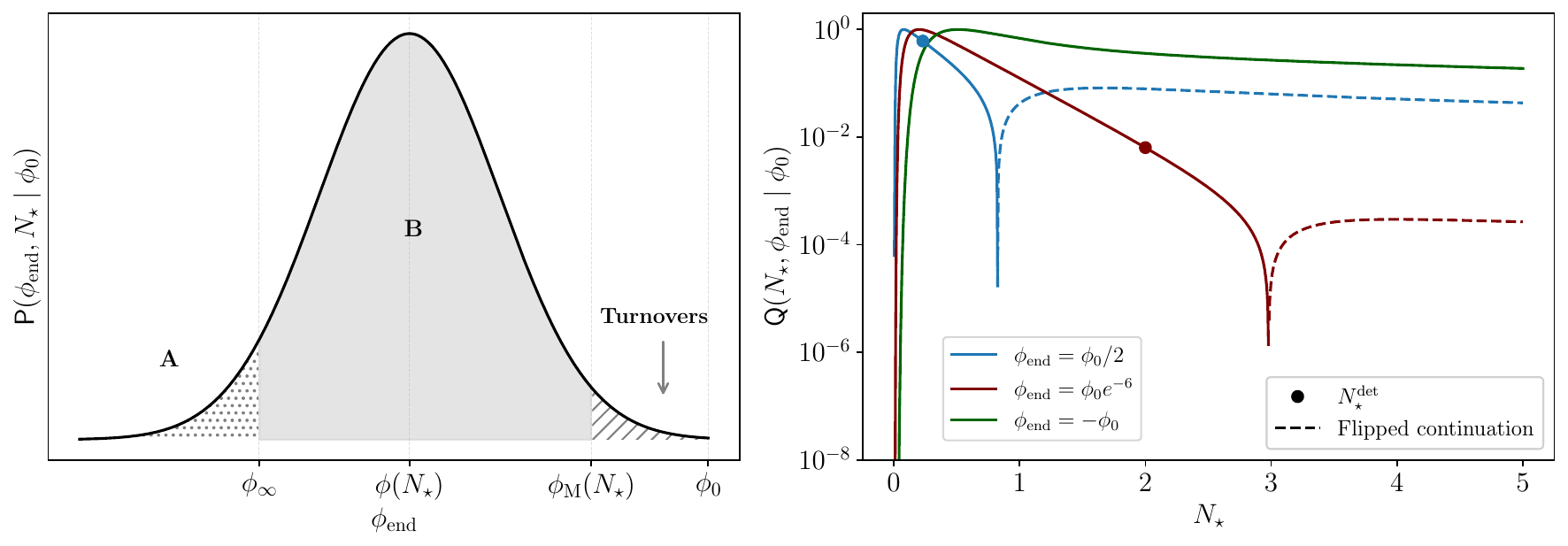}
    \caption{\semibold{Left:} Illustration of $\ProbP(\phiend, \Nstar \divider \phiinit)$ in an USR phase, as determined by Eq.~\eqref{eq:SMSR_USR}. The dotted region is classically inaccessible, whereas the diagonally hatched region corresponds to second crossings. $\phi_{\rm M}(\Nstar) = (1+3\Nstar)\phiinit e^{-3\Nstar}$ and $\phiinit$ delimit this zone.
    \semibold{Right:} $\ProbQ(\Nstar, \phiend \divider \phiinit)$, as defined in Eq.~\eqref{eq:Q_usr}. Each curve has been rescaled so that its peak equals $1$. Dashed lines indicate a sign flip due to turnovers. Markers show the time the field would have reached $\phiend$ under noiseless evolution. Note that these are \emph{not} normalized distributions, and they are only meant to illustrate the exponential scaling behaviour captured by the instanton.} 
    \label{fig:ProbPQ_usr}
\end{figure}

With our conventional assumption that $\qa$
evolves towards $\phiasm$
from larger values,
we distinguish two cases.
\begin{itemize}
    \item \semibold{Case A:}
    $\phiend < \phiasm = 0 < \phiinit$.
    In this case,
    because $\phiasm$
    corresponds to the asymptotic field value under
    noiseless evolution, the field \emph{cannot}
    reach $\phiend$ without assistance
    from fluctuations.
    In this regime, no turnovers are observed for any
    value of $\Nstar$.
    Using Eq.~\eqref{eq:sol_qa_usr},
    we obtain 
    an explicit formula for $\d \qa / \d N$,
    \begin{equation}\label{eq:qa_prime_usr}
        \frac{\d \qa}{\d N}
        =
        -3\phiinit \e{-3N}
        +
        \frac{\phiend - \phiinit \e{-3\Nstar}}{\Nstar}
        .
    \end{equation}
    Each of these terms is negative,
    so that $\d \qa / \d N < 0$ throughout the evolution.
    Therefore $\qa$
    evolves monotonically toward
    $\phiend$,
    as illustrated in the left panel of
    Fig.~\ref{fig:USR_traj}. 

    We now specialize Eq.~\eqref{eq:SMSR_USR}
    to the rare limit $\Nstar \gg 1$, yielding
    \begin{equation}
        \ln \ProbQ(\Nstar)
        \approx
        -
        \frac{1}{2}
        \bigg(
        \frac{H}{2\pi}
        \bigg)^{-2}
        \bigg(
            \frac{\phiend}{\Nstar}
        \bigg)^2
        \Nstar
        \equiv
        -
        \frac{1}{2}
        \frac{\Nstar}{\tilde{\dimP}_{\zeta}}
        .
    \end{equation}
    In the final step we have written
    the tail estimate
    in terms of a ``fake''
    power spectrum $\tilde{\dimP}_\zeta$,
    defined by
    $\tilde{\dimP}_\zeta = (H/2\pi)^2 / (2 \Mp^2 \epsilon)$.
    To do so,
    we have identified
    $2 \Mp^2 \epsilon = (\d \qa / \d N)^2
    \approx (\phiend/\Nstar)^2$.
    This ``fake'' $\tilde{\dimP}_\zeta$
    does \emph{not} give the correct expression
    for the power spectrum produced by a USR
    epoch, because it invokes the slow-roll
    approximation.
    However,
    it is notable
    that the tail is governed by a form characteristic
    of slow-roll dynamics.
    We will see a more general manifestation of this
    behaviour in the next section.

    \item \semibold{Case B:} $0 = \phiasm < \phiend < \phiinit$.
    In this configuration, with our
    choice $\phiasm = 0$,
    both $\phiinit$ and $\phiend$
    are positive.
    Therefore,
    Eq.~\eqref{eq:sol_qa_usr} shows
    that the field can overshoot its target
    and subsequently return to $\phiend$.
    This behaviour is illustrated in the right panel of Fig.~\ref{fig:USR_traj},
    where the non-monotonic trajectory is evident.
    In the limit of large $\Nstar$,
    Eq.~\eqref{eq:qa_prime_usr} gives
    $\d \qa / \d N \approx \phiend/\Nstar > 0$
    at $N = \Nstar$.
    The conclusion is that turnovers are generic
    in this limit.

     Setting this issue aside for the moment, we write
     the MSR action~\eqref{eq:SMSR_USR} as
    \begin{multline}
        \im
        \SMSR
        =
        -
        \frac{1}{2}
        \left(
            \frac{H}{2\pi}
        \right)^{-2}
        \frac{\phiinit^2 \e{-6 \Nstar}}{\Nstar}
        \left(
            1 - \e{3(\Nstar - \Nend)}
        \right)^2
        \\
        =
        -
        \frac{1}{18 \dimP_{\zeta}(\Nstar) \Nstar}
        \left(
            1 - \e{3(\Nstar - \Nend)}
        \right)^2
        ,
    \end{multline}
    where we have used $2\Mp^2 \epsilon(N) = \pi(N)^2= \qb(N)^2$
    to express the result in terms of the
    (true) USR power spectrum $\dimP_{\zeta}(\Nstar)$.
    The left panel of Fig.~\ref{fig:ProbPQ_usr}
    illustrates the unrestricted transition probability
    $\ProbP(\phiend, \Nstar \divider \phiinit)$
    obtained via the instanton approximation;
    see also Eq.~\eqref{eq:SMSR_USR}.
    In this plot, we highlight
    the region that gives rise to turnovers. 
\end{itemize}

\para{Influence of turnover trajectories in $\ProbPre$}
As we have observed, turnovers are generic when
the target is accessible during noiseless evolution.
The underlying physical picture is that the field is rolling
too rapidly for the stochastic noise
fields $\XEmom_i$ to arrest its motion over a long interval
$\Nstar$. In realistic scenarios, the field would typically exit the USR
phase at a value beyond $\phiend$, which presumably
modifies the interpretation of
the turnover-and-return phase.

Clearly, there is a question regarding the status of
turnover trajectories,
since they enter the forbidden region $\phi < \phiend$.
These cannot be saddle-point solutions of the restricted
path integral for $\ProbPre$,
since the domain of integration is restricted to
$\phi(N) > \phiend$.
Therefore, it is legitimate to wonder whether these
instantons can represent physical first-passage
transitions.

However, this conclusion seems unjustified.
If a method-of-images formula exists, then
the restricted transition probability
$\ProbPre$
is built from a weighted,
linear combination of unrestricted probabilities $\ProbP$.
Provided the instanton gives a good
approximation to the physical
transition probability $\ProbP$,
it does not matter if the instanton
trajectory enters an unphysical region at intermediate times.
This would be undesirable from the perspective
of an easy physical interpretation, but would not actually
make the solution invalid.
However, for trajectories with turnover behaviour,
it does appear that we must abandon attempts to give a
physical interpretation of
the solution $\qa(N)$
and its supporting
noise realizations.

Notice also that there is no reason to
expect that the instanton
approximation to $\ProbP$, when used in a method-of-images
formula, will yield exactly the same answer as a direct instanton
approximation to $\ProbPre$.
It would be interesting to understand what the saddle-point
trajectory for $\ProbPre$ looks like, in a case where
the corresponding instanton for $\ProbP$ has overshoot behaviour.

\para{First-passage distribution}
Unfortunately, a simple implementation of the images method
is not available in this case. Thus, we can only 
estimate the exponential behaviour of $\ProbQ$ by 
taking $\CurrentRe \sim \Current$ in 
Eq.~\eqref{eq:first-passage-probability-current}.
The unrestricted 
current $\Current$ can be readily obtained from
Eq.~\eqref{eq:usrins1}, which is
a proxy for the Langevin equation.
Using Eq.~\eqref{eq:sol_qb_usr}, 
this gives
\begin{multline}
    \ProbQ(\Nstar, \phiend \divider \phiinit)
    \sim
    -
    \Current(\Nstar)
    =
    \left.
        \left(
            3\phiinit \e{-3\Nstar} \ProbP
            +
            \D
            \frac{\partial \ProbP}{\partial \phi}
        \right)
    \right|_{\phi = \phiend}
    \\
    \sim
    \Normalization(\Nstar)
    \frac{(1 + 6\Nstar) \phiinit \e{-3\Nstar}
    -
    \phiend}{\Nstar}
    \exp
    \left(
        -\frac{
            (\phiend - \phiinit \e{-3\Nstar})^2
        }{
            4\D \Nstar
        }
    \right)
    ,
    \label{eq:Q_usr}
\end{multline}
where $\Normalization(\Nstar)$
is the normalization factor of the Gaussian
distribution defined by $\ProbP$. Notice that we
could obtain an equally good estimate
by dropping the undifferentiated $\ProbP$ term in
$\Current$, since this would vanish for $\ProbPre$.
This gives a pre-factor of the form
$(\phiinit e^{-3\Nstar} - \phiend)\Nstar^{-1}$.

We plot Eq.~\eqref{eq:Q_usr},
for a selection of different parameters,
in the right panel of Fig.~\ref{fig:ProbPQ_usr}.
This shows that,
whenever the combination
of parameters allows for a second crossing,
the sign of the distribution changes.
This is a result of the approximation
$\CurrentRe \sim \Current$. 
As a consequence, our estimate for $\ProbQ$ inherits the ``pathologies''
of $\Current$, including a sign reversal whenever the field crosses
the boundary from the opposite side, as occurs in the case of 
turnovers.
This can be understood
intuitively from the velocity of the
field~\eqref{eq:qa_prime_usr},
which (as noted earlier) is negative
at crossing time if $\phiend < (1+ 3\Nstar) \phiinit e^{-3\Nstar}$.
When this condition is not satisfied, the field reaches 
the target from the opposite direction, leading to a sign 
reversal in the current. As discussed
in~\S\ref{sec:path-integral-flux-formula}, the velocity alone does
not yield the correct conserved current. Nevertheless, it provides a heuristic
picture for why a sign flip in $\ProbQ$, as estimated from $\Current$, is
generically expected for turnover trajectories. 
The net effect is an overall
misprediction of the prefactor of $\ProbQ$.

However, the long-time scaling of the functions in cases with 
turnovers closely resembles that of the no-turnover example,
suggesting that the full version of $\ProbQ$ will
exhibit similar asymptotic (exponential) behaviour.

\subsection{Constant-roll inflation}\label{sec:CR}

In~\S\ref{sec:tomberg-langevin-method}
we introduced the constant-roll model of inflation.
This provides another useful
scenario to test the instanton method.
One reason is to compare with the Langevin analysis
given by Tomberg~\cite{Tomberg:2022mkt}.
Another is that, in this model,
no components of the noise matrix
$\D_{ij}$ can be neglected.
This leads to qualitatively new features in the structure of the
instanton solution, which distinguish it from both the
slow-roll and ultra-slow-roll scenarios.
A third reason is that
this model provides an example where the
tail of
the first-passage distribution $\ProbQ$ is not the
same as the tail of the unrestricted transition probability
$\ProbP$,
and indeed we will not be able to calculate it in general.
However, there is one particular case
(first passage to the equilibrium location in the stable
scenario)
that admits an explicit calculation of $\ProbPre$ and $\ProbQ$.
In this case, we will be able to show that the instanton
correctly reproduces a formula known in the mathematical
literature.

\para{Potential formulation}
We recall from~\S\ref{sec:tomberg-langevin-method}
that the
slow roll parameter
$\epsilon_2$ satisfies $\epsilon_2 \equiv \epsilon'/\epsilon$;
see Eq.~\eqref{eq:epsilon2-def}.
A constant roll era is defined by a fixed value of $\epsilon_2$,
which we write
$\epsilon_2 = 2\epc$.
The noiseless solution is given in
Eqs.~\eqref{eq:noiseless-constant-roll}--\eqref{eq:unperturbed-constant-roll},
and can be summarized by the relations
\begin{equation}\label{eq:CR_bckg}
    \phi(N)
    =
    \phiinit \e{\epc N}
    ,
    \quad
    \pi(N)
    =
    \epc \phi(N)
    ,
    \quad
    \epsilon(N)
    =
    \frac{\pi(N)^2}{2\Mp^2}
    =
    \frac{\epc^2 \phi(N)^2}{2\Mp^2}
    .
\end{equation}
For reasons to be discussed later we focus on the regime $\epsilon_2 \geq -3$.
We comment briefly on the case $\epsilon_2 < -3$ at the end
of this section.

In~\S\ref{sec:tomberg-langevin-method}
the Langevin
analysis was performed without specifying a potential
that realizes the scenario.
However, to apply the instanton method (at least in our current
framework), a potential is required.
A suitable choice that can realize the constant-roll background is
\begin{equation}
    V(\phi)
    =
    \Lambda
    \bigg(
        6
        -
        \frac{\sigma^2\phi^2}{\Mp^2}
    \bigg)
    \exp
    \bigg(
        {-\frac{\sigma \phi^2}{2\Mp^2}}
    \bigg)
    ,
\end{equation}
where $\Lambda$ is a normalization that we will not
need to specify explicitly.
We must also impose the initial condition
$\pi(0) = \epc \phi(0)$.
Otherwise, the system does not naturally evolve towards
the constant roll trajectory~\eqref{eq:CR_bckg}.
Indeed,
one reason to choose $\epsilon_2 \geq -3$ is the ensure that
the relation $\pi = \epc \phi$
continues to hold even under stochastic evolution.

We now consider the instanton
equations~\eqref{eq:inst_ps_2}--\eqref{eq:inst_ps_1}.
For this purpose we require the derivatives of
$V(\phi)$.
Using the constant roll background
and expanding
to leading order in $\epsilon$, we find
\begin{subequations}
\begin{align}
    \label{eq:CR-v-prime}
    \frac{V'(\phi)}{H^2}
    & =
    (3 - \epsilon)\Mp^2
    \frac{V'(\phi)}{V(\phi)}
    \approx
    -
    \epc (3 + \epc) \phi
    +
    \Or(\epsilon^{3/2})
    ,
    \\
    \label{eq:CR-v-prime-prime}
    \frac{V''(\phi)}{H^2}
    & =
    (3 - \epsilon)\Mp^2
    \frac{V''(\phi)}{V(\phi)}
    \approx
    -
    \epc (3 + \epc)
    +
    \Or(\epsilon)
    ,
\end{align}
\end{subequations}
We have discarded terms of order $\Or(\epsilon)$
or higher. In this sense, we are working in a quadratic
potential approximation.

\para{Relation to Ornstein--Uhlenbeck process}
There is a well-studied stochastic process with a quadratic
potential,
which is the Ornstein--Uhlenbeck process~\cite{uhlenbeck1930theory}.
This model was originally
introduced to study the Brownian
motion of a
overdamped particle moving in a rarefied gas,
subject to a frictional force proportional to the pressure.
It also has significant applications in mathematical
finance.
The model is said to be \emph{stable}
if $\sigma < 0$, in which case $\phi = 0$
is a
stable equilibrium;
\emph{unstable} if $\sigma = 0$;
and \emph{explosive} (or tachyonic)
if $\sigma > 0$.
In the explosive case, the noiseless drift
rapidly expels the field from the vicinity of $\phi=0$.

When specialized to Eqs.~\eqref{eq:CR-v-prime}--\eqref{eq:CR-v-prime-prime},
the instanton equations
have a wider state space than the Ornstein--Uhlenbeck
process, and therefore are not exactly equivalent.
However, we show below
that they are equivalent on the constant roll trajectory
$\pi = \sigma \phi$.
We can therefore study the constant roll
model by repurposing results from the mathematical literature.

The first-passage problem for the Ornstein--Uhlenbeck
process is known to be mathematically rich
and highly challenging.
It was introduced by Darling \& Siegert
(1959)~\cite{darling1953first},
but significant progress was not made until
relatively recently.
Alili, Patie \& Pederson (2005)
gave three representations
for the first-passage probability~\cite{Alili01102005}:
one in terms of a series of parabolic cylinder functions;
a second as an indefinite integral
involving
special functions;
and a third
in terms of a Bessel bridge (a stochastic process
derived from the Bessel process).
These representations are exact
and can be used for numerical analysis,
but cannot easily be used for analytic developments
or to obtain asymptotic formulas.
Later,
Lipton \& Kaushansky (2018)
were able to re-express the problem
in terms of a heat equation with moving boundary~\cite{2018arXiv181002390L}.
It can be solved in terms of a Volterra integral
equation of the second kind.
This was numerically more favourable than previously
known representations, but still did not lead
to simple analytic results for the rare tail.
Most recently,
Martin, Kearney \& Craster (2019)
gave explicit asymptotic formulas~\cite{Martin_2019}
for the stable case.
Their results show the emergence of an exponential
tail,
as we would expect based on our
experience
in~\S\ref{sec:slow-roll-instanton}
in~\S\S\ref{sec:SR}--\ref{sec:USR}.
However, as we explain below, this tail has a
highly nontrivial
dependence on the target field value
$\phiend$.

To demonstrate that
the constrained constant-roll
model is equivalent to an Ornstein--Uhlenbeck process,
we begin from the
forward Kolmogorov equation
in phase space.
This should be obtained from the
Fokker--Planck Hamiltonian, Eq.~\eqref{eq:HFP}.
Interpreting this as an operator
following 
a process
analogous
to
that described in~\S\ref{sec:FPH},
we obtain 
\begin{equation}
    \frac{\partial \ProbW}{\partial N}
    =
    -
    \frac{\partial}{\partial \phi}
    (\pi \ProbW)
    +
    \frac{\partial}{\partial \pi}
    \bigg[
        \bigg(
            \pi + \frac{V'(\phi)}{H^2}
        \bigg)
        \ProbW
    \bigg]
    +
    \D_{11}
    \bigg[
        \frac{\partial^2}{\partial\phi^2}
        +
        2 \epc \frac{\partial}{\partial\phi}\frac{\partial}{\partial\pi}
        +
        \epc^2 \frac{\partial^2}{\partial\pi^2}
    \bigg]
    \ProbW
    .
\end{equation}
The noise amplitude $\D_{11}$ is evaluated 
for this model in
Eq.~\eqref{eq:CR-D11} below.
We use the symbol $\ProbW$
to distinguish the transition probability in phase space
(which is a function of $\phi$ and $\pi$)
from the probability $\ProbP$ (which depends
only on $\phi$ and should be understood to be
marginalized over $\pi$).
To obtain an evolution equation for $\ProbP$
we carry out this marginalization.
In general, this would be a non-trivial procedure.
However,
we have the CR trajectory constraint.
On this trajectory we have
\begin{equation}
    \label{eq:ProbW-ProbP-constraint}
    \ProbW(\phi, \pi, N \divider \phiinit)
    =
    \delta(\pi - \epc \phi)
    \ProbP(\phi, N\divider\phiinit)
    .
\end{equation}
As explained above, the trajectory
$\qb(N) = \epc \qa(N)$ \emph{is} an attractor, but it is only
exactly enforced through a special choice
of initial condition.
Therefore, Eq.~\eqref{eq:ProbW-ProbP-constraint}
applies only for this choice.
The forward Kolmogorov equation for $\ProbP$
follows immediately, 
\begin{equation}\label{eq:Kolm_CR}
    \frac{\partial \ProbP}{\partial N}
    =
    -
    \sigma
    \frac{\partial}{\partial \phi} (\phi \ProbP)
    +
    \D_{11}
    \frac{\partial^2 \ProbP}{\partial\phi^2}
    \equiv
    - \frac{\partial \Current_\phi}{\partial \phi}
    .
\end{equation}
Total derivatives in $\pi$ disappear after marginalization.
Eq.~\eqref{eq:Kolm_CR}
is precisely the forward Kolmogorov
equation for an Ornstein--Uhlenbeck process.%
    \footnote{It is also an approximation for the
    overdamped evolution in a quadratic potential,
    over a small time period where $H$ can be taken
    approximately constant.
    This seems to have been first noticed by
    Stewart~\cite{Stewart:1991dy}.
    However, this approximation is likely to be inadequate
    to determine the statistics of rare events.}

\para{Instanton equations and trajectory}
We now determine the phase space instanton
describing a transition
from $\phiinit$ to $\phistop$.
In doing so we do not yet assume
any relation
to an Ornstein--Uhlenbeck process.
However, the
transition probability we
obtain will exhibit the equivalence.

The noise matrix components
are determined from Eq.~\eqref{eq:PS_matrix}
and take the form
\begin{equation}\label{eq:Dcr}
    \D_{11} = \frac{H^2}{8\pi^2} , \qquad 
    \D_{12} = \frac{H^2}{8\pi^2} \left(\nu - \frac{3}{2}\right) , \qquad 
    \D_{22} = \frac{H^2}{8\pi^2} \left(\nu - \frac{3}{2}\right)^2 .
\end{equation}
Here, $\nu - 3/2$
should be set equal to $\epc$
for $\epsilon_2\geq -3$,
or equal to $-3-\epc$
otherwise;
see Eq.~\eqref{eq:Hankel_index}.
Under these conditions,
the instanton equations can be written
\begin{subequations}
\begin{align}
    \label{eq:CR-D11}
    \frac{\d \pa}{\d N}
    & =
    -\epc(3+\epc)\pb
    \\
    \frac{\d \pb}{\d N}
    & =
    -
    \pa
    +
    3\pb
    ,
    \\
    \frac{\d \qa}{\d N}
    & =
    \qb
    +
    2 \D_{11} \Big(
        \pa
        +
        (\nu - \frac{3}{2}) \pb
    \Big)
    =
    \qb
    + \xi_1
    ,
    \label{eq:inst_cr_1}
    \\
    \frac{\d \qb}{\d N}
    & =
    -
    3 \qb
    +
    \epc(3+\epc) \qa
    +
    2 \D_{11} (\nu - \frac{3}{2}) \Big(
        \pa
        +
        (\nu - \frac{3}{2}) \pb
    \Big)
    =
    -
    3 \qb
    +
    \epc(3+\epc) \qa
    +
    \xi_2
    . \label{eq:inst_cr_2}
\end{align}
\end{subequations}
In contrast to the idealized
scenarios of slow-roll
and ultra-slow-roll,
the instanton equations for \emph{both} $\qa$ and $\qb$
now
include noise contributions,
which we write as $\xi_1$ and $\xi_2$
in Eqs.~\eqref{eq:inst_cr_1}--\eqref{eq:inst_cr_2}.
These satisfy the relation
$\xi_2 / \xi_1 = \nu - 3/2$.
For $\epsilon_2 \geq -3$
this ratio becomes $\xi_2 / \xi_1 = \sigma = \dot{\pi} / \dot{\phi}$,
where an overdot means a derivative with respect to $N$.
It follows that these stochastic
effects preserve the ratio $\qb/\qa$,
and therefore do not push the system away from the constant-roll
trajectory.
The conclusion is that only adiabatic perturbations are
generated.
The ensures conservation of the curvature
perturbation $\zeta$,
as emphasized by Tomberg~\cite{Tomberg:2022mkt}.

The equations for the noise fields
$\pa$ and $\pb$ are decoupled from those of the
primary fields.
They admit the general solutions
\begin{subequations}
\begin{align}
    \pa(N)
    & =
    \frac{
        (3 + \epc) (p_1 + \epc\, p_2 ) \e{-\epc N}
        +
        \epc \Big(
            p_1
            -
            (3 + \epc) p_2
        \Big)
        \e{(3 + \epc)N}
    }{
        3
        +
        2 \epc
    }
    ,
    \label{eq:pa_cr_gen}
    \\
    \pb(N)
    & = 
    \frac{
        (p_1 + \epc p_2) \e{-\epc N}
        -
        \Big(
            p_1
            -
            (3 + \epc) p_2
        \Big)
        \e{(3 + \epc)N} 
    }{
        3 + 2 \epc
    }
    .
    \label{eq:pb_cr_gen}
\end{align}
The integration constants $p_1$ and $p_2$
are chosen so that
$\pa(0) = p_1$ and $\pb(0) = p_2$.
Then, for $\epsilon_2 \geq -3$, the solutions for the
primary fields are
\begin{align}
    \qa(N)
    & =
    \alpha_1 \e{\epc N}
    +
    \alpha_2
    \e{-(3+\epc)N}
    -
    \frac{\D_{11}}{\epc}
    (p_1 + \epc\, p_2)
    \e{-\epc N}
    ,
    \label{eq:qa_cr_gen}
    \\
    \qb(N)
    & =
    \epc \alpha_1 \e{\epc N}
    -
    (3+\epc)\alpha_2 \e{-(3+\epc)N}
    -
    \D_{11}
    (p_1 + \epc\, p_2)
    \e{-\epc N}
    .
    \label{eq:qb_cr_gen}
\end{align}
\end{subequations}
On
substitution into Eq.~\eqref{eq:HFP}
we obtain the Fokker--Planck Hamiltonian,
\begin{equation}\label{eq:HFP_cr}
    \HFPtilde
    =
    \alpha_1 \epc (p_1 + \epc\, p_2)
    -
    \alpha_2 ( 3+\epc) \Big( p_1 - (3+\epc)p_2 \Big)
    .
\end{equation}
Evaluating the MSR action at the saddle point
using Eq.~\eqref{eq:Smps}, we find
\begin{multline}
    \im \SMSR
    =
    -
    \int_0^{\Nstar}
    \d N \,
    \Big(
        \pa \frac{\d \qa}{\d N}
        +
        \pb \frac{\d \qb}{\d N}
        -
        \HFPtilde
    \Big)
    \\
    =
    -
    \int_0^{\Nstar}
    \d N \,
    \D_{11} (p_1 + \epc\, p_2)^2 \e{-2\epc N}
    =
    -
    \frac{\D_{11}}{2\epc}
    (p_1 + \epc\, p_2)^2
    \Big(
        1 - \e{-2 \epc \Nstar}
    \Big)
    .
    \label{eq:SMSR_cr}
\end{multline}

In Eq.~\eqref{eq:SMSR_cr},
and also in Eqs.~\eqref{eq:pa_cr_gen}--\eqref{eq:qb_cr_gen}
for the
instanton trajectory,
the noise terms always appear in the combination
$(p_1 + \epc\, p_2) \e{-\epc N}$.
This is exactly the combination
$\pef \equiv \pa + \epc \pb$,
which may be confirmed
directly from
Eqs.~\eqref{eq:pa_cr_gen}--\eqref{eq:pb_cr_gen}.
This structure arises naturally, because
(assuming $\epsilon_2 \geq -3$)
the stochastic contributions
enter
the instanton equations~\eqref{eq:inst_cr_1}--\eqref{eq:inst_cr_2}
in precisely this combination.
It follows that
$\pef$
is the single physically relevant degree of freedom
associated with the noise sector.

To finish construction of the instanton
we should impose appropriate boundary conditions.
First, we place the system on the CR phase-space trajectory
by imposing $\qb(0) = \epc\,\qa(0)$.
Applying this condition to
Eqs.~\eqref{eq:qa_cr_gen}--\eqref{eq:qb_cr_gen}
yields $\alpha_2 = 0$,
which also eliminates the second term in
Eq.~\eqref{eq:HFP_cr}. The structure of the
MSR action in Eq.~\eqref{eq:SMSR_cr} remains unchanged.
In principle we should also impose the CTP condition
$\qb(\Nstar) = 0$, although this is not actually needed
in practice.

Next, we fix the endpoints of the trajectory
by imposing
$\qa(0) = \phiinit$ and $\qa(\Nstar) = \phistop$.
This yields
\begin{equation}
    \alpha_1
    =
    \frac{
        \phistop \e{\epc \Nstar} - \phiinit
    }{
        \e{2\epc \Nstar}-1
    }
    ,
    \qquad
    \frac{
        \D_{11}
    }{
        \epc
    }
    (p_1 + \epc p_2 )
    =
    \frac{
        \e{\epc\Nstar}
    }{
        \e{2\epc \Nstar}-1
    }
    (\phistop - \e{\epc \Nstar}\phiinit)
    .
\end{equation}
The resulting field trajectory reads:
\begin{equation}
    \qa(N)
    =
    \csch (\epc \Nstar)
    \Big(
        \phistop \sinh(\epc N)
        -
        \phiinit \sinh\big[ \epc(N - \Nstar) \big]
    \Big)
    .
\end{equation}
Substitution into Eq.~\eqref{eq:SMSR_cr} yields the saddle-point action
\begin{equation}\label{eq:SMSR_cr2}
    \im \SMSR
    =
    -
    \frac{\epc}{2\D_{11}}
    \frac{
        (\phistop - \phiinit \e{\epc \Nstar})^2
    }{
        \e{2\epc\Nstar}-1
    }
    .
\end{equation}

\para{Transition probability}
Finally, still working in the instanton
approximation, we obtain the normalized,
unrestricted transition probability
from~\eqref{eq:SMSR_cr2}
\begin{equation}\label{eq:Prob_CR}
    \ProbP(\phistop, \Nstar \divider \phiinit)
    =
    \bigg(
        \frac{
            \sigma
        }{
            2\pi \D_{11} (\e{2\epc\Nstar}-1)
        }
    \bigg)^{1/2}
    \exp
    \bigg(
        {-
        \frac{\epc}{2\D_{11}}}
        \frac{
            (\phistop - \phiinit \e{\epc \Nstar})^2
        }{
            \e{2\epc\Nstar}-1
        }
    \bigg)
        .
\end{equation}
As in previous examples,
we have found a Gaussian distribution
for $\phistop$ centred around the classical value
of the field at $\Nstar$.
Eq.~\eqref{eq:Prob_CR}
is valid for either sign of $\epc$,
and in fact coincides exactly with the known
expression for
the transition probability of an Ornstein--Uhlenbeck
process.
It was derived at least as early as
Smoluchowski (1915)~\cite{smoluchowski1916brownsche},
and later
independently by
Ornstein \& Uhlenbeck (1930)~\cite{uhlenbeck1930theory}.%
    \footnote{A readable summary of the history,
    including earlier contributions by
    Bachelier, Markov, and even Laplace, is given in the
    paper by Jacobson~\cite{jacobsen1996laplace}.}
The instanton approximation
reproduces the exact
result because the saddle-point approximation
becomes exact for Gaussian integrands.

\begin{figure}[ht]
    \centering
    \includegraphics[width=\textwidth]{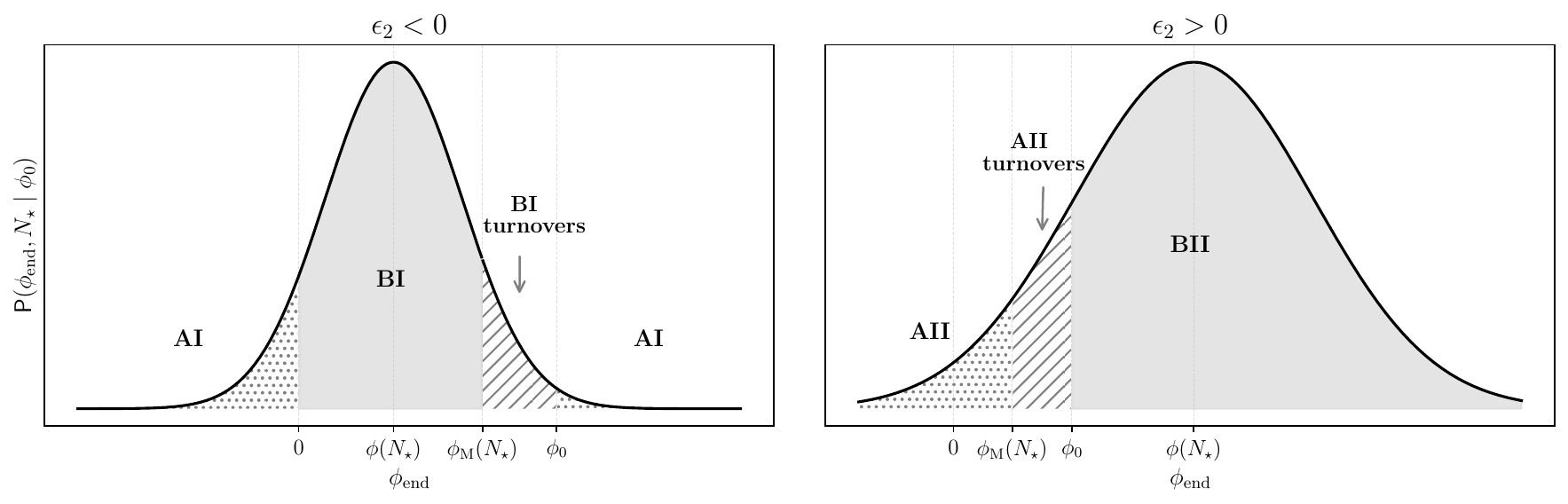}
    \caption{$\ProbP(\phiend, \Nstar \divider \phiinit)$ in a CR phase, as given by Eq.~\eqref{eq:Prob_CR}. Dotted regions are classically inaccessible; diagonally hatched regions correspond to second crossings. $\phi_{\rm M}(\Nstar) = \phiinit\, \mathrm{sech}(\epc \Nstar)$ and $\phiinit$ delimit these zones.
    \semibold{Left:} $\epsilon_2 < 0$. The background field moves right to left, and the distribution peaks left of $\phiinit$, shifting toward zero as $\Nstar$ grows. \semibold{Right:} $\epsilon_2 > 0$. The field moves left to right, with the peak right of $\phiinit$ and reduced impact from the turnover region.} 
    \label{fig:CR_Prob}
\end{figure}

\para{Trajectory behaviour}
In a similar way to the USR
model discussed above,
the behaviour of~\eqref{eq:Prob_CR}
varies depending whether the
target field value
$\phiend$ can be reached under purely
noiseless evolution.
See Fig.~\ref{fig:CR_Prob}.

\begin{itemize}
    \item \semibold{Case AI:} $\epsilon_2 < 0$,
    and either $\phiend < 0 < \phiinit$ or $\phiinit < \phiend$.
    This is the stable case,
    for which
    $\phi$ rolls from larger to smaller values
    but cannot pass zero.
    Therefore the target $\phiend$ cannot be reached
    under noiseless evolution.
    The field can therefore arrive at the target
    only due to fluctuations.
    Using Eq.~\eqref{eq:qa_cr_gen}, the derivative of the field is given by
    \begin{equation}\label{eq:field_vel_cr}
        \qa'(N)
        =
        \epc
        \csch(\epc\Nstar)
        \Big[
            \phiend \cosh(\epc N)
            -
            \phiinit \cosh\big( \epc(N - \Nstar) \big)
        \Big]
        \;.
    \end{equation}
    For $\epsilon_2 < 0$, this expression is negative at $N = \Nstar$, ensuring that the field crosses $\phiend$ at that point for the first time when $\phiend <0$. Conversely, $\qa'(\Nstar)>0$ for $\phiend>\phiinit$, so the field also reaches the target for the first time for this parameter regime. These two scenarios contribute to the two extreme
    regions shown in the left panel of Fig.~\ref{fig:CR_Prob}.

    \item \semibold{Case AII:} $\epsilon_2 > 0$ and $\phiend < \phiinit$.
    This is the explosive case.
    With this choice of initial conditions,
    $\phi$ rolls from smaller
    to larger positive values.
    The target is still inaccessible on a noiseless trajectory
    because it is smaller than the initial point.
    A necessary condition to avoid turnovers is $\qa'(\Nstar) < 0$,
    which holds when $\phiend \cosh(\epc \Nstar) < \phiinit$.
    For $\phiend > 0$,
    this imposes an upper bound on $\Nstar$.
    The corresponding regions associated with first 
    and second passages (in the unrestricted transition probability)
    for this setup are illustrated in the right panel of Fig.~\ref{fig:CR_Prob}.

    \item \semibold{Case BI:} $\epsilon_2 < 0$ and $0 < \phiend < \phiinit$.
    This is the stable case.
    The
    noiseless motion is from larger to smaller values of $\phi$.
    The target lies in this direction of flow and is therefore
    accessible.
    However,
    as in the USR \semibold{Case B},
    large values of $\Nstar$ are associated with turnover
    trajectories.
    To avoid such
    requires $\phiend \cosh(\epc \Nstar) < \phiinit$,
    so that $\qa'(\Nstar) < 0$.
    This can be rewritten in terms of $\Nend$,
    \begin{equation}
        \Nstar
        +
        \frac{1}{\epc}
        \ln
        \bigg(
            \frac{2}{1+\e{2\epc \Nstar}}
        \bigg)
        <
        \Nend
        .
    \end{equation}
    For example, with $\epc = -1$ and $\Nend = 2$,
    to avoid a turnover requires
    $\Nstar < 2.69$. In contrast to \semibold{Case AII},
    a constraint in terms of $\NstarDet$ can be obtained here because
    $\phiend$ is classically accessible.
    For fixed $\Nstar$, the values of $\phiend$ corresponding to
    trajectories the avoid turnovers
    delimit the shaded region in the left panel of Fig.~\ref{fig:CR_Prob}.

    \item \semibold{Case BII:} $\epsilon_2 > 0$ and $0 < \phiinit < \phiend$.
    This is the explosive case, and
    noiseless flow is from smaller to larger values of $\phi$.
    The target lies in the direction of the
    background flow, and is therefore reachable
    on a noiseless trajectory.
    Moreover, we do not observe overshoot behaviour
    for any value of $\Nstar$.
    However, to attain \emph{large} values of $\Nstar$,
    stochastic fluctuations
    are required to initially
    move the field in the opposite direction to the noiseless flow.
    The field then turns around
    before
    and reaching the target (for the first time)
    at time $\Nstar$.
    In practice, the exponential growth of the field
    velocity will eventually violate the slow-roll condition.
    This case corresponds to the shaded region of the right plot of Fig.~\ref{fig:CR_Prob}.
\end{itemize}

\para{Comparison with Tomberg formalism}
First, we consider the
comparison to Tomberg's formalism,
described in~\S\ref{sec:tomberg-langevin-method}.
This assumes $\sigma > 0$, and also that
the target value
$\phiend$ is accessible during noiseless evolution.
In the limit $\Nstar \gg 1$,
working to exponential accuracy,
the transition probability~\eqref{eq:Prob_CR}
evaluated at $\phistop \rightarrow \phiend$
becomes
\begin{align}
    \ln \ProbP(\phiend, \Nstar \divider \phiinit)
    \sim
    -
    \frac{2}{\epsilon_2
    \dimP_{\zeta}(\Nstar)}
    \frac{\left( 1 - \e{-\epsilon_2(\Nstar - \Nend)/2} \right)^2}{\e{\epsilon_2 \Nstar} - 1} - \frac{\epsilon_2}{2}\Nstar
    .
    \label{eq:P_CR_Tomberg}
\end{align}
We have used
$\pi(N)^2 = \epc^2 \phi(N)^2 = 2\Mp^2 \epsilon(N)$
to express the result in terms of the power spectrum
$\dimP_{\zeta}(\Nstar)$,
and written it in terms of $\epsilon_2$
to aid comparison with
Tomberg's
result~\eqref{eq:Tomberg-constant-roll-P}.
In this form the resemblance is already clear,
but can be made more precise as follows.
The first term comes from the
MSR action $\im \SMSR$.
We have seen several times
that this is a Gaussian in the noise
fields.
Here, as explained above,
that means the combination
$\pef$.
Interpreting $\exp(\im \SMSR)$
as a distribution on
$\pef(0)$, we identify
$\Gamma$ and $\pef(0)$.
Then, computing the variance $\sstar^2$
using Eq.~\eqref{eq:PGamma}
and the noiseless trajectory for $\phi(N)$,
this contribution exactly reproduces
Tomberg's formula
$\ProbP(\Gamma) \sim \e{-\Gamma^2/(2\sstar^2)}$.

The conclusion is that
Tomberg's formula
for the transition probability
should be regarded the same as ours,
but obtained
by working in the Onsager--Machlup formalism
rather than using the MSR path integral.
The key difference
is only that in our formalism
Eq.~\eqref{eq:P_CR_Tomberg}
should be understood as a density with respect to the final field
configuration.

Further,
focusing on unrestricted transition probabilities,
Tomberg's procedure 
leading to Eq.~\eqref{eq:prob-nstar-jacobian-tomberg}
can apparently be given an interpretation in our framework,
although its physical meaning is somewhat unclear.
The equivalent procedure would be
to consider~\eqref{eq:P_CR_Tomberg} as
a density with respect to $\pef$
and use the instanton solution to relate $\pef$ to $\Nstar$.
Remarkably,
asymptotically,
this yields the same
second term, $\epsilon_2 \Nstar / 2$.
It would be interesting to understand whether this
change of variables can be
given a clear justification.
On the other hand, in Eq.~\eqref{eq:P_CR_Tomberg},
the $\epsilon_2 \Nstar/2$ 
term comes from properly normalizing the transition probability
with respect to $\phistop$.
There is no analogue of this for Eq.~\eqref{eq:PGamma}.

\para{Comparison with known results for $\ProbQ$}
Next,
although Ref.~\cite{Tomberg:2023kli}
did not aim to compute $\ProbQ$,
we consider how
the tail of Eq.~\eqref{eq:P_CR_Tomberg}
translates to the first passage distribution.
for this purpose we require the
restricted transition probability $\ProbPre$.
Unfortunately, there is no simple way to
obtain this
in the constant roll model;
the method of images does not apply, except in one special case
(see below), and the Bromwich inversion needed
for the renewal formula apparently
cannot be done analytically.
In~\S\ref{sec:SP_restr_trans}
we suggested that, when
an explicit relation between $\ProbPre$ and
$\ProbP$ is not available,
the best alternative is to estimate
$\ProbQ$
from the tail of $\partial \ProbP / \partial \phi$.
Unfortunately,
we will see that for the constant roll model
this does not yield an accurate result.

We first consider the tail behaviour of $\ProbP$
in the limit $\Nstar \gg 1$.
For $\sigma > 0$ we have
\begin{subequations}
\begin{align}
    \label{eq:P-CR-tail-explosive}
    \ProbP
    &
    \sim
    \exp\big( {-\sigma} \Nstar \big)
    ,
    \\
    \label{eq:dP-CR-tail-explosive}
    \frac{\partial \ProbP}{\partial \phistop}
    &
    \sim
    \exp\big( {-2} \sigma \Nstar \big)
    ,
\end{align}
\end{subequations}
whereas for $\sigma < 0$ we have
\begin{subequations}
\begin{align}
    \label{eq:P-CR-tail-stable}
    \ProbP
    &
    \sim
    \exp(0)
    ,
    \\
    \label{eq:dP-CR-tail-stable}
    \frac{\partial \ProbP}{\partial \phistop}
    &
    \sim
    \left\{
        \begin{array}{l@{\hspace{4ex}}l}
            \exp(0) & \phistop \neq 0 \\
            \exp(-|\sigma| \Nstar) & \phistop = 0
        \end{array}
    \right.
    .
\end{align}
\end{subequations}

In the special case
$\sigma < 0$ and $\phiend = 0$,
the reflection symmetry of the
drift velocity
means that
it is possible to obtain
the restricted
transition probability $\ProbPre$ by the method of images.
This yields
\begin{equation}
    \label{eq:Pre-CR-images-formula}
    \ProbPre(\phistop,\Nstar\divider \phiinit)
    =
    \ProbP(\phistop, \Nstar \divider\phiinit)
    -
    \ProbP(\phistop, \Nstar \divider {-\phiinit})
    .
\end{equation}
$\ProbPre$ defined in this way
solves the forward Kolmogorov equation, Eq.~\eqref{eq:Kolm_CR},
and clearly satisfies the absorbing
boundary condition at $\phiend$,
$\ProbPre(0, \Nstar \divider \phiinit) = 0$.
It produces the first-passage distribution
\begin{multline}
    \ProbQ(\Nstar,0|\phiinit)
    =
    \left.
        \D
        \frac{\partial \ProbPre}{\partial \phi}
    \right|_{\phi = 0}
    \\
    =
    \bigg( \frac{\epc^3}{4\pi \D \sinh^3(\epc \Nstar)} \bigg)^{1/2}
    \phiinit\exp
    \bigg(
        {-
        \frac{\epc}{4\D}}
        \frac{\phiinit^2 \e{\epc \Nstar}}{\sinh(\epc \Nstar)}
        -
        \frac{\epc \Nstar}{2}
    \bigg)
    .
    \label{eq:P_CR_Tomberg_im}
\end{multline}
This result
was given by Pitman \& Yor (1981)~\cite{10.1007/BFb0088732}
and is exact.
It corresponds to Eq.~(2.6)
of Alili {\etal}~\cite{Alili01102005}.
The instanton approximation is able to reproduce
this exact formula because it follows directly from
$\ProbP$,
and
we have already noted that the instanton
approximation for $\ProbP$
is exact because the MSR integrand is Gaussian.
The dominant tail behaviour is
$\sim \e{-|\sigma| \Nstar}$.
This matches Eq.~\eqref{eq:dP-CR-tail-stable},
and shows that the estimate $\sim \partial \ProbP / \partial \phistop$
actually would work here, even if we did not have the
formula~\eqref{eq:Pre-CR-images-formula}.

When $\sigma < 0$
but $\phiend \neq 0$ there is no known
formula comparable to Eq.~\eqref{eq:P_CR_Tomberg_im}.
However, Martin {\etal} gave the asymptotic
estimate (in our notation)~\cite{Martin_2019}
\begin{equation}
    \label{eq:Q-Martin-tail}
    \ProbQ(\Nstar) \sim
    \exp
    \Big(
        {-\lambda(\phiend)} |\sigma| \Nstar
    \Big) ,
\end{equation}
where $\lambda(0) = 1$.
They were able to characterize the multiplier $\lambda(\phiend)$
in terms of the position of the right-most singularity
of the Laplace transform solution to the
forward Kolmogorov equation, Eq.~\eqref{eq:Kolm_CR},
and gave an explicit (but complex)
algorithm to compute it in terms of
a certain limiting ratio.
Their Table~1 lists some representative values
of $\lambda$ in terms of the dimensionless
variable $y_+ = (-2\sigma)^{1/2} \phiend / \D_{11}$.
In this case, the guess
$\sim \partial \ProbP / \partial \phistop$ clearly does not yield
a good estimate, because~\eqref{eq:dP-CR-tail-stable} does not
decay for $\phistop \neq 0$.
Note that this cannot be attributed to the instanton
approximation not capturing the derivative
$\partial \ProbP / \partial \phistop$
with sufficient fidelity,
since we know that Eq.~\eqref{eq:Prob_CR}
is exact.
The necessary decay law must be embedded in the transformation
from $\ProbP$ to $\ProbPre$.
The key conclusion from this analysis is that the
tail~\eqref{eq:Q-Martin-tail}
depends on the position of the boundary in a way
that cannot be replicated without a detailed
expression for $\ProbPre$.

This leaves open the question of what happens in the explosive
$\sigma > 0$ case.
The results reported by Martin~{\etal} do not appear to apply
to this scenario~\cite{Martin_2019}.
It is possible that their procedure
to compute $\lambda$ could be generalized
to $\sigma > 0$,
but this will require some effort to verify.
On physical grounds, it seems that one could associate
the difference in tail behaviour between
$\ProbP$
and $\ProbQ$ (or $\ProbPre$)
with the large or small drift velocities
at $|\phi| \gg 0$
and $\phi \sim 0$, respectively.
It is plausible that these could change the
relative importance of backflow events.
In particular, if there is a large drift velocity
in the direction of motion through the terminal boundary,
we could expect the tails of $\ProbP$ and $\ProbQ$
to be very similar.
Conversely, if there is a large drift velocity
anti-aligned with the motion, this would typically
strongly enhance the number of backflow events.%
    \footnote{We thank Eemeli Tomberg for pointing this out to us.}
At present,
there does not seem
any clear rationale to choose between the tail
behaviours of~\eqref{eq:P-CR-tail-explosive}
and~\eqref{eq:dP-CR-tail-explosive}.
In our view,
further work is required to clarify this situation.
Although the flux formula involves $\partial \ProbPre / \partial \phi$,
it does not seem impossible that corrections
from a method-of-images relation could yield
a contribution to $\CurrentRe$ that scales like
$\ProbP$ rather than $\partial \ProbP / \partial \phi$.
If these have the same tail behaviour there is no issue,
but that is not the case here.

For this model, it is possible that the spectral method
has advantages because it enables a direct
construction of $\ProbPre$, via the boundary conditions
imposed on the eigenfunction expansion).
$\lambda(\phiend)$ should presumably be identified
with the lowest-lying eigenvalue of the
adjoint Fokker--Planck operator $\LFPadj$.
It would be very interesting to check whether
this procedure reproduces the estimates reported by
Martin~{\etal}~\cite{Martin_2019},
but
we leave this interesting question to future work.

\subsubsection*{The $\epsilon_2 < -3$ regime}
Finally, let us elaborate more about the $\epsilon_2 < -3$ regime. There, the solutions of the equations of motion give:
\begin{subequations}
\begin{align}
    \qa (N)
    & =
    \alpha_1 \e{\epc N}
    +
    \alpha_2 \e{-(3+\epc)N}
    +
    \frac{\D_{11}}{3+\epc}
    \Big( p_1 - (3+\epc) p_2 \Big)
    \e{(3+\epc)N}
    ,
    \label{eq:qa_cr_v2}
    \\
    \qb (N)
    & =
    \epc \alpha_1 \e{\epc N}
    -
    (3+\epc)\alpha_2 \e{-(3+\epc)N}
    -
    \D_{11} \Big( p_1 - (3+\epc) p_2 \Big)
    \e{-(3+\epc)N}
    .
    \label{eq:qb_cr_v2}
\end{align}
\end{subequations}
From these expressions, it is evident that imposing the initial condition $\qa(0) = \sigma\, \qb(0)$ eliminates the stochastic contribution to the evolution, particularly if this relation is required to hold at all times. Conversely, if the initial conditions permit a nonzero stochastic component, the $\alpha_1\, e^{\epc N}$ terms will decay most rapidly. Then, at late times, the dominant terms enforce $\qb(N) \approx -(3+\epc) \qa(N)$, which mirrors the form found earlier but now with an effective second slow-roll parameter $\epsilon_2^{\rm eff}/2 = -(3+\epc)$, which corresponds to that of the Wands dual of the original one. In other words, the system naturally evolves toward a phase space trajectory associated with its Wands dual, going from the non--attractor to the attractor region. We already saw a manifestation of this behaviour for USR, where, given sufficient time, the noise drove the field toward a SR trajectory, corresponding to the Wands dual of the original background.

\subsection{Exponentially decaying noise}
\label{sec:decaying-noise}
In our final example,
we consider a simple model
with highly time-dependent noise.
In scenarios of this type it does not appear
easy to obtain
tail estimates
using any other formalism.
In particular, strong
time dependence invalidates the formal
solution~\eqref{eq:formal-Kolmogorov-soln}
which plays a critical role in the spectral method.

The Starobinsky--Langevin equation~\eqref{eq:Starobinsky-Langevin}
and its associated
Kolmogorov forward equation~\eqref{eq:kolmogorov-forwards}
describe the distribution of field values only on a scale
close to the horizon.
However,
as explained in~\S\ref{sec:Est_tail},
we often wish to understand the distribution
for the Fourier mode $\zeta_k$ associated with
a fixed
comoving wavenumber $k$,
which can be regarded roughly as $\zeta$
smoothed over a volume $V$ of comoving scale $L = 2\pi/k$.
To determine the evolution of the
smoothed field $\phi$ interior to $V$, we argue as follows.
$\phi$ receives an
inflationary perturbation as $V$ exits the horizon.
For a short period afterwards, subsequent perturbations
emerge on a similar scale,
and therefore disturb $\phi$
coherently.
During this period the Starobinsky--Langevin
equation~\eqref{eq:Starobinsky-Langevin},
or its generalization beyond slow-roll,
likely provides an adequate description.
But soon, the scale $L$ becomes much larger than the current
comoving horizon, $R_H \sim 1/(aH)$.
Hence, the volume $V$ will sample
of order $(L/R_H)^3 \sim (a H / k)^3$
realizations
of the emerging perturbations.
In the Gaussian approximation that all these realizations
are independent,
we expect the variance of the smoothed
perturbation in $V$ to be suppressed by a
``central limit theorem'' factor
$(k/aH)^3 \sim \e{-3N_k}$,
where $N_k$ is the number of e-folds since horizon exit of
the mode $k$.

This kind of volume suppression is a well-understood effect
encountered in calculations of 1-loop backreaction~\cite{Riotto:2023hoz,Iacconi:2023ggt}.
The result is that the amplitude of the noise experienced
by the smoothed field in $V$
will decay exponentially outside the horizon.
As described in~\S\ref{sec:Est_tail},
in stochastic calcuations
this is often modelled by taking $\phiend$
to correspond to the time when $V$ exits the horizon.
After this time, the evolution of the field in $V$
is assumed to be dominated by noiseless drift.
For example,
using this prescription,
Figueroa~{\etal} found that in their numerical
studies, averaging over kicks
after horizon exit had the same effect as
simply switching off the noise altogether~\cite{Figueroa:2021zah}.

In this section we use the instanton
formalism to confirm this expectation,
by using it to estimate the distribution of the field
fluctuation \emph{after} $V$ exits the horizon.
In~\S\ref{sec:instanton-linear-model} and~\S\ref{sec:SR}
we saw that the least unlikely noise
realization~\eqref{eq:SR-instanton-soln-P}, \eqref{eq:sol_p_sr}
acts at a constant rate, at a level calibrated to just cancel the
deterministic motion.
The constant ``noise cost'' of this realization per e-fold
is independent of $\Nstar$.
After integration $\int \d N$
over the instanton trajectory, it
produces the $\Nstar$ scaling of $\ln \ProbQ$
in Eq.~\eqref{eq:instanton-slow-roll-exp-tail}.
In contrast,
if the amplitude of the noise decays strongly,
to achieve the same outcome
we must accept a much more extreme noise realization.
The extra ``cost'' of this extreme noise
produces a depopulated tail.

\para{Instanton equations}
We take the instanton equations for this model
to be those of the slow-roll model with linear potential,
Eqs.~\eqref{eq:srins1}--\eqref{eq:srins2},
with the noise amplitude taken to decay
as $\e{-3N}$,
\begin{subequations}
\begin{align}
    \label{eq:decaying-noise-instanton-eq1}
    \frac{\d \pa}{\d N}
    &=
    0
    ,
    &
    \frac{\d \pb}{\d N}
    &=
    -
    \pa
    +
    3\pb
    ,
    \\
    \label{eq:decaying-noise-instanton-eq2}
    \frac{\d \qa}{\d N}
    &=
    \qb
    +
    2 D
    \e{-3N}
    \pa
    ,
    &
    \frac{\d \qb}{\d N}
    &=
    -
    3 \qb
    -
    3 v
    ,
\end{align}
\end{subequations}
where $v$ and $D$ have the same meanings as in~\S\ref{sec:SR}.
Noise with any other time-dependent amplitude can be handled
in a similar way.
In this model, only one element of the noise matrix
$\D_{ij}$ survives.
In more general scenarios, separate time-dependence
could be specified for each element, if required.
Even if the resulting equations cannot be solved analytically,
it may be possible to solve them numerically.

The boundary conditions match those for the linear potential.
We take $\qa(0) = \phiinit$
and $\qa(\Nstar) = \phiend$.
We also impose the velocity
condition
$\qb(0) = -v + \beta$,
with $\beta$ left arbitrary.
As in~\S\ref{sec:SR},
we expect the leading tail estimate to be independent of $\beta$;
we shall see that it enters at first subleading order
in the tail expansion.
Finally, there is the CTP boundary condtion
$\qb(\Nstar) = 0$,
which entails $\pb(\Nstar) = 0$.
The solutions are
\begin{subequations}
\begin{align}
    \label{eq:decaying-noise-pa-soln}
    \pa
    & =
    \frac{1}{2D}
    \frac{\Nstar}{\e{3\Nstar} - 1}
    \left(
        \frac{\beta}{\Nstar}
        +
        \e{3 \Nstar}
        \Big(
            3 v
            +
            \frac{3 \Delta \phi - \beta}{\Nstar}
        \Big)
    \right)
    ,
    \\
    \pb
    & =
    \frac{\Nstar}{6D}
    \frac{1 - \e{3N-3\Nstar}}{\e{3\Nstar} - 1}
    \left(
        \frac{\beta}{\Nstar}
        +
        \e{3\Nstar}
        \Big(
            3 v
            +
            \frac{3 \Delta \phi - \beta}{\Nstar}
        \Big)
    \right)
    ,
    \\
    \qa
    & =
    \begin{multlined}[t]
        \frac{\Nstar}{\e{3\Nstar} - 1}
        \bigg\{
            v \frac{N}{\Nstar}
            -
            \frac{\phiinit}{\Nstar}
            +
            \e{3\Nstar}
            \Big[
                v \Big(1 - \frac{N}{\Nstar}\Big)
                +
                \frac{\phiend}{\Nstar}
            \Big]
            \\
            -
            \e{3\Nstar - 3N}
            \Big(
                v + \frac{\Delta \phi}{\Nstar}
            \Big)
        \bigg\}
    \end{multlined}
    \\
    \qb
    & =
    - v
    + \beta \e{-3 N}
    .
\end{align}
\end{subequations}
Notice that the instanton equations still find a constant solution
for the noise field $\pa$.
However, because of the decaying amplitude, the
physical noise realization
that couples to $\qa$
is exponentially decaying.
We conclude that most of the stochastic motion occurs soon after
horizon exit, where the noise amplitude is still appreciable
and large kicks are less expensive.
For this reason,
the solution for $\qa$
is no longer an even linear
progression from $\phiinit$ to $\phiend$.
Further,
in order to arrive at the final position
$\phiend$
in $\Nstar$ e-folds, we must have proportionately larger
stochastic events at early times.
This is responsible for the overall
factor of $\Nstar$
in~\eqref{eq:decaying-noise-pa-soln},
which scales the noise in proportion to the transition
duration.
As noted above,
this factor is absent in Eq.~\eqref{eq:srins1}.
It is this $\Nstar$
scaling of the noise amplitude,
combined with the $\int \d N$ integral over $\SMSR$,
that yields a light tail
for $\ProbQ$.

\para{MSR action and tail estimate}
We now evaluate $\SMSR$ on this instanton trajectory. That yields
\begin{equation}
    \im \SMSR
    =
    -
    \frac{
        17
        +
        6 \e{-3\Nstar}
        (
            \Nstar
            -
            3
        )
        +
        \e{-6 \Nstar}
    }{
        108 D (\e{3\Nstar} - 1)^2
    }
    \bigg(
        \beta
        +
        \e{3\Nstar}
        \Big(
            3 v \Nstar
            + 3 \Delta \phi
            -
            \beta
        \Big)
    \bigg)^2
    .
\end{equation}
We drop the normalization factor $\Normalization$
and fluctuation determinant, which do not contribute to 
exponential accuracy.
To extract the tail, we
evaluate the behaviour of $\im \SMSR$
in the limit $\Nstar \gg 1$.
This finally yields
\begin{equation}
    \ln \ProbQ(\Nstar)
    \approx
    - \frac{17}{6} \frac{\Nstar^2}{\dimP_\zeta}
    +
    \frac{34\pi}{9}
    \frac{\beta - 3 \Delta \phi}{H \dimP_\zeta^{1/2}}
    \Nstar
    + \Or(1) .
    \label{eq:decaying-noise-tail-estimate}
\end{equation}
Clearly we have reverted to a light, Gaussian tail.
As advertised, the initial velocity
(represented by $\beta$)
appears only in the subleading
linear term.

Note that this tail behaviour
cannot be obtained
in the spectral formalism described in~\S\ref{sec:ezquiaga-spectral-method}.
As explained there, the reason is that the formal
solution~\eqref{eq:formal-Kolmogorov-soln}
applies only when the differential operator
$\LFPadj$ is time independent;
in the time-dependent case, Eq.~\eqref{eq:formal-Kolmogorov-soln}
must be replaced by an ordered exponential.
For this scenario, the time dependence of the noise
is sufficiently strong to break the prediction of an exponential tail.
However, note that even an exponentially strong decay profile is only
sufficient to change the leading tail behaviour in $\ln \ProbQ$
from $\Or(\Nstar)$ to $\Or(\Nstar^2)$.
This shows that the tail behaviour is rather
robust to very significant
changes in the noise amplitude.
For example, if we change the decay profile from
$\e{-3\Nstar}$ to $\e{-\alpha \Nstar}$,
for a positive constant $\alpha$,
we retain the Gaussian form of~\eqref{eq:decaying-noise-tail-estimate}.
The only difference is that the coefficient of the leading
$\Or(\Nstar^2)$
term becomes $-17 \alpha / 18$.

In contrast, as expected, noise with
a growing amplitude produces a substantially
heavier tail.
In this case, the exact structure of the tail depends on
model-dependent details of when the growth switches off.

\para{Superposition of tails}
This analysis applies only
after horizon exit.
As already explained, evolution in this regime is
often neglected
in stochastic
calculations.
The Gaussian tail~\eqref{eq:decaying-noise-tail-estimate}
can be regarded as a validation of this procedure.

To understand the
effect of combining Eq.~\eqref{eq:decaying-noise-tail-estimate}
with earlier stochastic evolution, we proceed as follows.
We model the number of e-folds
experienced by the region $V$
as $\Nstar = X + Y$.
Here, $X$ roughly
measures the number of e-folds
up to
a time around horizon exit,
and should be estimated
by computing the first passage distribution
$\ProbQ_X(\Nstar)$
with noise determined by
the matrix $\D_{ij}$ in the normal way,
up to to a boundary
beyond which backflow events are unlikely,
due to the $\e{-3N_k}$ decay of the amplitude.
Meanwhile, $Y$
measures the number of e-folds from this
boundary to the end of inflation,
and should be computed
by estimating the first passage distribution
$\ProbQ_Y(\Nstar)$
with exponentially decaying noise
amplitude, as in this section.
The use of first passage distributions allows us to
cleanly separate the two regimes, provided
we choose the boundary with sufficient care that there is
not significant backflow,
as discussed in~\S\ref{sec:Est_tail}.

The distribution $\ProbP(\Nstar)$
of $\Nstar$
is the convolution of $\ProbQ_X$ and $\ProbQ_Y$.
It follows that the tails of $\ProbQ_X$ and $\ProbQ_Y$
approximately co-add.
Since the light, Gaussian tail of $\ProbQ_Y$ will generally be
highly suppressed compared to the heavy, exponential tail of $\ProbQ_X$,
this exponential tail is effectively
inherited by $\ProbP(\Nstar)$.
One can interpret $Y$ as a normal inflationary perturbation
generated at horizon exit, with variance $\sigma^2 \sim \dimP_\zeta$
from comparison with Eq.~\eqref{eq:decaying-noise-tail-estimate},
superposed on the fluctuation $X$ produced by noise on larger scales.%
    \footnote{This also provides an interpretation
    of the fact that
    changing the rate in the exponential decay law leaves the Gaussian tail intact.
    Effectively, no matter how strong the decay, the noise
    gets a single opportunity to act,
    during an $\Or(1)$ e-fold window around horizon
    exit. This can be regarded
    as the superposition of a single
    perturbation,
    with ordinary
    statistical properties,
    at the exit time.}

This conclusion was expected.
However,
it is worth considering that scenarios producing
a large power spectrum amplitude
(and hence where stochastic effects are relevant)
typically also generate large non-Gaussianities.
Recent work on 1-loop back reaction shows that the stochastic
kick to the field in $V$ need not decay like $\e{-3N}$
if there is a significant local-type non-Gaussianity~\cite{Iacconi:2023ggt}.
In this case, the realizations of the horizon-scale
noise process sampled by the superhorizon region $V$
need not be independent.
The conclusion is that there can be a
residual non-decaying effect,
effectively suppressed by
the amplitude of three-point correlations
on a squeezed configuration between
long-scale $k$ and short-scale $aH$ (the current horizon scale).
In this scenario, the tail can interpolate between a Gaussian and exponential,
depending on the relative amplitude of the non-decaying part of the noise.
It would be very interesting to evaluate the correct noise amplitude
for this non-Gaussian, non-decaying component,
and hence the contribution to the tail of $\ProbP(\Nstar)$.

\section{Discussion and conclusions}
\label{sec:conclusions}

In this paper we have introduced the technique of stochastic
instantons
for the calculation of
rare first-passage distributions
(or other extreme value statistics)
in stochastic inflation.
These first-passage statistics
are needed to obtain the
distribution of the inflationary
curvature perturbation
when stochastic effects are significant.
As part of our analysis,
we have emphasized the distinction between
the different transition probabilities
$\ProbP$ and $\ProbPre$,
and the first-passage distribution
$\ProbQ$.
(We caution, however, that whether the first
passage distribution $\ProbQ$ is the appropriate
observable can depend on what is being computed,
as discussed briefly in~\S\ref{sec:Est_tail}
and~\S\ref{sec:decaying-noise}.)
We give a Feynman--Kac-like
formula for $\ProbQ$,
Eq.~\eqref{eq:Feynman-Kac-Q},
which is the basis for our path integral treatment.

We argue that the instanton method has a number
of advantages compared to techniques
presently in use.
Although it involves the technically
sophisticated setting of path integrals,
the instanton equations themselves
are simple.
The method
generalizes easily
to models with multiple fields,
and does not require the slow-roll approximation.
It could also be generalized to account
for a curved field-space or noncanonical kinetic terms.
The instanton gives access to the full range of tail
behaviours, including those involving
strong time dependence where the tail may not be simply exponential.
Further, there is a systematic procedure to
incorporate microphysical
details such as
finite correlation times or scales,
and non-Markovian memory effects.

Our framework is based on
the Martin--Siggia--Rose path integral,
or an interpretation
of the Schwinger--Keldysh path integral
in MSR language.
The MSR formalism is a standard tool
used to describe the dynamics
of stochastic systems, including those governed by
Langevin equations, using field-theory methods.
In~\S\ref{sec:instanton-phase}
we explained how
MSR-like actions can be derived from the Schwinger--Keldysh
path integral after evaluation of a
suitable influence
functional.
Indeed,
this is a standard technique in applications
of the Schwinger--Keldysh method to transport
phenomena in condensed matter systems.
From this perspective, the
so-called ``quantum'' fields of the Keldysh basis
act as the response fields.
Meanwhile, the ``classical'' fields act as the
primary field variables.
Taken together, these fields can be reorganized
to produce a set of conjugate variables
for a ``fake'' Fokker--Planck phase space.
It is the Schr\"{o}dinger equation for this ``fake''
phase space that reproduces the Fokker--Planck equation
governing the stochastic evolution.
In this paper, we use the Schwinger--Keldysh
framework, together with an appropriate
influence functional
accounting for degrees of freedom emerging from the
horizon,
to go beyond the overdamped slow-roll limit.
We argue that this provides a framework
in which to address many of the difficulties
encountered when moving beyond slow-roll;
see, e.g., the summary by Vennin \& Wands~\cite{Vennin:2024yzl}.

\para{The instanton approximation}
As in any path integral formulation,
rare events can be studied by evaluating the integral
in a saddle point approximation.
In our terminology, a \emph{stochastic instanton}
is a trajectory corresponding to such a saddle point.
The use of these
instantons to describe rare events
was introduced in the context of turbulent fluid flow
by Falkovitch {\etal}~\cite{Falkovich:1995fa}
and Guarie \& Migdal~\cite{Gurarie:1995qc}.
They have since found application to a large number
of stochastic systems, including
magnetic field reversals,
chemical reactions, option-pricing fluctuations,
genetic switching,
and climate modelling.

The location of the saddle is determined by
finding critical points of the
effective MSR action, whether this is obtained by applying the MSR
construction to a Langevin equation
and building ``up'' to the corresponding
path integral,
or by starting with a Schwinger--Keldysh formulation and working ``down''.
The critical points
are determined by a set of coupled, partial differential equations.
These \emph{instanton equations}
not only describe the evolution of the inflationary
fields,
but also of the conjugate noise variables.
The relevant instanton solutions
obey specific boundary conditions,
and interpolate between the required initial and final field
values
over a specified transition time.
Typically,
it is the boundary condition that the transition
completes in a specified number of e-folds $\Nstar$
that activates support from the noise fields.
We describe such transitions as \emph{noise supported}.

There is a relation between
the instanton
and \emph{Langevin bridges}.
These are constrained stochastic processes
with specified initial and final configurations,
and a fixed transition time,
in a similar way to the instanton.
Such ``bridge'' processes can be derived from the original
Langevin dynamics via a Doob
transformation~\cite{doob1957conditional,2011JChPh.134q4114O,2015JSMTE..06..039M}.%
\footnote{We thank Vincent Vennin for drawing
our attention to this possibility.} 
This formulation was applied to stochastic
inflation by Tokeshi \& Vennin~\cite{Tokeshi:2023swe}.
A very similar construction
(but not identical)
was studied by Aguilar {\etal}~\cite{PhysRevE.105.064138},
who found that
solutions to the bridge reduced to the
instanton in the limit of weak noise.
They argued that important information
was encoded in fluctuations around the instanton,
which is captured by realizations of the bridge.
(In the saddle point approximation to the path integral,
information about Gaussian fluctuations
around the instanton is
encoded in the fluctuation
determinant, cf. Eq.~\eqref{eq:instanton-approximation}.
However, fluctuations in the bridge process are not limited
to a Gaussian approximation.)
It would be useful to develop this connection further.

Access to the time history
supplied by the instanton
trajectory is very useful.
It automatically gives information about the
least-unlikely noise realization needed
to support the transition,
such as Eqs.~\eqref{eq:SR-instanton-soln-P}
and~\eqref{eq:decaying-noise-pa-soln}.
Access to
these details
is important to understand the
spacetime history of the transitions that populate the
rare tail of $\ProbQ$.
This information is not easily
accessible in other formalisms,
although as explained in~\S\ref{sec:tomberg-langevin-method},
it may be possible to partially reconstruct it.
As another example, it was already noted
by Tomberg that the most probable
transition trajectory
can be used as a bias in importance
sampling~\cite{Tomberg:2022mkt,Tomberg:2023kli}.
This is a technique
to accelerate accurate reconstruction of the tail
in numerical simulations~\cite{Jackson:2022unc,Jackson:2024aoo}.
The instanton solution can be used for
exactly this purpose.
Further details
can be found
in the papers by Ebener {\etal}~\cite{ebener2019instanton}
and Bouchet, Rolland \& Wouters~\cite{10.1063/1.5120509}.
See also the review of
stochastic instantons
by Grafke \& Vanden-Eijnden~\cite{Grafke_2019}.

In this paper, in order to make simple
statements, we have generally presented
tail estimates
as a
series expansion for
$\ln \ProbQ$, essentially an asymptotic Laurent
expansion in $1/\Nstar$.
However, it should not be thought that this is a
limitation of the method.
Indeed, for physical applications such as PBH
abundance calculations, the asymptotic
region may be inadequate.
For example, if PBHs form at a density
contrast of order unity, when $\zeta_k \sim \Delta N \sim 1$,
then we wish to know the probability
distribution in the ``transition'' region
between the centre (where $|\zeta_k| \ll 1$)
and the asymptotic tails (where $|\zeta_k| \gg 1$).
For this purpose, it is important that
the instanton approximation is
not simply a series expansion in the
sense of giving the leading terms in
$\ln \ProbQ$ when $|\Nstar| \gg 1$.
In particular, the full $\Nstar$
dependence of the leading term is trustable.
An example is the complicated $\Nstar$ dependence
of Eq.~\eqref{eq:Prob_CR}
or
Eq.~\eqref{eq:P_CR_Tomberg_im},
which we have shown to match
known exact results.
In particular, the $\Nstar$
dependence of the leading term
does not need
to be used only in the asymptotic limit,
or Kramers regime,
$\Nstar \gg 1$ (cf. Eq.~\eqref{eq:eigenvalue-tail}).
On the other hand,
the leading correction to the saddle point
approximation is expected to be of
order $\Or(\Delta N^{-1})$.
This correction is not yet known, but
should be computed,
possibly in addition to higher ones,
either to
confirm that they can be neglected,
or to include their contribution
in a matching calculation near $\Delta N \sim 1$.

\para{Applications of the method}
In~\S\ref{sec:Apps}
we demonstrate that the instanton method can be used to
recover a number
of results that have already been reported in the literature.
In some cases these have assumed the slow-roll
approximation,
which is not required in the instanton framework.
We find exact or very close agreement.
We are also able to confirm
some of our results by comparison with
the mathematical literature.
This includes examples of
constant-roll inflation, which
we show to be closely related to the famous Ornstein--Uhlenbeck
stochastic process.
We show that the instanton approximation
is exactly able to reproduce the known transition
probability for this process.
Unfortunately, converting this to an estimate
for the first passage distribution
is a very challenging undertaking.
Some aspects of it
apparently
remain open problems
in the mathematical literature.

In~\S\ref{sec:decaying-noise}
we are able to study a problem with strongly time-dependent
noise,
modelling the effect of inflationary
perturbations on a spacetime region with fixed
comoving volume
\emph{after} horizon exit.
In this case use of the instanton
method is essential,
because
it does not appear possible to
obtain a tail estimate using
any existing technique.
We find the tail generated
by this decaying noise is light and
Gaussian, which is consistent with
prior expectations and numerical
results reported by, e.g., Figueroa {\etal}~\cite{Figueroa:2021zah}.
Therefore, in a first approximation, a heavy tail does not form
outside the horizon.
As explained in~\S\ref{sec:decaying-noise}, this does not mean that
a heavy tail does not form at all.
Rather, the fluctuation interior to the region
is composed of an ordinary Gaussian perturbation generated
at horizon exit, superposed on a exponential-tailed
stochastic contribution
generated at earlier times.

Models capable of producing large stochastic effects,
through an enhanced amplitude of the power spectrum,
typically also generate significant non-Gaussianity.
In such scenarios
recent progress on one-loop backreaction has shown
that we should actually expect both decaying \emph{and} non-decaying
contributions outside the horizon~\cite{Iacconi:2023ggt}.
The decaying noise has larger amplitude, but soon
becomes irrelevant.
Meanwhile, the non-decaying noise would be
mediated by non-vanishing three-point correlations
that evade the central limit theorem.
It typically has smaller amplitude, but acts for longer.
The balance between these two contributions
will govern the
detailed properties of the tail formed outside the horizon.
The ability to import
detailed microphysical information
using the Feynman--Vernon influence function
is then potentially very useful.
In principle, it seems that a one-loop calculation of
the influence functional would enable us to
capture both of these effects.
There is clearly value in understanding
whether existing methods used
to calculation correlation functions
at one-loop can be generalized to the influence functional.

An attractive feature of the instanton is that it gives
a clear intuition for the weight of the tail.
In a model such as the linear potential,
the noise realization is given by Eq.~\eqref{eq:SR-instanton-soln-P}
or~\eqref{eq:sol_p_sr}.
The key feature is that it is a constant,
independent of $\Nstar$.
The fixed ``cost'' of realizing this noise,
summed over $\Nstar$ e-folds,
produces the linear $\Nstar$ dependence of the exponential tail.
In contrast, in a model with decaying exponential noise,
the realization is given by Eq.~\eqref{eq:decaying-noise-pa-soln}.
The amplitude now scales with $\Nstar$,
because it switches off quickly outside the
horizon.
Summing this $\Nstar$-dependent ``cost''
over the transition time
produces a quadratic Gaussian tail.

\para{Numerical considerations}
The primary drawback of the method
is that the instanton
satisfies a boundary value problem rather than an initial value
problem.
Where the instanton can be constructed analytically,
as we have done in this paper, this change
introduces very little extra complexity.
On the other hand, if we were to attempt to build
the instanton solution numerically
(perhaps for very precise computations),
this may require specialized methods,
such as a shooting technique.
Such methods can be expensive, due to the need for an iterative
refinement step,
and they may become inefficient in high dimensions where
the parameter space is large.
Further, in limited numerical
experiments we have found that the instanton equations
are frequently stiff.
Some mitigations of this problem
have been explored in the literature.
Strategies for eliminating the future boundary
condition,
meaning that the instanton trajectory can be
solved simply as an initial value problem,
are summarized in the review by
Grafke, Grauer \& Sch\"{a}fer~\cite{Grafke_2015}
and
Grafke \& Vanden-Eijnden~\cite{Grafke_2019}.
These reviews also describe other efficient numerical
algorithms.
It would be extremely
valuable to understand whether these
algorithms can also be applied in the context of
stochastic inflation,
but we leave this for future work.

However, these problems should not be over-emphasized.
The change from initial value problem
to boundary value problem
parallels an analogous difficulty in the spectral
method, where it is necessary to construct eigenfunctions
of a particular differential operator
satisfying prescribed boundary conditions.
One therefore has the same numerical problem.
Although
there is no similar issue in the Tomberg formalism,
this can (currently)
be applied
only to a limited range of models.

\para{Future directions}
Clearly,
our analysis can be expanded in a number of directions.

First, it would be useful to have a
direct saddle point approximation
for the restricted transition probability $\ProbPre$
rather than $\ProbP$.
This would evade the intermediate step of
expressing $\ProbPre$ in terms of $\ProbP$
using the method of images, or the renewal equation.
Determining this relationship
is currently one of the most
important limitations on the
instanton technique.
For example, it is this step that prevented us from
giving an estimate for $\ProbQ$ in the constant roll model
except when $\sigma < 0$ and $\phiend = 0$.
Unfortunately, such a formulation is unlikely to be simple,
because it would have to reproduce the
scaling estimate $\exp(-\lambda(\phiend) |\sigma| \Nstar)$
reported by Martin {\etal}~\cite{Martin_2019}
for $\sigma < 0$.
Their algorithm to compute $\lambda$ already requires
some sophisticated mathematics.
It may be possible to combine the instanton method with
a spectral analysis. A similar strategy was
already suggested by Martin {\etal}~\cite{Martin_2019}.

Second, in this paper we have
generally attempted only to work to exponential
accuracy,
although some of our results go beyond it.
The subexponential prefactor is sometimes
important.
This involves computation of the fluctuation determinant in
Eq.~\eqref{eq:instanton-approximation}.
In this paper, we have not discussed the
details of these computations,
although we have evaluated the determinants in some cases.
For the Gaussian models considered in this paper,
the determinants can be computed exactly
but are always field-independent.
For more general cases the position is not clear.
In applications of stochastic
instantons in other fields
there has been progress in evaluating
such fluctuation determinants
based on generalizations of the Gelfand--Yaglom formula
and matrix Riccati equations.
See, e.g., the discussions given by
Schrolepp, Grafke \& Grauer~\cite{schorlepp2021gel,Schorlepp_2023}
and Bouchet \& Reygner~\cite{bouchet2022path}.
However, these methods are generally restricted
to second-order differential operators, and it is not yet clear
whether they can be adapted to work for the first-order
operators encountered in our formalism.
Another complication,
not discussed in this paper at all,
involves fixing the overall normalization of the tail.
In our analysis this remains undetermined.
It should be obtained by matching to a complementary
(typically non-instanton)
calculation that captures the centre of the distribution.

Third, as explained above,
further microphysical information can
be brought into the calculation by improvements
in the computation of the Feynman--Vernon
influence functional.
These would provide a more accurate statistical
characterization of the noise,
reflecting nonzero correlation times
and memory effects.
It would be interesting to understand the effect
of these details.
On the one hand, it is 
plausible that the least unlikely noise realization
may be sensitive to such features.
On the other, we have seen in~\S\ref{sec:decaying-noise}
that the tail estimate is rather robust even to
dramatic changes in the time dependence of the noise.
It would be interesting to understand how far the
structure of the tail depends on a detailed
understanding of the correlation properties
of the noise.

Fourth, although we explain how the
flux formula~\eqref{eq:first-passage-probability-current-higher-dim}
and
Feynman--Kac
formula~\eqref{eq:Feynman-Kac-Q}
generalize to multiple field models, we have not considered
such models in this paper.
An example was considered by Ach\'{u}carro \etal~\cite{Achucarro:2021pdh}.
It seems possible that
first-passage problems
in the multiple-field context
can be addressed more easily
in the instanton framework than in the context
of the forward Kolmogorov equation~\eqref{eq:kolmogorov-forwards}.
Further,
in such models we might expect that there would
be more than one saddle point,
and hence more than one instanton solution.
There are known to be interesting critical phenomena
where the dominant contribution to the path integral
switches between these saddles;
see, e.g., the discussion in
the book by Kamenev~\cite{Kamenev_2011}.
The use of general complex contours
in the path integral has been discussed
by Keski-Vakkuri \& Kraus~\cite{Keski-Vakkuri:1996lbi},
Bramberger, Lavrelashvili \& Lehners~\cite{Bramberger:2016yog},
and Feldbrugge, Lehners \& Turok~\cite{Feldbrugge:2017kzv}.
In Ref.~\cite{Feldbrugge:2017kzv}
use was made of the concepts of Picard--Lefshetz
theory and Lefshetz thimbles.
These can be used to define the integration
contour in the complexified field space.
Such techniques are useful in other contexts, 
including lattice QCD, where they are used to deal with rapidly oscillating integrals
(the so--called ``sign problem''),
and also in resurgence theory.
While these methods were not employed in our analysis,
they may become relevant in future studies.

Fifth,
the instanton method
can in principle be used to incorporate spatial information.
For instance, the Starobinsky--Langevin
equation~\eqref{eq:Starobinsky-Langevin}
assumes that each horizon volume evolves independently,
neglecting correlations both in their initial conditions
and in the noise fluctuations experienced by different volumes~\cite{SalopekBond1991}.
See Refs.~\cite{Animali:2024jiz,Briaud:2025ayt}
for recent advances in these areas.
It also omits gradient couplings between nearby
patches~\cite{SalopekBond1990,Rigopoulos:2005xx}.
In certain circumstances these
can play an important role.
For example, if an extreme fluctuation arises in a single patch or
a small cluster of nearby patches,
it seems plausible that
gradient interactions will induce a ``pull-back''
effect,
as seen in numerical simulations by Clough \etal~\cite{Clough:2016ymm}
and also Caravano \etal~\cite{Caravano:2024tlp,Caravano:2024moy,Caravano:2025diq}.
This would suppress the probability of such events.
Very recently,
the effect was studied analytically by Briaud~{\etal}~\cite{Briaud:2025ayt}
using a higher-dimensional Langevin framework.
The instanton approach offers a complementary tool.
Focusing more narrowly on the PBH applications,
one could use spatial information
to write a path integral for the compaction function,
and apply an instanton approximation to it directly.
This would avoid the need to go through the $\zeta$
distribution as an intermediate step.

In addition, to relate the first passage distribution
$\ProbQ(\Nstar)$ to $\zeta$,
something must be known about the typical expansion history
experienced in a larger volume,
cf. Eq.~\eqref{eq:prob-zeta-estimate}.
This enables us to define the curvature
perturbation in each
small patch via $\zeta = \Nstar - \langle \Nstar \rangle$.
In this paper we have assumed
$\langle \Nstar \rangle = \NstarDet$.
Although this seems reasonable,
it would be worth clarifying that
spatial correlations in the large volume do not
appreciably change $\langle \Nstar \rangle$
when we condition on the presence of large fluctuations
in at least one patch.
We defer all these interesting questions for further work.

\acknowledgments

We would like to thank
S\'{e}bastien Renaux-Petel,
Eemeli Tomberg and Vincent Vennin
for helpful discussions, and comments on a draft version of this paper.
JCF and DS
are funded by the UK Science and Technology Facilities Council (STFC)
under grant number ST/X001040/1.

\appendix

\section{Influence functional for stochastic inflation}\label{sec:Inf_func}

In this appendix, we briefly review the derivation of the Feynman--Vernon influence functional underlying the standard noise correlations in Starobinsky's formulation of stochastic inflation. Beyond its well-established merits, the stochastic framework serves here as a useful benchmark for comparison with existing treatments of rare event probabilities in the literature. Our presentation follows, in part, the approach of Andersen \etal~\cite{Andersen:2021lii}, though the essential structure of the derivation—particularly the results most relevant to our discussion—can be traced back to the early work of Morikawa~\cite{Morikawa:1989xz}. Subsequent developments and refinements appear in, e.g., Refs.~\cite{Matarrese:2003ye,Lombardo:2004fr,Moss:2016uix,Pinol:2020cdp}.

To derive the influence functional, we work in the Schwinger--Keldysh formalism, which allows to track the so-called ``response'' fields within the effective action. As discussed in the main text, these fields can be identified with the canonical momenta appearing in the Fokker--Planck Hamiltonian formalism. This formulation not only recovers the standard overdamped stochastic dynamics, but also naturally generalizes to a full phase space description valid beyond the slow-roll regime. In addition, it provides a systematic framework for incorporating at a fundamental level further physical ingredients such as dissipation, memory effects, etc.

The generating functional in the closed-time-path (CTP) formalism is written as%
    \footnote{Note that we are omitting the terms that specify the initial quantum state, as well as those that enforce the matching of the field configurations on the forward and backward branches at the turning point of the CTP contour.}
\begin{equation}
    Z[\mathbb{J}] = \int [\d \Phi] \exp \left[\im \int_x \left( \frac{1}{2}\Phi^T \mathbb{G}^{-1} \Phi - V(\phi) + \mathbb{J}^T \Phi \right)\right]\,,
\end{equation}
where $\Phi = (\phi_+, \phi_-)^T$ contains the fields on the forward and backward branches of the CTP contour, $\mathbb{G}^{-1}$ is the free inverse propagator, $V(\phi) = V(\phi_+)-V(\phi_-)$ is the potential, and $\mathbb{J} = (J_+, -J_-)^T$ are the sources. The shorthand $\int_x \equiv \int \d^4x \sqrt{-g}$ is used throughout.

We now switch to the Keldysh basis, introducing the `classical' and `quantum' (or response) fields,
\begin{equation}
    \cl{\phi} = \frac{1}{2}(\phi_+ + \phi_-), \qquad \qu{\phi} = \phi_+ - \phi_-\,,
\end{equation}
in which the propagator becomes
\begin{equation}
    \mathbb{G}(x,x') = \begin{bmatrix}
        -\im G_F(x,x') & G_R(x,x') \\
        G_A(x,x') & 0
    \end{bmatrix}\,,
\end{equation}
with the statistical function and the retarded propagator respectively given by
\begin{align}
    G_F(x,x') = \frac{1}{2} \langle \{ \hat{\phi}(x), \hat{\phi}(x') \} \rangle\,, \qquad 
    G_R(x,x') = -\im \Theta(x^0 - x'^0) \langle [ \hat{\phi}(x), \hat{\phi}(x') ] \rangle\,, 
\end{align}
whereas the advanced propagator is determined by $G_A(x,x') = G_R(x',x)$.

Following the philosophy of the stochastic approach, we isolate the long-wavelength dynamics by decomposing the field into UV and IR components using a window function $W$ ($\overline{W}$) that projects onto the UV (IR), such that
\begin{equation}
    \phi^{_\mathrm{UV}}(x) = \int_{x'} W(x,x') \phi(x')\,, \qquad \phi^{_\mathrm{IR}}(x) = \int_{x'} \overline{W}(x,x') \phi(x')\,,
\end{equation}
with $W(x,x') + \overline{W}(x,x') = \delta(x,x')/\sqrt{-g}$. The window functions are assumed to be time-local, i.e., 
\begin{equation}
W(x, x') = \frac{\delta(x^0 - x'^0)}{\sqrt{-g}} \int \frac{\d^3 k}{(2\pi)^3} W_k(x^0) e^{\im \mathbf{k} \cdot (\mathbf{x} - \mathbf{x}')}\,,
\end{equation}
and similarly for $\overline{W}(x,x')$. In general, it is expected that $W_k(x^0) \approx 1$ for $k \gg aH$ (UV modes) and $\overline{W}_k(x^0) \approx 1$ for $k \ll aH$ (IR modes). This behaviour is usually implemented through a step function in Fourier space, making $W$ and $\overline{W}$ orthogonal projection operators. However, it has been argued that a more careful treatment that recovers the expected long-distance correlations of the field time derivatives requires the use of smooth window functions, which no longer renders them orthogonal. For a more in depth discussion, see e.g., Refs.~\cite{Andersen:2021lii,Moss:2016uix}.

Integrating out the UV modes leads to an effective generating functional,
\begin{equation}
    Z[\mathbb{J}] = \int [\d \Phi_{\mathrm{IR}}] \, e^{\im S[\Phi_{\mathrm{IR}}] + \im \mathbb{J}^T \Phi_{\mathrm{IR}}} \, \mathcal{F}[\Phi_{\mathrm{IR}}]\,,
\end{equation}
where $\mathcal{F}$ is the influence functional, which captures the effect of the environment (the UV modes) on the system (the IR modes). At leading order, we neglect the mixed potential term $V(\Phi_{\mathrm{UV}}, \Phi_{\mathrm{IR}})$, even though it provides the full system-environment coupling.

Setting the source to zero, the influence functional becomes
\begin{equation}
    \mathcal{F}_0[\Phi_{\mathrm{IR}}] = \int [\d \Phi_{\mathrm{UV}}] \exp\left(\im \int_x \left[ \frac{1}{2} \Phi^T_{\mathrm{UV}} \mathbb{G}^{-1} \Phi_{\mathrm{UV}} + \Phi^T_{\mathrm{IR}} \mathbb{G}^{-1} \Phi_{\mathrm{UV}} \right] \right)\;.
\end{equation}
This is a Gaussian path integral of the type
$$
\int \d^n x \, e^{-\frac{1}{2} x^T A x + b^T x} \propto \exp\left( \frac{1}{2} b^T A^{-1} b \right)\;.
$$
Identifying $A = -\im \mathbb{G}^{-1}$ and $b^T = \im \Phi_{\mathrm{IR}}^T \mathbb{G}^{-1}$, we obtain
\begin{equation}
    \mathcal{F}_0[\Phi_{\mathrm{IR}}] = C \exp\left( -\frac{\im}{2} \int_{x,x'} \Phi_{\mathrm{IR}}^T(x) \, \mathbb{G}^{-1}(x) \, \mathbb{G}_{\mathrm{UV}}(x,x') \, \mathbb{G}^{-1}(x') \, \Phi_{\mathrm{IR}}(x') \right)\,,
\end{equation}
where $C$ is a normalization factor. The UV-filtered propagator is defined by
\begin{equation}
    G_{\mathrm{UV}}(x,x') = \int_{y, y'} W(x,y) \, G(y,y') \, W(y',x')\;.
\end{equation}
Assuming spatial translational invariance, this admits a Fourier representation,
\begin{equation}
    G_{\mathrm{UV}}(x,x') = \int \frac{\d^3 k}{(2\pi)^3} W_k(t) \, G_k(t,t') \, W_k(t') \, e^{\im \mathbf{k} \cdot (\mathbf{x} - \mathbf{x}')}\;.
\end{equation}
Acting with $G^{-1}(x)$ on $G_{\mathrm{UV}}$ gives
\begin{equation}
    G^{-1}(x) G_{\mathrm{UV}}(x,x') = \int \frac{\d^3 k}{(2\pi)^3} e^{\im \mathbf{k} \cdot (\mathbf{x} - \mathbf{x}')} \left[ \mathcal{Q}_t(k) + W_k(t) G^{-1}_t \right] G_k(t,t') W_k(t')\,,
\end{equation}
where
\begin{equation}
    \mathcal{Q}_t(k) = -\left[ \ddot{W}_k(t) + 3 H \dot{W}_k(t) + 2 \dot{W}_k(t) \partial_t \right]\;.
\end{equation}
In e-fold time, this is equivalent to 
\begin{equation}
    \mathcal{Q}_N(k) = -H^2 \left[ W_k''(N) + (3 - \epsilon) W_k'(N) + 2 W_k'(N) \partial_N \right]\;.
\end{equation}

Integrating by parts, a similar expression arises from acting with $G^{-1}(x')$ on the right. At leading order, we neglect terms where $W_k$ acts directly on IR fields, retaining in this way only the $\mathcal{Q}_t \mathcal{Q}_{t'}$ (or $\mathcal{Q}_N \mathcal{Q}_{N'}$) contributions. Then, the influence functional in the Keldysh basis becomes
\begin{align}\label{eq:F0}
    \mathcal{F}_0[\Phi_{\mathrm{IR}}] = C \exp \Big( \im \int_x \int_{x'} \Big[ \frac{\im}{2} \pq(x) \, & \mathrm{Re}[\Pi(x,x')] \, \pq(x') \nonumber \\
    & - \Theta(x^0 - x'^0) \, \pq(x) \, \mathrm{Im}[\Pi(x,x')] \, \pc(x') \Big] \Big)\,,
\end{align}
where the self-energy kernel is given by
\begin{equation}
    \Pi(x,x') = \int \frac{\d^3 k}{(2\pi)^3} e^{\im \mathbf{k} \cdot (\mathbf{x} - \mathbf{x}')} \, \mathcal{Q}_{x^0}(k) \mathcal{Q}_{x'^0}(k) \, \phi_k(x^0) \phi_k^*(x'^0)\;.
\end{equation}
Here, as a leading-order approximation, we take $\phi_k(x^0)$ as the Bunch--Davies solutions of the Mukhanov--Sasaki equation.

The term proportional to $(\pq)^2$ encodes the statistical properties of the noise exerted on the system, while the term proportional to $\pq \pc$ is associated to  dissipative dynamics. In equilibrium, these are related via the fluctuation--dissipation theorem. At this level of approximation, taking a step function as the window function, the dissipation term vanishes and only the noise survives. Nonetheless, dissipative effects are expected and have been studied in this context since the early quantum field theoretic treatments of stochastic inflation~\cite{Morikawa:1989xz}.

Next, we shall focus our discussion for $x^0 = N$, but the case $x^0 = t$ is completely analogous. First, in order to obtain a phase space description, where the noise acting along both directions can be identified, we use that
\begin{equation}
    -\frac{a^3_N}{H_N} {\cal Q}_N \phi_k (N) = \partial_N [a^3_N H_N W_k'(N) \phi_k(N)] + a^3_N H_N W_k'(N) \phi_k'(N)\,,
\end{equation}
to then integrate by parts in the first term (with those multiplying it), therefore picking up a time derivative of $\pq$. Then, using $\ppq = (\pq)'$, we obtain 
\begin{align}
   \int_{x,x'} \pq(x) {\rm Re}[\Pi(x,x')]& \pq(x')  \nonumber \\
   & = \int_{x,x'}  ( -\ppq (x)\,, \pq(x) ) {\cal M}(x,x') (-\ppq (x') \,, \pq(x'))^T\,,
\end{align}
where the matrix encoding the noise correlations is given by
\begin{align}\label{eq:M2}
 {\cal M}(x,x') & = {\rm Re} \int \frac{\d^3 k}{(2\pi)^3} e^{\im \vt{k}\cdot (\vt{x}-\vt{x'})} H^2_N W_N' \left[\begin{array}{cc}
 \phi_k (N) \phi_k^* (N') & \phi_k (N) \pi_k^* (N') \\
\pi_k (N) \phi_k^* (N') & \pi_k (N) \pi_k^* (N')
\end{array}\right] H^2_{N'} W'_{N'}  \,, 
\end{align}
where we have denoted $\pi_k(N) = \phi_k' (N)$.

This has already determined the influence functional in phase space. Since there should no longer be opportunities
for confusion, we will drop the `IR' superscript, which will be implicit on the coarse grained fields. 

\subsection{Langevin-like equations}\label{sec:Lang_app}

Even though it is not strictly necessary to write down the emerging Langevin equations from our calculations above, which were meant to recover the well-known results of the stochastic picture, it is instructive to derive them in order to appreciate the relation between the `quantum' components of the field and their relation to the noise in the stochastic description. A crucial part for this is the so-called Hubbard--Stratonovich transformation
\begin{align*}
    \exp\bigg(-\frac{1}{2}\int_{x, x'} \qu{\phi} (x) & M(x,x') \qu{\phi} (x') \bigg) \nonumber \\
    & = \int [\d \xi] \exp\bigg( -\frac{1}{2} \int_{x,x'} \xi (x) M^{-1} (x,x') \xi (x') + \im \int_x \xi (x) \qu{\phi} (x)\bigg) \,,
\end{align*}
where $\xi(x)$ denotes the noise field, with correlation given by $\langle \xi(x) \xi(x') \rangle = M(x,x')$. The phase space version of this expression leads to the influence functional in terms of the noise fields
\begin{align}
    \FeynmanVernon_0 & [\Phi_{\rm IR}] = C \int [\d \xi_{\phi}\, \d\xi_{\pi}] \exp \bigg( \int_{x,x'} \Big[ -\frac{1}{2} [\xi_{\phi}(x)\,, \xi_{\pi}(x)] {\cal M}^{-1} (x,x') [\xi_{\phi}(x')\,,\xi_{\pi}(x')]^T \nonumber \\
    & - \im \Theta(x^0-x'^0) \qu{\phi}(x) {\rm Im}[\Pi(x,x')] \cl{\phi}(x') \Big] + \im \int_x [-\qu{\pi}(x)\,, \cl{\phi}(x)][\xi_{\phi}(x)\,,\xi_{\pi}(x)]^T \bigg)\;.
\end{align}
Then, obtaining the Langevin equations is straightforward. For this, we focus on the effective action containing up to linear order terms in $\qu{\phi}$ and $\qu{\pi}$,
\begin{align}
    S \supset & \int_x H^2 \Big[ \cl{\pi} (\qu{\phi})' + \qu{\pi} (\cl{\phi})' - \cl{\pi} \qu{\pi} - (aH)^{-2} (\nabla \cl{\phi} ) \cdot (\nabla \qu{\phi}) - \qu{\phi} V'(\cl{\phi})  \nonumber \\\
    & - H^{-2}\qu{\pi} \xi_{\phi} + H^{-2} \qu{\phi} \xi_{\pi} \Big] - \int_{x,x'} \Theta(x^0 - x'^0) \qu{\phi} (x) {\rm Im}[\Pi(x,x')]\cl{\phi} (x')\,,
\end{align}
where we have used that, in the Keldysh basis, the potential can be expanded as
\begin{equation}\label{eq:Vcq}
    V(\phi) = \qu{\phi} \frac{\partial V(\cl{\phi})}{\partial\cl{\phi}} + \sum_{m = 1}^{\infty} \frac{V^{(2m+1)}(\cl{\phi})}{2^m (2m+1)!} (\qu{\phi})^{2m+1}\;.
\end{equation}
Finally, the Langevin equations for the `classical' fields are found by varying the action with respect to the `quantum' components, which yields
\begin{subequations}
\begin{align}
    \left.\frac{\delta S}{\delta \qu{\pi}}\right|_{\qu{\pi} = 0} = 0 \qquad & \implies \qquad \cl{\phi}' = \cl{\pi} + \frac{\xi_{\phi}}{H^2}\,, \label{Eq:Lng1}\\
    \left.\frac{\delta S}{\delta \qu{\phi}}\right|_{\qu{\phi} = 0} = 0 \qquad & \implies \qquad \cl{\pi}' + (3-\epsilon) \cl{\pi} - \frac{\nabla^2 \cl{\phi}}{a^2 H^2} + \frac{V'(\cl{\phi})}{H^2} = \frac{\xi_{\pi}}{H^2}\,, \label{Eq:Lng2}
\end{align}
\end{subequations}
where we have omitted the dissipative term in the last equation due to its subleading contribution. On the other hand, notice that the awkward appearance of $H^{-2}$ with the noise terms is a result of our choice of variables together with the Jacobian terms present throughout the process. These can be reabsorbed by defining rescaled noise fields, rendering the equations consistent with the standard stochastic formalism. This rescaling is effectively applied in Eq.~\eqref{eq:constant-roll-evolution}. 

\subsection{Noise correlations}

To obtain explicit expressions for the noise correlations, we must specify a window function that separates the UV and IR sectors. We choose a step function of the form
\begin{equation}
    W_k (N) = \Theta(k - \mu aH)\,,
\end{equation}
where $\mu \ll 1$ is a small, dimensionless parameter that defines the coarse-graining scale, effectively delimiting system (IR) and environment (UV) degrees of freedom.

With this choice, the integrals in the noise kernel ${\cal M}(x,x')$, Eq.~\eqref{eq:M2},  can be evaluated explicitly. The result is given by
\begin{equation}\label{eq:M21}
    {\cal M}_{ij}(x,x')
    =
    \Big(
        1 - \epsilon(N)
    \Big)
    \,
    H_N^2 H_{N'}^2
    \,
    \frac{
        \sin(k_\mu |\vt{x} - \vt{x}'|)
    }{
        k_\mu |\vt{x} - \vt{x}'|
    }
    \,
    \dimP_{ij}(k_{\mu})
    \,
    \delta(N - N')
    \,
    ,
\end{equation}
where $k_\mu \equiv \mu aH$, and the indices $i,j \in \{\phi, \pi\}$ (or $\{1,2\}$) label the entries of the noise matrix, corresponding to the fields and their momentum variables. The matrix $\dimP_{ij}(k_\mu)$ encodes the power spectra evaluated at the coarse-graining scale,
\begin{equation}\label{eq:PS_matrix}
    \dimP_{ij}(k_\mu) \approx \left(\frac{H}{2\pi}\right)^2 \mu^{3 - 2\nu}
    \begin{bmatrix}
        1 & \nu - \tfrac{3}{2} \\
        \nu - \tfrac{3}{2} & (\nu - \tfrac{3}{2})^2
    \end{bmatrix} \equiv 2\D_{ij}\;.
\end{equation}

This expression follows from the asymptotic form of the mode functions in the superhorizon regime
\begin{equation}
    \phi_k = \frac{H}{\sqrt{2k^3}} \left( \frac{k}{aH} \right)^{3/2 - \nu}\,, \qquad
    \pi_k = \left( \nu - \tfrac{3}{2} \right) \delta\phi_k\,,
\end{equation}
where $\nu$ is the Hankel index characterizing solutions of the Mukhanov--Sasaki equation. Expressed in terms of the slow-roll parameters, $\nu$ satisfies
\begin{equation}\label{eq:Hankel_index}
    \nu^2 = \frac{9}{4} + \frac{3}{2} \epsilon_2 + \frac{1}{4} \epsilon_2^2 + \Or(\epsilon_1)\,,
\end{equation}
where $\epsilon_1 \equiv \epsilon$, and $\epsilon_{i+1} \equiv \d \ln \epsilon_i / \d N$. We neglect $\Or(\epsilon_1)$ corrections under the assumption of quasi-de Sitter evolution, with $H \sim \text{const.}$, and $\epsilon_1$ negligible. However, we leave open the possibility that $\epsilon_2$ may remain sizeable, as required in USR and CR inflationary models.

Finally, the correlations of the stochastic forces that appear in the Langevin equations~\eqref{Eq:Lng1} and~\eqref{Eq:Lng2} are directly determined by the entries of ${\cal P}_{ij}$. Explicitly, we find
\begin{align}
    \left\langle \frac{\xi_i (x)}{H^2}, \frac{\xi_j (x')}{H^2} \right\rangle & = (1 - \epsilon_1)\, \frac{\sin(k_\mu |\vt{x} - \vt{x}'|)}{k_\mu |\vt{x} - \vt{x}'|} \, \dimP_{ij}(k_\mu)\, \delta(N - N')\,, 
\end{align}
which, together with Eq.~\eqref{eq:PS_matrix}, recovers the standard result in the literature.

\section{Renewal equations and the Laplace transform}\label{sec:FPT_LT}
The discussion in the main text
characterized the first-passage probability density
$\ProbQ$
in terms of the survival probability
$\Survival$.
In this Appendix we consider an alternative characterization
in terms of \emph{renewal equations}.

\subsection{The renewal equation}\label{sec:App_ren_eq}
It sometimes happens that we have a solution for a transition
probability $\ProbP$ computed using
boundary conditions that
are incompatible with $\Survival$,
as used in~\eqref{eq:survival-probability-def}.
In certain circumstances, it may still be possible to determine
the first-passage distribution $\ProbQ$
from $\ProbP$
via a
\emph{renewal equation}.
The following presentation is based on Balakrishnan~\cite{Balakrishnan2021}.

Consider the transition probability from $\phiinit$ to some other value
$\phistop$,
taken to occur between times $\Ninit$ and $\Nstop$.
As in the main text,
we continue to assume the field rolls from right to left, and pick
an arbitrary intermediate value $\phi$ satisfying
$\phistop < \phi < \phiinit$.
The stochastic process must first cross $\phi$
at some time $N$ satisfying
$\Ninit \leq N \leq \Nstop$.
After this, the process \emph{renews},
or restarts with a new initial condition.
Further, each of these first-passage events is exclusive.
Therefore,
no matter what boundary conditions we choose for $\ProbP$,
we must have
\begin{equation}
    \label{eq:pdf_conv00}
    \ProbP( \phistop, \Nstop \divider \phiinit, \Ninit )
    =
    \int_{\Ninit}^{\Nstop}
    \ProbP( \phistop, \Nstop \divider \phi, N )
    \ProbQ( \phi, N \divider \phiinit, \Ninit )
    \,
    \d N
    .
\end{equation}
This is an example of a renewal equation. Similar equations can be written
for many stochastic processes, and provide an alternative way to characterize
first-passage distributions.

In general, Eq.~\eqref{eq:pdf_conv00}
is very difficult to solve.
However, if the stochastic process is Markovian,
then $\ProbP( \phistop, \Nstop \divider \phi, N )$
depends only on the time difference $\Nstop - N$,
and Eq.~\eqref{eq:pdf_conv00}
has the structure of a convolution.
We express $\ProbP$
as a Bromwich integral,
\begin{equation}
    \ProbP( \phistop, \Nstop \divider \phiinit, \Ninit )
    =
    \int_{\gamma - \im \infty}^{\gamma + \im \infty}
    \frac{\d s}{2\pi \im}
    \ProbP( \phistop, s \divider \phiinit )
    \e{s (\Nstop - \Ninit)}
    \,,
\end{equation}
where $\gamma$ is a real number
chosen so that the integration contour lies to the right
of any singularities
of the Laplace transform $\ProbP( \phistop, s \divider \phiinit )$.
$\ProbQ$ can be given a similar Bromwich representation.
It follows that $\ProbQ( \phi, s \divider \phiinit )$
can be written
\begin{equation}
    \label{eq:Q_LT0}
    \ProbQ( \phi, s \divider \phiinit )
    =
    \frac{\ProbP(\phistop, s \divider \phiinit)}{\ProbP(\phistop, s \divider \phi)} \;.
\end{equation}
For Markovian processes, the dependence on $\phistop$
will cancel on the right-hand side.
For non-Markovian processes, or where the
required Laplace transforms $\ProbP(\phistop, s \divider \phiinit)$
cannot be computed explicitly,
the renewal equation is of limited utility.
However, even in such cases,
Eq.~\eqref{eq:pdf_conv00} remains valid.

\bibliography{refs} 
\bibliographystyle{JHEP}

\end{document}